\pdfoutput=1
\documentclass[%
 reprint,
superscriptaddress,
footinbib,
nobibnotes,
 amsmath,amssymb,
 aps,
 prx,
floatfix
]{revtex4-2}

\usepackage{graphicx}
\usepackage{dcolumn}
\usepackage{bm}
\usepackage{amsthm, ascmac, amsfonts, mathrsfs, physics, mathtools, enumerate}
\usepackage{natbib}
\usepackage[resetlabels]{multibib}
\newcites{SM}{References}
\usepackage[%
setpagesize=false,%
bookmarksnumbered=true,%
bookmarksopen=true,%
colorlinks=true,%
linkcolor=magenta,%
citecolor=blue,%
]{hyperref}
\hypersetup{pdfauthor={Takanao Ishii and Masahito Ueda},pdftitle={Quantum i.i.d. Steady States in Open Many-Body Systems}}

\renewcommand\paragraph[1]{%
\par\emph{#1---}\kern2pt\relax\ignorespaces}

\theoremstyle{definition}
\newtheorem{theorem}{Theorem}

\newtheorem{lemma}[theorem]{Lemma}

\newtheorem{corollary}[theorem]{Corollary}
\newtheorem{example}{Example}
\newcommand{\Z}{\mathbb{Z}}
\newcommand{\R}{\mathbb{R}}
\newcommand{\C}{\mathbb{C}}
\newcommand{\mymat}[1]{\begin{pmatrix} #1 \end{pmatrix}}

\makeatletter
\def\frontmatter@maketitle{%
  \@author@finish
  \title@column\titleblock@produce
  \suppressfloats[t]%
  \let\abstract\@undefined\let\endabstract\@undefined
  \titlepage@sw{%
   \vfil
   \clearpage
  }{}%
  \onecolumn@grid@setup
  \def\set@footnotewidth{\set@footnotewidth@one}%
}%
\makeatother

\begin{document}

\preprint{APS/123-QED}

\title{Quantum i.i.d. Steady States in Open Many-Body Systems}

\author{Takanao Ishii}
\email{ishii@cat.phys.s.u-tokyo.ac.jp}
\affiliation{Department of Physics, The University of Tokyo, 7-3-1 Hongo, Bunkyo-ku, Tokyo 113-8654, Japan
}%


\author{Masahito Ueda}
\affiliation{Department of Physics, The University of Tokyo, 7-3-1 Hongo, Bunkyo-ku, Tokyo 113-8654, Japan
}
\affiliation{Institute for Physics of Intelligence, The University of Tokyo, 7-3-1 Hongo, Bunkyo-ku, Tokyo 113-0033, Japan
}
\affiliation{RIKEN Center for Emergent Matter Science (CEMS), Wako, Saitama 351-0198, Japan
}

\date{\today}

\begin{abstract}
    Understanding how a quantum many-body state is maintained stably as a nonequilibrium steady state is of fundamental and practical importance for exploration and exploitation of open quantum systems. 
    We establish a general equivalent condition for an open quantum many-body system governed by the Gorini-Kossakowski-Sudarshan-Lindblad dynamics under local drive and/or dissipation to have a quantum independent and identically distributed (i.i.d.) steady state. We present a sufficient condition for a system to have a quantum i.i.d. steady state by identifying a set of operators that commute with arbitrary quantum i.i.d. states. In particular, a set of quantum i.i.d. states is found to be an invariant subset of time evolution superoperators for systems that satisfy the sufficient condition. These findings not only identify a class of models with exactly solvable steady states but also lead to a no-go theorem that precludes quantum entanglement and spatial correlations in a broad class of quantum many-body steady states in a dissipative environment.
\end{abstract}

\maketitle

\section{Introduction}\label{sec1}
Recent experimental advances in artificial quantum matter, such as ultracold atoms \cite{RevModPhys.80.885, PhysRevX.8.041054, PhysRevX.8.041055}, trapped ions \cite{PhysRevLett.74.4091, Blatt2008, Barreiro2011nature}, Rydberg atoms \cite{PhysRevLett.85.2208, PhysRevLett.87.037901, Weimer2010nature}, and cavity and circuit QED systems \cite{Hartmann2006nature, PhysRevA.76.031805, PhysRevX.7.011016}, have enabled one to control open quantum many-body systems with unprecedented controllability. Among the recent accomplishments is the experimental realization of a dissipative quantum phase transition (DQPT) \cite{Baumann2010, PhysRevLett.107.140402, doi:10.1073/pnas.1306993110, doi:10.1073/pnas.1417132112, PhysRevLett.113.020408, PhysRevX.7.011016, PhysRevX.7.011012}, which is a dramatic change in the steady state as a function of parameters of a system due to the competition between interaction, external driving, and dissipation \cite{Nagy2008, PhysRevX.5.031028, PhysRevA.97.013825}. Together with these experimental developments, extensive theoretical research has been conducted on open quantum systems in search of novel phenomena. Two fundamental theoretical frameworks for describing open quantum many-body systems are the Keldysh formalism \cite{Keldysh:1964ud} and quantum master equations \cite{nakajima1958quantum, zwanzig1960ensemble, REDFIELD19651, 10.1063/1.522979, 10.1007/BF01608499}. Although the foundations of these frameworks date back to the 1960s and 1970s, their extensive applications to open quantum many-body systems only began in the late 2000s. In the present work, we adopt the latter approach, specifically the Gorini-Kossakowski-Sudarshan-Lindblad (GKSL) formalism, which features some important properties, such as Markovianity, complete positivity, and trace preservation \cite{10.1063/1.522979, 10.1007/BF01608499, breuer2002theory}.

In the present work, we focus on spatial correlations and quantum entanglement in the steady state with particular emphasis on when the steady state becomes uncorrelated or disentangled. Recent studies have shown under some assumptions that the sudden death of entanglement \cite{PhysRevLett.93.140404, PhysRevLett.97.140403, doi:10.1126/science.1167343} occurs in systems subjected solely to local dissipation (without Hamiltonian dynamics) \cite{gong2024any}. However, the conditions under which the steady state of an open quantum many-body system under local dissipation (with Hamiltonian dynamics) exhibits the absence of spatial correlations or quantum entanglement remain largely unexplored.

\begin{figure}
    \centering
    \includegraphics[width=0.8\linewidth]{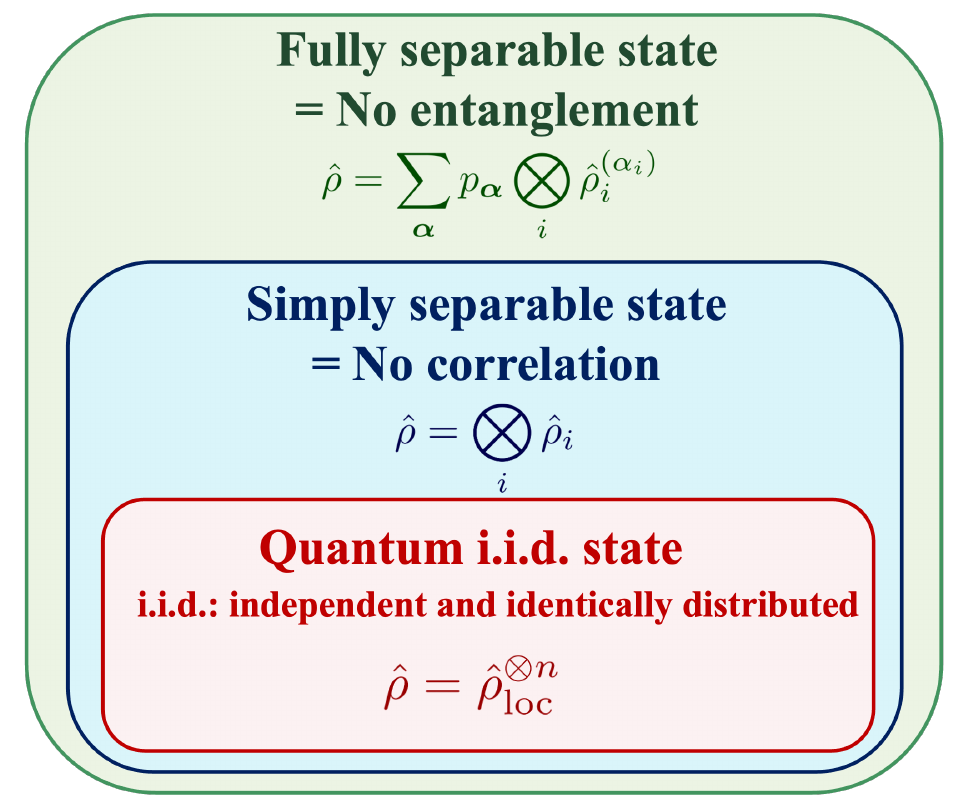}
    \caption{Relationship between three classes of density matrices: fully separable states (Eq.~\eqref{eq1-1}), simply separable states (a.k.a. tensor product states, Eq.~\eqref{eq1-2}), and quantum i.i.d. states (Eq.~\eqref{eq1-3}). In the present paper, we investigate conditions for an open quantum many-body system to have a quantum i.i.d. steady state.}
    \label{fig_classofstates}
\end{figure}

The state without quantum entanglement is equivalent to a fully separable state written as
\begin{align}\label{eq1-1}
    \hat{\rho} = \sum_{\bm{\alpha}}p_{\bm{\alpha}}\qty(\hat{\rho}_{1}^{(\alpha_{1})}\otimes \cdots \hat{\rho}_{n}^{(\alpha_{n})}), \quad \sum_{\bm{\alpha}}p_{\bm{\alpha}} = 1,
\end{align}
and the state without spatial correlations for any physical quantities is equivalent to a simply separable state (a.k.a. a tensor product state) written as
\begin{align}\label{eq1-2}
    \hat{\rho} = \hat{\rho}_{1}\otimes \cdots \otimes \hat{\rho}_{n},
\end{align}
where $\hat{\rho}_{i}^{(\alpha_{i})}$'s and $\hat{\rho}_{i}$'s are density matrices. Since it is highly nontrivial to analyze the condition that the steady state becomes fully or simply separable, we focus on the special class of fully or simply separable state, which is a quantum independent and identically distributed (i.i.d.) state (see Fig.~\ref{fig_classofstates})-- a state whose total density matrix $\hat{\rho}$ is expressed as a tensor product of identical local density matrices $\hat{\rho}_{\rm loc}$:
\begin{align}\label{eq1-3}
    \hat{\rho} = \hat{\rho}_{\rm loc}^{\otimes n}.
\end{align}

In the present paper, we establish a general equivalent condition (see Theorem \ref{thm:equivalent_SS}) and some sufficient conditions (see Lemma \ref{lem:equivalent_SS} and Theorem \ref{thm:sufficient}) for an open quantum many-body system under local drive or dissipation to have a quantum i.i.d. steady state. These theorems provide a practical criterion for determining whether a given open quantum many-body system has a quantum i.i.d. steady state or not. 
They can also be interpreted as a no-go theorem for steady-state quantum entanglement and spatial correlations. While the condition for quantum entanglement and spatial correlations to vanish in the steady state is known for a spin-1/2 XXZ model \cite{PhysRevB.110.155129}, we fundamentally extend this no-go theorem by making it independent of specific details of the models.

The second key contribution of our work is that we provide a class of open quantum many-body systems whose steady states can be obtained exactly.
Exact solutions of simple models that capture the bare essentials of physical systems have significantly deepened our understanding of isolated quantum many-body systems \cite{10.1143/PTP.56.1454, PhysRevB.14.1165, Sachdev_1999, suzuki2012quantum, doi:10.1098/rspa.1963.0204, 10.1143/ptp/5.4.544, 10.1063/1.1704046, 10.1063/1.1704281, KITAEV20062}, and they have also been playing a crucial role in the study of open quantum many-body systems governed by the GKSL dynamics. 
In particular, for DQPTs, which are characterized by a dramatic change in the steady state, obtaining exact steady-state solutions is highly beneficial.
Previous studies have reported exact solutions of steady states, such as spin-1/2 models with boundary drive and dissipation \cite{PhysRevLett.101.105701, Znidaric_2010, PhysRevE.83.011108, PhysRevE.88.062118, PhysRevLett.110.047201, PhysRevLett.107.137201, PhysRevLett.124.160403, PhysRevX.14.021028, popkov2025exactness}, spin-1/2 models with collective dissipation \cite{Stannigel_2012, PhysRevLett.131.190403}, spin-1/2 models with bulk dissipation\cite{PhysRevLett.131.190403, PhysRevLett.113.237203, PhysRevLett.128.033602}, dissipative Hubbard models with two-body loss \cite{PhysRevLett.109.230501, PhysRevA.105.L051302, PhysRevA.107.033332}, and dissipative Hubbard models with boundary dissipation \cite{PhysRevLett.112.030603}. Beyond model-specific results, several systematic analytical methods have established solvable classes of models. For example, the method of third quantization \cite{Prosen_2008} gives exact eigenvalues and eigenvectors of a Lindblad superoperator for quadratic and quasi-free fermionic and bosonic systems. When a system satisfies the Yang-Baxter equation \cite{PhysRevLett.19.1312}, the Bethe ansatz can be utilized to derive analytical expressions of eigenvalues of the Lindblad superoperator \cite{10.21468/SciPostPhys.8.3.044, Buča_2020, PhysRevLett.126.110404, PhysRevLett.126.240403, 10.21468/SciPostPhysCore.8.1.011}. For a system subject to the quantum detailed balance, a systematic approach exists for finding analytical expressions of nontrivial steady states \cite{PRXQuantum.2.020336}. This paper proposes yet another important class of systems for which exact steady states, which are quantum i.i.d. steady states, can be obtained. \\

The rest of this paper is organized as follows. In Section \ref{sec2A}, we present rigorous theorems on open quantum many-body systems that have quantum i.i.d. steady states. In Section \ref{sec2B} we show that the theorem regarding the existence of quantum i.i.d. steady state leads to a no-go theorem for spatial correlations and the quantum entanglement of the steady state. In Section \ref{sec3}, we focus on the dynamical stability of a set of quantum i.i.d. steady states. We show that in some open quantum many-body systems, the density matrix maintains the quantum i.i.d. form throughout the time evolution, if the initial state is a quantum i.i.d. state. We refer to this property as the dynamical stability of a set of quantum i.i.d. steady states. In Section \ref{sec3A}, we derive an equivalent condition under which a system exhibits this dynamical stability. In Section \ref{sec3B}, we demonstrate that dynamical properties, including time-correlation functions and response functions, can be calculated analytically for such a system. In Section \ref{sec4}, we discuss examples of various systems, including spin systems, Fermi systems, and Bose systems to illustrate the theorems derived in the preceding sections. In Section \ref{sec5}, we conclude this paper and present an outlook. Some technical details of the proofs and lemmas are relegated to appendices to avoid digressing from the main subject. In Appendix \ref{sec_SM_A}, we provide a supplementary lemma to Corollary \ref{cor:equivalent_SS_2dim}
 for spin-1/2 systems. In Appendix \ref{sec_SM_B}, we prove Lemma \ref{lem:commutable}. In Appendix \ref{sec_SM_C}, we describe the relationship between the sufficient condition in Theorem \ref{thm:sufficient} and the equivalent condition in Theorem \ref{thm:equivalent_sec3}.
 
\section{Quantum i.i.d. steady state}\label{sec2}
\subsection{Conditions to have a quantum i.i.d. steady state}\label{sec2A}
We consider a lattice model with $n$ sites, each of which has the same dimension of the local Hilbert space. The time evolution of the density matrix $\hat{\rho}(t)$ of the system is governed by the following GKSL equation:
\begin{align}\label{eq2-1}
    \dv{\hat{\rho}(t)}{t} &= \mathcal{L}(\hat{\rho}(t)) = -{\rm i}\left[ 
\hat{H}, \hat{\rho}(t) \right]  + \sum_{k}\mathcal{D}_{\hat{L}_{k}}(\hat{\rho}(t)),
\end{align}
where $\hat{H}$ is the Hamiltonian of the system, $[\bullet, \bullet]$ is a commutator, and $\mathcal{D}_{\hat{L}_{k}}(\bullet)\coloneq\hat{L}_{k}\bullet \hat{L}_{k}^{\dagger}-\frac{1}{2}\left\{\hat{L}_{k}^{\dagger}\hat{L}_{k}, \bullet\right\}$ is the dissipator with $\{\bullet, \bullet\}$ being an anticommutator and $\hat{L}_{k}$'s being Lindblad operators which describe the effects of interactions with an environment. The superoperator $\mathcal{L}$ is referred to as a Lindbladian. The steady state is defined such that the density matrix $\hat{\rho}_{\rm SS}$ satisfies $\mathcal{L}\qty(\hat{\rho}_{\rm SS}) = 0$.
We denote the single-site Hilbert space as $\mathcal{H}_{\rm loc}$ and its dimension as $d$, which we assume to be finite, and the Hilbert space of the system as $\mathcal{H} = \mathcal{H}_{\rm loc}^{\otimes n}$. Let $\Lambda = \{1, 2, \dots, n\}$ and $\Lambda_{2} = \{(i, j)\ |\ 1\leq i < j \leq n \}$ be a set of labels of sites and that of pairs of two sites. 
We assume that dissipation is local, i.e., all Lindblad operators act on single sites. When an operator acts on $n$ sites, which need not be adjacent to each other, we call it $n$-local. We denote a set of labels of the Lindblad operators that act on site $i$ as $\Gamma_{i}$.
The Hamiltonian $\hat{H}$ is the sum of at most 2-body Hermitian operators. Here, we define $\hat{H}_{ij}$ as an irreducible part of the Hamiltonian acting on sites $i$ and $j$, i.e., the term of $\hat{H}$ that acts on sites $i$ and $j$ whose partial trace on either site vanishes. $\hat{H}_{ij}$ is uniquely determined from $\hat{H}$ by the following equation:
\begin{align}\label{eq2-2}
    \hat{H}_{ij}  = \mathcal{T}_{ij}(\hat{H})\coloneq &\frac{{\rm Tr}_{\overline{ij}}[ \hat{H} ]}{d^{n-2}} - \frac{{\rm Tr}_{\overline{i}}[ \hat{H}]}{d^{n-1}}  - \frac{{\rm Tr}_{\overline{j}}[\hat{H}]}{d^{n-1}} + \frac{{\rm Tr}[\hat{H}]}{d^n},
\end{align}
where ${\rm Tr}_{\overline{ij}}$ represents the trace over all the sites other than sites $i$ and $j$, and $\mathcal{T}_{ij}$ is a superoperator that takes $\hat{H}$ as an input and returns $\hat{H}_{ij}$. Let 
$\hat{H}_{i}$ represent the traceless part of the Hamiltonian that acts solely on site $i$; $\hat{H}_{i}$ is uniquely determined from $\hat{H}$ as
\begin{align}\label{eq2-3}
    \hat{H}_{i}  = \mathcal{T}_{i}(\hat{H})&\coloneq \frac{{\rm Tr}_{\overline{i}}[ \hat{H}]}{d^{n-1}} - \frac{{\rm Tr}[\hat{H}]}{d^n},
\end{align}
where $\mathcal{T}_{i}$ is a superoperator that takes $\hat{H}$ as an input and returns $\hat{H}_{i}$. Then, $\hat{H}$ can be expressed in terms of $\hat{H}_{ij}$ and $\hat{H}_{i}$ as
\begin{align}\label{eq2-4}
    \hat{H} = \sum_{(i, j) \in \Lambda_{2}}\hat{H}_{ij} + \sum_{i\in \Lambda}\hat{H}_{i} + d^{-n}{\rm Tr}[\hat{H}].
\end{align}
Note that no restriction is imposed on the pair $(i, j)$. Thus, interactions can be nearest-neighbor, long-ranged, and all-to-all. When these assumptions are satisfied, the GKSL equation in Eq.~\eqref{eq2-1} can be rewritten as
\begin{align}
    \mathcal{L}\qty(\hat{\rho}(t)) &= -{\rm i}\sum_{(i,j) \in \Lambda_{2}}\left[\hat{H}_{ij}, \hat{\rho}(t)\right] + \sum_{i\in \Lambda}\mathcal{L}_{i}\qty(\hat{\rho}(t)),\label{eq2-5-1}\\
    \mathcal{L}_{i}\qty(\bullet) &\coloneq -{\rm i}\left[ \hat{H}_{i}, \bullet \right] + \sum_{\alpha \in \Gamma_{i}}\mathcal{D}_{\hat{L}_{i}^{(\alpha)}}\qty(\bullet),\label{eq2-5-2}
\end{align}
where $\mathcal{L}_{i}$ is a single-site superoperator that acts on site $i$ and $\hat{L}_{i}^{(\alpha)}$'s ($\alpha \in \Gamma_{i}$) are Lindblad operators that act on site $i$. We also define an effective Hamiltonian $\hat{H}_{\rm eff}$ as follows:
\begin{align}\label{eq2-6}
    \hat{H}_{\rm eff} \coloneq \hat{H} - \frac{\rm i}{2}\sum_{i\in \Lambda}\sum_{\alpha \in \Gamma_{i}}\hat{L}_{i}^{(\alpha)\dagger}\hat{L}_{i}^{(\alpha)}.
\end{align}
\begin{figure}
    \centering
    \includegraphics[width=\linewidth]{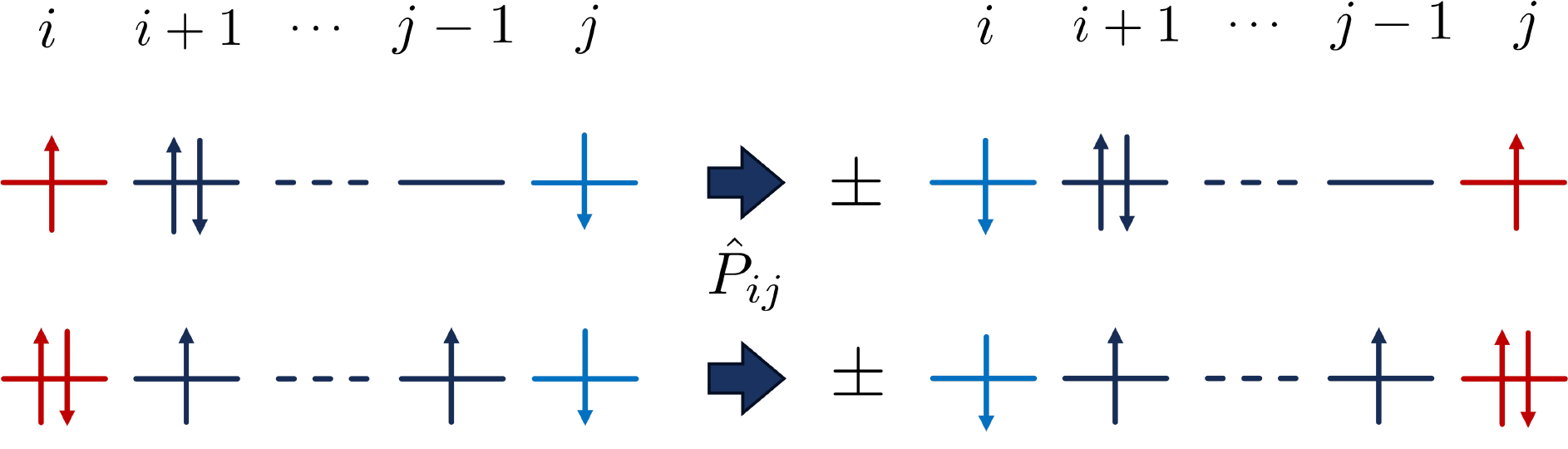}
    \caption{Schematic illustration of the action of the permutation operator on states for spin-1/2 fermionic systems. The permutation operator $\hat{P}_{ij}$ swaps sites $i$ and $j$. The change in sign associated with the action of $\hat{P}_{ij}$ depends on the number of particles between sites $i$ and $j$, as shown in Eq.~\eqref{eq2-8}.}
    \label{fig_permutation}
\end{figure}

Let $\mathfrak{S}_{n}$ be a symmetric group of degree $n$. The permutation operator $\hat{P}_{\sigma}$ for $\sigma \in \mathfrak{S}_{n}$ is defined as a unitary operator that satisfies
\begin{align}\label{eq2-7}
    \hat{P}_{\sigma}\hat{X}_{i}\hat{P}_{\sigma}^{\dagger} = \hat{X}_{\sigma(i)},
\end{align}
where $\hat{X}_{i} \in \mathcal{B}\qty(\mathcal{H}_{\rm loc})$ is an operator that acts on the $i$-th site. For a transposition $\sigma = (i, j)$, we simply denote the permutation operator as $\hat{P}_{ij}$. It is easy to show that the permutation operator satisfies $\hat{P}_{\sigma}^{\dagger} = \hat{P}_{\sigma}^{-1} = \hat{P}_{\sigma^{-1}}$. For example, in Fermi systems, the permutation operator $\hat{P}_{ij}$ satisfies $\hat{P}_{ij}\hat{c}_{is} = \hat{c}_{js}\hat{P}_{ij}$, where $\hat{c}_{is}$ denotes the fermionic annihilation operator on site $i$ with internal degree of freedom labeled by $s$. Therefore, the action of the permutation operator $\hat{P}_{ij}$ on a state is written as 
\begin{align}\label{eq2-8}
    \quad &\hat{P}_{ij}\prod_{i}\prod_{s}\qty(\hat{c}_{is}^{\dagger})^{n_{is}}\ket{0} \notag \\
    &= (-1)^{n_{i}n_{j} + n_{i\sim j}\qty(n_{i} + n_{j})}\prod_{i}\prod_{s}\qty(\hat{c}_{is}^{\dagger})^{\tilde{n}_{is}}\ket{0},
\end{align}
where $\ket{0}$ is the vacuum state, $\hat{n}_{is} \in \{0, 1\}$ denotes the eigenvalue of the number operator $\hat{n}_{is} = \hat{c}_{is}^{\dagger}\hat{c}_{is}$, $n_{i} = \sum_{s}n_{is}$ is the number of particles on site $i$, and $n_{i \sim j} = \sum_{i \lessgtr k \lessgtr j}\sum_{s} n_{ks}$ is the number of particles between sites $i$ and $j$. $\tilde{n}_{ks}$ is defined from $n_{ks}$ as follows:
\begin{align}\label{eq2-9}
    \tilde{n}_{ks} = \left\{
    \begin{aligned}
        n_{ks}& &  (k\neq i, j);\\
        n_{js}& & (k = i);\\
        n_{is}& & (k = j).
    \end{aligned}
    \right. 
\end{align}
Equation \eqref{eq2-8} implies that the sign depends on the number of particles between sites $i$ and $j$. An example of the action of the permutation of $\hat{P}_{ij}$ for a spin-1/2 fermion system is illustrated in Fig.~\ref{fig_permutation}.

An operator $\hat{X} \in \mathcal{B}(\mathcal{H})$ is said to be symmetric under permutation $\sigma$ if 
    $\hat{P}_{\sigma}\hat{X}\hat{P}_{\sigma}^{\dagger} = \hat{X}\ \Leftrightarrow \ [\hat{P}_{\sigma}, \hat{X}] = 0$
is satisfied. Moreover, the operator $\hat{X}$ is said to have permutation symmetry if it is symmetric under all permutations in $\mathfrak{S}_{n}$.\\

In short, we make the following four assumptions on the $n$-site lattice model throughout this paper:
\begin{enumerate}
    \item The local Hilbert space $\mathcal{H}_{\rm loc}$ is finite dimensional.
    \item The dynamics is governed by the GKSL equation \eqref{eq2-1}.
    \item Lindblad operators are 1-local, i.e., dissipation and/or drive occur locally.
    \item The Hamiltonian $\hat{H}$ consists of at most 2-body terms.
\end{enumerate}

We now formulate the equivalent condition to have a quantum i.i.d. state $\hat{\rho}_{\rm loc}^{\otimes n}$ as a steady state in the most general form. For a density matrix on a local Hilbert space $\hat{\rho}_{\rm loc} \in \mathcal{B}(\mathcal{H}_{\rm loc})$, we denote an orthonormal basis that diagonalizes $\hat{\rho}_{\rm loc}$ as $\{\ket{\psi_{k}}\}_{1 \leq k \leq d}$ and the corresponding eigenvalues as $\{\lambda_{k}\}_{1 \leq k \leq d}$. We denote the rank of $\hat{\rho}_{\rm loc}$ as $r$ and assume that $\lambda_{k\leq r}\neq 0$ and $\lambda_{k>r} = 0$. Then, regarding $\hat{\rho}_{\rm loc}$ as a linear map from $\mathcal{H}_{\rm loc}$ to itself, the image of $\hat{\rho}_{\rm loc}$, ${\rm Im}(\hat{\rho}_{\rm loc})$, is ${\rm span}\{\ket{\psi_{1}}, \dots, \ket{\psi_{r}}\}$. Moreover, regarding $\hat{\rho}_{\rm loc}^{\otimes n}$ as a linear map from $\mathcal{H}$ to itself, we have ${\rm Im}(\hat{\rho}_{\rm loc}^{\otimes n}) = {\rm Im}(\hat{\rho}_{\rm loc})^{\otimes n}$. We denote a projection operator to ${\rm Im}(\hat{\rho}_{\rm loc}^{\otimes n})$ as $\hat{\Pi}$, which can be explicitly written as $\hat{\Pi} = \qty(\sum_{k = 1}^{r}\ketbra{\psi_{k}}{\psi_{k}})^{\otimes n}$.

\begin{theorem}\label{thm:equivalent_SS}
\textit{Equivalent condition to have a quantum i.i.d. steady state.}\\
    For open quantum many-body systems, the condition that the quantum i.i.d. state $\hat{\rho}_{\rm loc}^{\otimes n}$ is a steady state (i.e., $\mathcal{L}(\hat{\rho}_{\rm loc}^{\otimes n}) = 0$) is equivalent to the following four conditions being simultaneously satisfied:
    \begin{enumerate}[(i)]
        \item $\forall i\in \Lambda\ \forall \alpha\in \Gamma_{i}\ \hat{L}_{i}^{(\alpha)}{\rm Im}(\hat{\rho}_{\rm loc})\subset {\rm Im}(\hat{\rho}_{\rm loc})$.
        \item $\hat{H}_{\rm eff}{\rm Im}(\hat{\rho}_{\rm loc}^{\otimes n})\subset {\rm Im}(\hat{\rho}_{\rm loc}^{\otimes n})$.
        \item $\forall i\in \Lambda\ \  \mathcal{P}\mathcal{L}_{i}(\hat{\rho}_{\rm loc}) = 0$.
        \item $\forall (i, j)\in \Lambda_{2}\ \ \left[\mathcal{T}_{ij}\qty(\hat{\Pi}\hat{H}\hat{\Pi}), \hat{\rho}_{\rm loc}\otimes \hat{\rho}_{\rm loc}\right] = 0$.
    \end{enumerate}
    Here, $\mathcal{P}\mathcal{L}_{i}$ is a projected superoperator which is defined in terms of $\mathcal{L}_{i}$ by the following equation:
    \begin{align}\label{eq2-10}
        \mathcal{P}\mathcal{L}_{i}(\bullet) = -{\rm i}\left[ \mathcal{T}_{i}\qty(\hat{\Pi}\hat{H}\hat{\Pi}), \bullet \right] + \sum_{\alpha \in \Gamma_{i}}\mathcal{D}_{\hat{\Pi}\hat{L}_{i}^{(\alpha)}\hat{\Pi}}(\bullet).
    \end{align}
    Superoperators $\mathcal{T}_{ij}$ and $\mathcal{T}_{i}$ are defined in Eqs.~\eqref{eq2-2} and \eqref{eq2-3}. We here regard $\hat{\Pi}\hat{H}\hat{\Pi}$ as an operator on ${\rm Im}(\hat{\rho}_{\rm loc}^{\otimes n})$. Accordingly, $d$ in Eqs.~\eqref{eq2-2} and \eqref{eq2-3} should be replaced with $r$, which is the dimension of ${\rm Im}(\hat{\rho}_{\rm loc})$. $\hat{\rho}_{\rm loc}\otimes \hat{\rho}_{\rm loc}$ in condition (iv) means $\qty(\hat{\rho}_{\rm loc})_i  \otimes \qty(\hat{\rho}_{\rm loc})_{j}$, i.e., $\hat{\rho}_{\rm loc}$ acts on sites $i$ and $j$. As such, we sometimes omit these site labels.
\end{theorem}

Condition (i) means that ${\rm Im}(\hat{\rho}_{\rm loc})$ is an invariant subspace of all Lindblad operators $\hat{L}_{i}^{(\alpha)}$, whereas condition (ii) means that ${\rm Im}(\hat{\rho}_{\rm loc}^{\otimes n})$ is an invariant subspace of $\hat{H}_{\rm eff}$. Condition (iii) indicates that $\hat{\rho}_{\rm loc}$ is a steady state of the projected single-site Lindbladian $\mathcal{P}\mathcal{L}_{i}$, and condition (iv) states that irreducible parts of the projected Hamiltonian acting on two sites commute with $\hat{\rho}_{\rm loc}\otimes \hat{\rho}_{\rm loc}$.
\\

From Theorem \ref{thm:equivalent_SS}, one can judge whether or not a certain open quantum many-body system has a quantum i.i.d. steady state by examining if there exists $\hat{\rho}_{\rm loc}$ that simultaneously satisfy the four conditions (i)-(iv). First, we solve a mean-field equation, that is, ${\rm Tr}_{\overline{i}}[\mathcal{L}(\hat{\rho}_{\rm loc}^{\otimes n})] = 0$ \cite{jin2016cluster}, which is a necessary condition for $\mathcal{L}(\hat{\rho}_{\rm loc}^{\otimes n}) = 0$. The mean-field equation, ${\rm Tr}_{\overline{i}}[\mathcal{L}(\hat{\rho}_{\rm loc}^{\otimes n})] = 0$, is analytically tractable, since it is an equation over the local Hilbert space. Once we get $\hat{\rho}_{\rm loc}$ from the mean-field equation, we examine whether it satisfies the conditions (i)-(iv). If it satisfies all of them, then we have $\mathcal{L}(\hat{\rho}_{\rm loc}^{\otimes n}) = 0$, which means that the system has a quantum i.i.d. steady state. While $\mathcal{L}(\hat{\rho}_{\rm loc}^{\otimes n}) = 0$ is the condition on the global Hilbert space $\mathcal{H} = \mathcal{H}_{\rm loc}^{\otimes n}$, conditions (i), (iii), (iv) are those on the local one, which greatly facilitates the analytical tractability.\\

Theorem \ref{thm:equivalent_SS} can be proved by using the following lemma. 
\begin{lemma}\label{lem:equivalent_SS}
    We assume that $\hat{\rho}_{\rm loc}$ is a regular density matrix (i.e., all the eigenvalues are nonzero). Then, $\mathcal{L}(\hat{\rho}_{\rm loc}^{\otimes n}) = 0$ is equivalent to the following two conditions being simultaneously met:
    \begin{enumerate}[(i$^{\prime}$)]
    \setcounter{enumi}{2}
    \item $\forall i\in \Lambda\ \  \mathcal{L}_{i}(\hat{\rho}_{\rm loc}) = 0$.
    \item $\forall (i, j)\in \Lambda_{2}\ \ [\hat{H}_{ij}, \hat{\rho}_{\rm loc}\otimes \hat{\rho}_{\rm loc}] = 0$.
  \end{enumerate}
  Moreover, for arbitrary $\hat{\rho}_{\rm loc}$, regardless of whether or not the density matrix is regular, the simultaneous fulfillment of conditions (iii$^{\prime}$) and (iv$^{\prime}$) is a sufficient condition for the system to have $\hat{\rho}_{\rm loc}^{\otimes n}$ as a steady state.
\end{lemma}

The first part of Lemma \ref{lem:equivalent_SS} is a special case of Theorem \ref{thm:equivalent_SS} in which $\hat{\rho}_{\rm loc}$ is a regular density matrix (i.e., full-rank). We can see that Theorem \ref{thm:equivalent_SS} reduces to Lemma \ref{lem:equivalent_SS} in this case, as ${\rm Im}(\hat{\rho}_{\rm loc}) = \mathcal{H}_{\rm loc}$ and $\hat{\Pi}$ is equal to the identity operator if $\hat{\rho}_{\rm loc}$ is a regular matrix.\\

\noindent\textit{Proof of Lemma \ref{lem:equivalent_SS}}
\begin{enumerate}[(1)]
    \item $\mathcal{L}(\hat{\rho}_{\rm loc}^{\otimes n}) = 0\ \ \Rightarrow$ (iii$^{\prime}$), (iv$^{\prime}$).\\
    From ${\rm Tr}_{\overline{ij}}[\mathcal{L}(\hat{\rho}_{\rm loc}^{\otimes n})] = 0$, we obtain
    \begin{align}
      -{\rm i}[\hat{H}_{ij}, \hat{\rho}_{\rm loc}\otimes \hat{\rho}_{\rm loc}] + \hat{\sigma}_{i}\otimes \hat{\rho}_{\rm loc} + \hat{\rho}_{\rm loc}\otimes \hat{\sigma}_{j} = 0,\label{eq2-11-1}\\
      \hat{\sigma}_{i}\coloneq \mathcal{L}_{i}(\hat{\rho}_{\rm loc}) - {\rm i}\sum_{k\neq i, j}{\rm Tr}_{k}[\hat{H}_{ik}, \hat{\rho}_{\rm loc}\otimes \hat{\rho}_{\rm loc}],\label{eq2-11-2}\\
      \hat{\sigma}_{j}\coloneq \mathcal{L}_{j}(\hat{\rho}_{\rm loc}) - {\rm i}\sum_{k\neq i, j}{\rm Tr}_{k}[\hat{H}_{jk}, \hat{\rho}_{\rm loc}\otimes \hat{\rho}_{\rm loc}].\label{eq2-11-3}
    \end{align}
    Here, $\hat{\sigma}_{i}$ and $\hat{\sigma}_{j}$ are traceless operators. Applying $\bra{\psi_{m}}\otimes\bra{\psi_{k}}\bullet \ket{\psi_{n}}\otimes \ket{\psi_{k}}$ to Eq.~\eqref{eq2-11-1}, we have
    \begin{align}\label{eq2-12}
      &{\rm i}(\lambda_{m}-\lambda_{n})\lambda_{k}\bra{\psi_{m}}\otimes\bra{\psi_{k}}\hat{H}_{ij} \ket{\psi_{n}}\otimes\ket{\psi_{k}} \notag \\
      & + \lambda_{k}\mel{\psi_{m}}{\hat{\sigma}_{i}}{\psi_{n}} + \delta_{mn}\lambda_{m}\mel{\psi_{k}}{\hat{\sigma}_{j}}{\psi_{k}} = 0.
    \end{align}
    If $m\neq n$, the above equation reduces to
    \begin{align}\label{eq2-13}
      \mel{\psi_{m}}{\hat{\sigma}_{i}}{\psi_{n}} = {\rm i}(\lambda_{n}-\lambda_{m})\bra{\psi_{m}}\otimes\bra{\psi_{k}}\hat{H}_{ij} \ket{\psi_{n}}\otimes\ket{\psi_{k}},
    \end{align}
    where the condition $\lambda_{k}\neq 0$, which is a consequence of $\hat{\rho}_{\rm loc}$ being a regular density matrix, is used.
    By taking the summation from $k = 1$ to $d$ of Eq.~\eqref{eq2-13}, we obtain
    \begin{align}\label{eq2-14}
      \mel{\psi_{m}}{\hat{\sigma}_{i}}{\psi_{n}} = \frac{{\rm i}(\lambda_{n}-\lambda_{m})}{d} \mel{\psi_{m}}{{\rm Tr}_{j}[\hat{H}_{ij}]}{\psi_{n}} = 0.
    \end{align}
    If $m=n$, taking the summation of Eq.~\eqref{eq2-12} from $k = 1$ to $d$, we obtain
    \begin{align}\label{eq2-15}
      \mel{\psi_{m}}{\hat{\sigma}_{i}}{\psi_{m}} = -\lambda_{m}{\rm Tr}[\hat{\sigma}_{j}] = 0.
    \end{align}
    From Eqs.~\eqref{eq2-14} and \eqref{eq2-15}, we have $\hat{\sigma}_{i} = 0$. Similarly, we have $\hat{\sigma}_{j} = 0$. Thus, Eq.~\eqref{eq2-11-1} is simplified to
    \begin{align}\label{eq2-16}
        [\hat{H}_{ij}, \hat{\rho}_{\rm loc}\otimes \hat{\rho}_{\rm loc}] = 0,
    \end{align}
    which is the condition (iv$^{\prime}$). We also obtain condition (iii$^{\prime}$) from ${\rm Tr}_{\overline{i}}[\mathcal{L}(\hat{\rho}_{\rm loc}^{\otimes n})] = 0$.
    \item (iii$^{\prime}$), (iv$^{\prime}$) $\Rightarrow\ \ \mathcal{L}(\hat{\rho}_{\rm loc}^{\otimes n}) = 0$.\\
    The proof is straightforward. We just need to write down $\mathcal{L}(\hat{\rho}_{\rm loc}^{\otimes n})$ explicitly and use conditions (iii$^{\prime}$) and (iv$^{\prime}$). This argument holds regardless of whether $\hat{\rho}_{\rm loc}$ is a regular matrix or not.\qed
\end{enumerate}

Example \ref{ex5} illustrates an open quantum many-body system for which there exists $\hat{\rho}_{\rm loc}$ that satisfies conditions (iii$^{\prime}$) and (iv$^{\prime}$) simultaneously, and thus has a quantum i.i.d. steady state.
Let us now use Lemma \ref{lem:equivalent_SS} to prove Theorem \ref{thm:equivalent_SS}.\\

\noindent\textit{Proof of Theorem \ref{thm:equivalent_SS}}
\begin{enumerate}[(1)]
    \item $\mathcal{L}(\hat{\rho}_{\rm loc}^{\otimes n}) = 0\ \Rightarrow$ (i), (ii), (iii), (iv).\\
    Using the effective Hamiltonian defined in Eq.~\eqref{eq2-6}, $\mathcal{L}(\hat{\rho}_{\rm loc}^{\otimes n}) = 0$ can be rewritten as follows:
    \begin{align}\label{eq2-17}
      -{\rm i}\hat{H}_{\rm eff}\hat{\rho}_{\rm loc}^{\otimes n} + {\rm i}\hat{\rho}_{\rm loc}^{\otimes n}\hat{H}_{\rm eff}^{\dagger} + \sum_{i \in \Lambda}\sum_{\alpha \in \Gamma_{i}}\hat{L}_{i}^{(\alpha)}\hat{\rho}_{\rm loc}^{\otimes n}\hat{L}_{i}^{(\alpha)\dagger} = 0.
    \end{align}
    Let $\ket{\Psi}\in {\rm Im}(\hat{\rho}_{\rm loc}^{\otimes n})$ and $\ket{\Psi^{\perp}} \in {\rm Im}(\hat{\rho}_{\rm loc}^{\otimes n})^{\perp}$. Since ${\rm Im}(\hat{\rho}_{\rm loc}^{\otimes n})^{\perp} = {\rm Ker}(\hat{\rho}_{\rm loc}^{\otimes n})$, $\hat{\rho}_{\rm loc}^{\otimes n}\ket{\Psi^{\perp}} = 0$ holds. Therefore, by applying $\mel{\Psi^{\perp}}{\bullet}{\Psi^{\perp}}$ to Eq.~\eqref{eq2-17}, we have
    \begin{align}\label{eq2-18}
        \sum_{i \in \Lambda}\sum_{\alpha \in \Gamma_{i}}\mel{\Psi^{\perp}}{ \hat{L}_{i}^{(\alpha)}\hat{\rho}_{\rm loc}^{\otimes n}\hat{L}_{i}^{(\alpha)\dagger} }{\Psi^{\perp}} = 0.
    \end{align}
    Let $\ket{\Psi^{\perp}}$ be
    \begin{align}\label{eq2-19}
        \ket{\psi_{1}} \otimes \cdots \otimes \underbrace{\ket{\psi_{m}}}_{i}\otimes \cdots \otimes \ket{\psi_{1}},
    \end{align}
    where $m>r$. Then, we have
    \begin{align}\label{eq2-20}
        \sum_{\alpha \in \Gamma_{i}}\mel{\psi_{m}}{\hat{L}_{i}^{(\alpha)}\hat{\rho}_{\rm loc}\hat{L}_{i}^{(\alpha)\dagger}}{\psi_{m}} = 0.
    \end{align}
    Substituting $\hat{\rho}_{\rm loc} = \sum_{k=1}^{r}\lambda_{k}\ketbra{\psi_{k}}{\psi_{k}}$ into Eq.~\eqref{eq2-20}, we obtain
    \begin{align}\label{eq2-21}
        \sum_{k = 1}^{r}\sum_{\alpha\in \Gamma_{i}}\lambda_{k}\|\mel{\psi_{m}}{\hat{L}_{i}^{(\alpha)}}{\psi_{k}}\|^2 = 0.
    \end{align}
    Thus, for $m > r$ and $k \leq r$, we have $\mel{\psi_{m}}{\hat{L}_{i}^{(\alpha)}}{\psi_{k}} = 0$, which means that ${\rm Im}(\hat{\rho}_{\rm loc})$ is an invariant subspace of $\hat{L}_{i}^{(\alpha)}$ (condition (i)). Since ${\rm Im}(\hat{\rho}_{\rm loc}^{\otimes n}) = {\rm Im}(\hat{\rho}_{\rm loc})^{\otimes n}$, ${\rm Im}(\hat{\rho}_{\rm loc}^{\otimes n})$ is also an invariant subspace of $\hat{L}_{i}^{(\alpha)}$, i.e., $\mel{\Psi^{\perp}}{\hat{L}_{i}^{(\alpha)}}{\Psi} = 0$. Thus, we have $\mel{\Psi}{\hat{L}_{i}^{(\alpha)\dagger}}{\Psi^{\perp}} = 0$, which means that ${\rm Ker}(\hat{\rho}_{\rm loc}^{\otimes n})$ is an invariant subspace of $\hat{L}_{i}^{(\alpha)\dagger}$. Therefore, by applying $\mel{\Psi^{\perp}}{\bullet}{\Psi}$ to Eq.~\eqref{eq2-17}, we obtain $\mel{\Psi^{\perp}}{\hat{H}_{\rm eff}}{\Psi} = 0$, which means that ${\rm Im}(\hat{\rho}_{\rm loc}^{\otimes n})$ is an invariant subspace of $\hat{H}_{\rm eff}$ (condition (ii)). Next, let us apply the projection operator $\hat{\Pi}$ from both left and right sides of the equation $\mathcal{L}(\hat{\rho}_{\rm loc}^{\otimes n}) = 0$. By using $\hat{\rho}_{\rm loc}^{\otimes n} = \hat{\Pi}\hat{\rho}_{\rm loc}^{\otimes n} = \hat{\rho}_{\rm loc}^{\otimes n}\hat{\Pi}$ and $\hat{L}_{i}^{(\alpha)}\hat{\Pi} = \hat{\Pi}\hat{L}_{i}^{(\alpha)}\hat{\Pi}$, we have
    \begin{align}\label{eq2-22}
        -{\rm i}[\hat{\Pi}\hat{H}\hat{\Pi}, \hat{\rho}_{\rm loc}^{\otimes n}] + \sum_{i, \alpha \in \Gamma_{i}}\mathcal{D}_{\hat{\Pi}\hat{L}_{i}^{(\alpha)}\hat{\Pi}}(\hat{\rho}_{\rm loc}^{\otimes n}) = 0.
    \end{align}
    This equation can be seen as the GKSL equation projected on ${\rm Im}(\hat{\rho}_{\rm loc}^{\otimes n})$. By decomposing the projected Hamiltonian $\hat{\Pi}\hat{H}\hat{\Pi}$ using $\mathcal{T}_{ij}$ and $\mathcal{T}_{i}$ defined in Eqs.~\eqref{eq2-2} and \eqref{eq2-3}, Eq.~\eqref{eq2-22} is rewritten as follows:
    \begin{align}\label{eq2-23}
        -{\rm i}\sum_{(i, j)}[\mathcal{T}_{ij}\qty(\hat{\Pi}\hat{H}\hat{\Pi}), \hat{\rho}_{\rm loc}^{\otimes n}] + \sum_{i}\mathcal{P}\mathcal{L}_{i}(\hat{\rho}_{\rm loc}^{\otimes n}) = 0, 
    \end{align}
    where $\mathcal{P}\mathcal{L}_{i}$ is a projected superoperator defined from $\mathcal{L}_{i}$ by Eq.~\eqref{eq2-10}.  Since $\hat{\rho}_{\rm loc}$ can be seen as a regular matrix by regarding it as an operator on ${\rm Im}(\hat{\rho}_{\rm loc})$, Lemma \ref{lem:equivalent_SS} can be applied to obtain
    \begin{enumerate}[(i)]
    \setcounter{enumii}{2}
        \item $\forall i\in \Lambda\ \  \mathcal{P}\mathcal{L}_{i}(\hat{\rho}_{\rm loc}) = 0$,
        \item $\forall (i, j)\in \Lambda_{2}\ \ \left[ \mathcal{T}_{ij}(\hat{\Pi}\hat{H}\hat{\Pi}), \hat{\rho}_{\rm loc}\otimes \hat{\rho}_{\rm loc}\right] = 0$.
    \end{enumerate}
    \item (i), (ii), (iii), (iv) $\Rightarrow \ \mathcal{L}(\hat{\rho}_{\rm loc}^{\otimes n}) = 0$.\\
    It suffices to show
    \begin{align}
        \mel{\Psi}{\mathcal{L}(\hat{\rho}_{\rm loc}^{\otimes n})}{\Psi} &= 0, \label{eq2-24-1}\\
      \mel{\Psi}{\mathcal{L}(\hat{\rho}_{\rm loc}^{\otimes n})}{\Psi^{\perp}} &= 0, \label{eq2-24-2}\\
      \mel{\Psi^{\perp}}{\mathcal{L}(\hat{\rho}_{\rm loc}^{\otimes})}{\Psi} &= 0, \label{eq2-24-3}\\
      \mel{\Psi^{\perp}}{\mathcal{L}(\hat{\rho}_{\rm loc}^{\otimes n})}{\Psi^{\perp}} &= 0 \label{eq2-24-4}
    \end{align}
    for arbitrary $\ket{\Psi}\in {\rm Im}(\hat{\rho}_{\rm loc}^{\otimes n})$ and $\ket{\Psi^{\perp}}\in {\rm Ker}(\hat{\rho}_{\rm loc}^{\otimes n})$. First, using $\ket{\Psi} = \hat{\Pi}\ket{\Psi}$, we have
    \begin{align}\label{eq2-25}
        &\quad \mel{\Psi}{\mathcal{L}(\hat{\rho}_{\rm loc}^{\otimes n})}{\Psi} = \mel{\Psi}{\hat{\Pi}\mathcal{L}(\hat{\rho}_{\rm loc}^{\otimes n})\hat{\Pi}}{\Psi}= \notag \\
      &\quad  \mel{\Psi}{\qty(-{\rm i}\sum_{(i, j)}\left[\mathcal{T}_{ij}\qty(\hat{\Pi}\hat{H}\hat{\Pi}), \hat{\rho}_{\rm loc}^{\otimes n}\right] + \sum_{i}\mathcal{P}\mathcal{L}_{i}(\hat{\rho}_{\rm loc}^{\otimes n}))}{\Psi}.
    \end{align}
    Due to conditions (iii) and (iv), the right-hand side (r.h.s.) of Eq.~\eqref{eq2-25} vanishes, which yields Eq.~\eqref{eq2-24-1}. Next, from the definition of $\ket{\Psi^{\perp}}$, we have $\hat{\rho}_{\rm loc}^{\otimes n}\ket{\Psi^{\perp}} = 0$. From conditions (i) and (ii), we find that ${\rm Ker}(\hat{\rho}_{\rm loc}^{\otimes n})$ is an invariant subspace of $\hat{L}_{i}^{(\alpha)\dagger}$ and $\hat{H}_{\rm eff}^{\dagger}$. Thus, we have $\hat{\rho}_{\rm loc}^{\otimes n}\hat{L}_{i}^{(\alpha)\dagger}\ket{\Psi^{\perp}} = 0$ and $\hat{\rho}_{\rm loc}^{\otimes n}\hat{H}_{\rm eff}^{\dagger}\ket{\Psi^{\perp}} = 0$, which yield
    \begin{align}\label{eq2-26}
        \mathcal{L}(\hat{\rho}_{\rm loc}^{\otimes n})\ket{\Psi^{\perp}} = 0.
    \end{align}
    Therefore, Eqs.~\eqref{eq2-24-2} and \eqref{eq2-24-4} are shown. Using property $\qty(\mathcal{L}(\hat{X}))^{\dagger} = \mathcal{L}(\hat{X}^{\dagger})$, Eq.~\eqref{eq2-24-3} follows from Eq.~\eqref{eq2-24-2}. \qed
\end{enumerate}

\begin{figure}
    \centering
    \includegraphics[width=\linewidth]{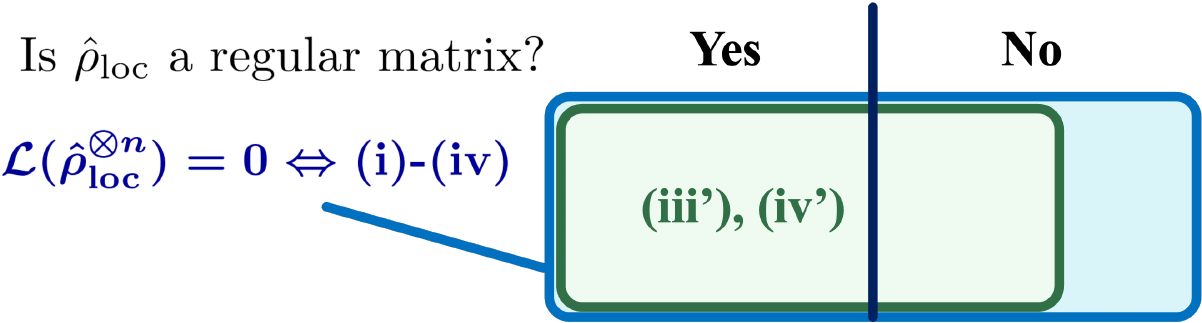}
    \caption{Relationship between the statements in Theorem \ref{thm:equivalent_SS} and Lemma \ref{lem:equivalent_SS}. For every single-site density matrix $\hat{\rho}_{\rm loc}$, $\mathcal{L}(\hat{\rho}_{\rm loc}^{\otimes n})$ is equivalent to the simultaneous satisfaction of conditions (i)-(iv). If $\hat{\rho}_{\rm loc}$ is a regular matrix (i.e., all the eigenvalues are nonzero), $\mathcal{L}(\hat{\rho}_{\rm loc}^{\otimes n}) = 0$ is equivalent to the simultaneous satisfaction of conditions (iii$^{\prime}$) and (iv$^{\prime}$). The simultaneous satisfaction of conditions (iii$^{\prime}$) and (iv$^{\prime}$) is the sufficient condition of $\mathcal{L}(\hat{\rho}_{\rm loc}^{\otimes n}) = 0$.}
    \label{fig_equivconditions}
\end{figure}

The relationship between the statements of Theorem \ref{thm:equivalent_SS} and Lemma \ref{lem:equivalent_SS} are summarized in Fig.~\ref{fig_equivconditions}.\\

Let us now consider the case in which the local Hilbert space is two-dimensional. The examples include spin-1/2 systems and spinless fermion systems. In this case, Theorem \ref{thm:equivalent_SS} reduces to the following corollary.

\begin{corollary}\label{cor:equivalent_SS_2dim}
    \textit{Equivalent condition to have a quantum i.i.d. steady state for $d = 2$.}\\
    Let us suppose that $d = {\rm dim}\mathcal{H}_{\rm loc} = 2$.\\
    Case 1. If $\hat{\rho}_{\rm loc}$ is not a regular density matrix, $\hat{\rho}_{\rm loc}$ can be expressed as $\ketbra{\psi}{\psi}$ where $\ket{\psi} \in \mathcal{H}_{\rm loc}$. Then, $\mathcal{L}(\hat{\rho}_{\rm loc}^{\otimes n}) = 0$ is equivalent to the following two conditions being simultaneously satisfied.
    \begin{enumerate}[(i)]
        \item $\ket{\psi}$ is the simultaneous eigenstate of all Lindblad operators $\hat{L}_{i}^{(\alpha)}$.
        \item $\ket{\psi}^{\otimes n}$ is the eigenstate of $\hat{H}_{\rm eff}$.
    \end{enumerate}
    Case 2. If $\hat{\rho}_{\rm loc}$ is a regular density matrix, $\mathcal{L}(\hat{\rho}_{\rm loc}^{\otimes n}) = 0$ is equivalent to the following two conditions being simultaneously satisfied.
    \begin{enumerate}[(i$^{\prime}$)]
    \setcounter{enumi}{2}
    \item $\forall i\in \Lambda\ \  \mathcal{L}_{i}(\hat{\rho}_{\rm loc}) = 0$.
    \item $\forall (i, j)\in \Lambda_{2}\ \ [\hat{H}_{ij}, \hat{\rho}_{\rm loc}\otimes \hat{\rho}_{\rm loc}] = 0$.
  \end{enumerate}
\end{corollary}

\noindent \textit{Proof of Corollary \ref{cor:equivalent_SS_2dim}}\\
Case 1. If $\hat{\rho}_{\rm loc}$ can be expressed as $\hat{\rho}_{\rm loc} = \ketbra{\psi}{\psi}$, ${\rm Im}(\hat{\rho}_{\rm loc})$ is one-dimensional, since ${\rm Im}(\hat{\rho}_{\rm loc}) = \C \ket{\psi}$. Then, the projection of an operator on ${\rm Im}(\hat{\rho}_{\rm loc})$ becomes a constant scalar rather than a matrix. Therefore, all operators commute with each other in the projected space, and therefore conditions (iii) and (iv) in Theorem \ref{thm:equivalent_SS} always hold. Moreover, conditions (i) and (ii) in Theorem \ref{thm:equivalent_SS} can be restated as in the corollary.\\
Case 2 is the same as Lemma \ref{lem:equivalent_SS}. \qed\\

In the case of $d = 2$, the set of $\hat{H}_{ij}$'s that commute with $\hat{\rho}_{\rm loc}\otimes \hat{\rho}_{\rm loc}$ can be explicitly given. For simplicity, suppose that $\hat{H}_{ij}$ must be symmetric with respect to sites $i$ and $j$, i.e. $[\hat{H}_{ij}, \hat{P}_{ij}] = 0$. When $\hat{\rho}_{\rm loc}$ is parametrized in terms of Pauli matrices as $\hat{\rho}_{\rm loc} = \frac{1}{2}\qty(\hat{I} + s_{x}\hat{X} + s_{y}\hat{Y} + s_{z}\hat{Z})\ \ (s_{x}, s_{y}, s_{z} \in \R,\ 0 \leq s_{x}^2 + s_{y}^2 + s_{z}^2 \leq 1)$, $\hat{H}_{ij}$ should satisfy
\begin{align}\label{eq2-27}
    &\hat{H}_{ij}\in {\rm span}_{\R}\left\{\hat{X}_{i}\hat{X}_{j} + \hat{Y}_{i}\hat{Y}_{j} + \hat{Z}_{i}\hat{Z}_{j}, \right.\notag \\
    &\left.\qty(s_{x}\hat{X}_{i} + s_{y}\hat{Y}_{i} + s_{z}\hat{Z}_{i})\qty(s_{x}\hat{X}_{j} + s_{y}\hat{Y}_{j} + s_{z}\hat{Z}_{j})\right\},
\end{align}
where $\hat{X}_{i}, \hat{Y}_{i}, \hat{Z}_{i}$ represent Pauli matrices that act on the local Hilbert space of site $i$, and ${\rm span}_{\R}$ represents the linear span over the field $\R$ (see Appendix \ref{sec_SM_A} for the proof). An application of Corollary \ref{cor:equivalent_SS_2dim} is provided in Example \ref{ex2}.\\

Next, let us state a sufficient condition for an open quantum many-body system to have a quantum i.i.d. steady state. First, we define $\mathcal{B}_{\rm com}$ as a set of operators on $\mathcal{H} = \mathcal{H}_{\rm loc}^{\otimes n}$ that commute with arbitrary quantum i.i.d. steady states $\hat{\rho}_{\rm loc}^{\otimes n}$. We note that some $\rho_{\rm loc}$'s in $\mathcal{S}(\mathcal{H}_{\rm loc})$ are prohibited due to the superselection rule \cite{RevModPhys.79.555}. When particles are massive, density matrices are restricted to those commuting with the total number operator $\hat{N}$. Since $[\hat{\rho}_{\rm loc}^{\otimes n}, \hat{N}] = 0 \Leftrightarrow [\hat{\rho}_{\rm loc}, \hat{n}] = 0$, where $\hat{n}$ is the single-site number operator on $\mathcal{H}_{\rm loc}$, $\hat{\rho}_{\rm loc}$ should commute with $\hat{n}$ for the quantum i.i.d. state $\hat{\rho}_{\rm loc}^{\otimes n}$ to be a physical state. The set $\mathcal{B}_{\rm com}$ with or without the number-superselection rule is provided in the following Lemma.

\begin{lemma}\label{lem:commutable}
    Let $\mathcal{B}_{\rm com}$ be a set of operators on $\mathcal{H} = \mathcal{H}_{\rm loc}^{\otimes n}$ that commute with every quantum i.i.d. state. For spin systems and massless boson systems, 
    \begin{align}\label{eq2-28}
        \mathcal{B}_{\rm com} = \{\hat{P}_{\sigma \in \mathfrak{S}_{n}}\}^{\prime\prime} = {\rm span}\{\hat{P}_{\sigma\in \mathfrak{S}_{n}} \},
    \end{align}
    while for massive fermion and boson systems subject to the number-superselection rule, 
    \begin{align}\label{eq2-29}
        \mathcal{B}_{\rm com} = \{\hat{P}_{\sigma \in \mathfrak{S}_{n}}, \hat{n}_{i\in \Lambda} \}^{\prime \prime},
    \end{align}
    where $\hat{n}_{i}$ is the number operator on site $i$. $\mathcal{A}^{\prime}$ is a commutant of the set of operators $\mathcal{A}$, i.e.,
  \begin{align}\label{eq2-30}
      \mathcal{A}^{\prime} = \left\{ \hat{X} \in \mathcal{B}(\mathcal{H})\ |\  \forall\hat{Y}\in \mathcal{A}\ [\hat{X}, \hat{Y}] = 0 \right\}.
  \end{align}
  The double commutant $\mathcal{A}^{\prime \prime} = \qty(\mathcal{A}^{\prime})^{\prime}$ for a set of operators $\mathcal{A}$
  constitutes a minimum von-Neumann algebra that includes $\mathcal{A}$, which is generated by scalar multiplication, addition, and multiplication of operators in the set $\mathcal{A}$, if $\mathcal{A}$ includes an identity operator $\hat{I}$ and is closed under Hermitian conjugation.
\end{lemma}

Lemma \ref{lem:commutable} is proven by using Schur-Weyl duality \cite{schur1901klasse, weyl1946classical}. See Appendix \ref{sec_SM_B} for detail. Now, we state the sufficient condition to have a quantum i.i.d. steady state using $\mathcal{B}_{\rm com}$.

\begin{theorem}\label{thm:sufficient}
    \textit{Sufficient condition to have a quantum i.i.d. steady state.}\\
    \indent Suppose that the 1-local Lindblad operators $\hat{L}_{i}^{(\alpha)}$'s are uniform, which means that $\Gamma_{i} = \Gamma_{j}$, and $\hat{L}_{i}^{(\alpha)}$ and $\hat{L}_{j}^{(\alpha)}$ ($i\neq j$) are equivalent to each other as operators on $\mathcal{H}_{\rm loc}$. We here omit the assumption that the Hamiltonian consists of at most 2-body terms.
    If the Hamiltonian can be written as
    \begin{align}\label{eq2-32}
        \hat{H} = \hat{H}_{\rm com} + \sum_{i}\hat{h}_{i},
    \end{align}
    where $\hat{H}_{\rm com} \in \mathcal{B}_{\rm com}$ , which is defined in Lemma \ref{lem:commutable}, and $\hat{h}_{i}$'s are uniform 1-local terms, the system has a quantum i.i.d. steady state.
    Since ${\rm Tr}_{\overline{i}}[\hat{H}_{\rm com}]$ is not always zero, $\hat{h}_{i}$ is not always equal to $\hat{H}_{i}$, which is uniquely defined in Eq.~\eqref{eq2-3}. We define a single-site Lindblad operator $\ell_{i}$ as 
    \begin{align}\label{eq2-33}
        \ell_{i}(\bullet) = -{\rm i}\left[\hat{h}_{i}, \bullet\right] + \sum_{\alpha \in \Gamma_{i}}\mathcal{D}_{\hat{L}_{i}^{(\alpha)}}(\bullet),
    \end{align}
    which is uniform since $\hat{h}_{i}$'s and $\hat{L}_{i}^{(\alpha)}$'s are uniform.
    The quantum i.i.d. steady state of such a system can be written as $\hat{\rho}_{\rm SS} = \hat{\rho}_{\rm loc}^{\otimes n}$, where $\hat{\rho}_{\rm loc}$ is a single-site density operator that satisfies $\ell_{i}(\hat{\rho}_{\rm loc}) = 0$.
\end{theorem}
\noindent \textit{Proof of Theorem \ref{thm:sufficient}}\\
An equation $\dv{\hat{\rho}(t)}{t} = \ell_{i}(\hat{\rho}(t))\ (\hat{\rho}(t)\in \mathcal{S}(\mathcal{H}_{\rm loc}))$ can be seen as a GKSL equation on a single site. Since the GKSL equation has at least one steady-state solution, there exists $\hat{\rho}_{\rm loc}\in \mathcal{S}(\mathcal{H}_{\rm loc})$ that satisfies $\ell_{i}(\hat{\rho}_{\rm loc}) = 0$. When the Hamiltonian can be written as Eq.~\eqref{eq2-32}, we have
\begin{equation}\label{eq2-34}
    \mathcal{L}\qty(\hat{\rho}_{\rm loc}^{\otimes n}) = -{\rm i}\left[ \hat{H}_{\rm com}, \hat{\rho}_{\rm loc}^{\otimes n}\right] + \sum_{i}\ell_{i}\qty(\hat{\rho}_{\rm loc}^{\otimes n}) = 0.
\end{equation}
Thus, the system has a quantum i.i.d. state $\hat{\rho}_{\rm SS} = \hat{\rho}_{\rm loc}^{\otimes n}$ as a steady state. \qed\\

Note that Theorem \ref{thm:sufficient} does not ensure the uniqueness of the steady state. Although there might be multiple steady states, the system has at least one steady state that is a quantum i.i.d. steady state. Some examples of Theorem \ref{thm:sufficient} for a spin-1/2 system, a spinless fermion system, a spin-1/2 fermion system, and a hardcore boson system are provided in Examples \ref{ex1}, \ref{ex3}, \ref{ex4}, and \ref{ex6}, respectively.\\

It has been shown \cite{PhysRevLett.116.030403} that applying a single-site Lindblad superoperator only on one site suffices to prepare a quantum i.i.d. steady state, if the Hamiltonian is the sum of permutation operators. The Lindbladian considered in Ref.~\cite{PhysRevLett.116.030403} can be written in our notation as follows:
\begin{align}
    \mathcal{L}(\bullet) = -{\rm i}\left[ \hat{H}_{\rm com}, \bullet \right] + \ell_{1}(\bullet),
\end{align}
where $\ell_{1}(\bullet)$ is a single-site Lindblad superoperator that acts solely on site $1$. The quantum i.i.d. steady state is given by $\hat{\rho}_{\rm loc}^{\otimes n}$, where $\hat{\rho}_{\rm loc}$ satisfies $\ell_{1}(\hat{\rho}_{\rm loc}) = 0$. Although Theorem \ref{thm:sufficient} does not directly encompass the above theorem, it can readily be extended to do so as follows:\\

\noindent \textbf{Theorem 5$^{\prime}$.}\\
\indent Suppose that the Lindbladian of an open quantum many-body system can be written in the following form:
\begin{align}\label{eq2-32-0}
    \mathcal{L}(\bullet) = -{\rm i}\left[ \hat{H}_{\rm com}, \bullet \right] + \sum_{i}\ell_{i}(\bullet),
\end{align}
where $\hat{H}_{\rm com}\in \mathcal{B}_{\rm com}$, which is defined in Lemma \ref{lem:commutable}, and $\ell_{i}$'s are the single-site Lindblad superoperators on each site that shares the same steady-state solution, i.e.,
\begin{align}\label{eq2-32-1}
    \exists \hat{\rho}_{\rm loc}\ {\rm s.t. }\ \forall i\in\Lambda \ \ell_{i}(\hat{\rho}_{\rm loc}) = 0. 
\end{align}
Then, the quantum i.i.d. state $\hat{\rho}_{\rm loc}^{\otimes n}$ is a steady state of the system.\\

Theorem 5$^{\prime}$ can be proved straightforwardly by substituting $\hat{\rho}_{\rm loc}^{\otimes n}$ into $\mathcal{L}$.

\subsection{Exponential decay of spatial correlations and quantum entanglement}\label{sec2B}

As discussed in Section \ref{sec1}, the state without quantum entanglement is equivalent to a fully separable state, and the state without spatial correlations for any physical quantities is equivalent to a simply separable state (a.k.a. a tensor product state). Since a quantum i.i.d. state is a special case of both fully and simply separable states, it has neither quantum entanglement nor spatial correlations. Therefore, one can prove that neither quantum entanglement nor spatial correlation is present in the steady state by showing that the system has a quantum i.i.d. steady state. For instance, Corollary \ref{cor:no-go correlation} follows from Theorem \ref{thm:sufficient}.

\begin{corollary}\label{cor:no-go correlation}
    \textit{No-go theorem for steady-state spatial correlations and quantum entanglement.}\\
    Suppose that $\hat{H} = \hat{H}_{\rm com} + \sum_{i}\hat{h}_{i}$, where $\hat{H}_{\rm com} \in \mathcal{B}_{\rm com}$ defined in Lemma \ref{lem:commutable} and $\hat{h}_{i}$'s are uniform 1-local terms, and the Lindblad operators $\hat{L}_{i}^{(\alpha)}$'s are uniform. Suppose also that the system has a unique steady state. Then, neither spatial correlations of arbitrary physical quantities nor quantum entanglement is present in the steady state.
\end{corollary}
Several equivalent and sufficient conditions discovered so far \cite{SPOHN1976189, 10.1007/BF00420668, 10.1007/BF01614091, doi10.1007, doi.org/10.1007/BF01196936, 10.1063/1.1424475, Nigro_2019, PhysRevA.109.022218} can be used to judge the uniqueness of the steady state of the GKSL equation. If we additionally assume the absence of purely imaginary eigenvalues for the Lindbladian, the following corollary holds.

\begin{figure}
    \centering
    \includegraphics[width=\linewidth]{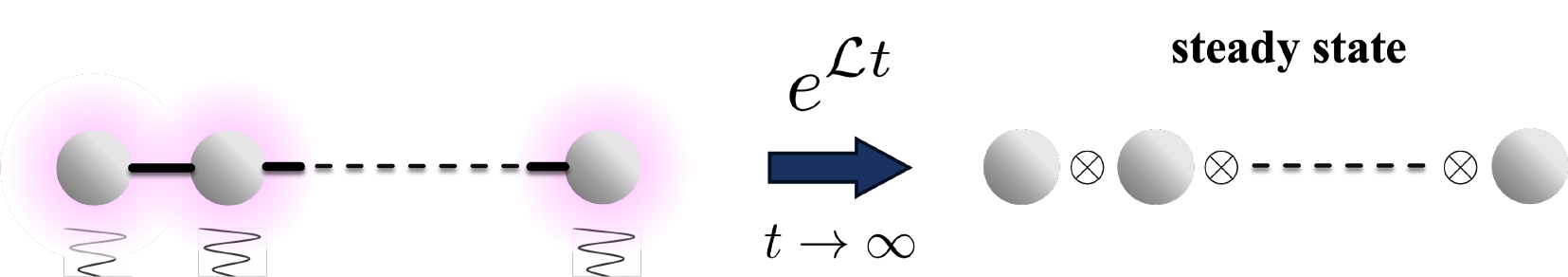}
    \caption{If an open quantum many-body system satisfies the sufficient condition in Theorem \ref{thm:sufficient}, it has a quantum i.i.d. steady state. If we additionally assume the uniqueness of the steady state and the absence of purely imaginary eigenvalues, the state exponentially decays to the quantum i.i.d. steady state, irrespective of the initial state (see Eq.~\eqref{eq2-35}). Therefore, spatial correlations and quantum entanglement decay exponentially as well.}
    \label{fig_corollary 7}
\end{figure}

\begin{corollary}\label{cor:expdecay}
    \textit{Exponential decay of spatial correlations and quantum entanglement.}\\
    Suppose that $\hat{H} = \hat{H}_{\rm com} + \sum_{i}\hat{h}_{i}$, where $\hat{H}_{\rm com} \in \mathcal{B}_{\rm com}$ defined in Lemma \ref{lem:commutable} and $\hat{h}_{i}$'s are uniform 1-local terms, and the Lindblad operators $\hat{L}_{i}^{(\alpha)}$'s are uniform. Suppose also that the system has a unique steady state, and the Lindblad superoperator does not have purely imaginary eigenvalues. Then, spatial correlations of arbitrary physical quantities and quantum entanglement decay exponentially in time (see Fig.~\ref{fig_corollary 7}). 
    We here refer to a function $f(t)$ as exponentially decaying, if it can be written in the following form, where $c_{j, k}\in \C$, $\lambda_{k}\in \C$, and $n_{k}\in\Z_{\geq 0}$.
    \begin{align}\label{eq2-35}
        f(t) = \sum_{k}\sum_{j\geq 0}^{n_{k}}c_{j, k}t^{j}e^{\lambda_{k}t}\ (\forall k\ {\rm Re}(\lambda_{k}) < 0).
    \end{align}
\end{corollary}

\noindent\textit{Proof of Corollary \ref{cor:expdecay}}\\
Let us assume that the Lindblad superoperator does not have purely imaginary eigenvalues in addition to the uniqueness of the steady state, where the eigenvalues of the Lindblad superoperator are defined by the eigenequation $\mathcal{L}(\hat{Q}) = \lambda\hat{Q}$. Let $\{ \hat{Q}_{k}^{(j)} \}_{0 \leq k \leq m, 1 \leq j \leq n_{k}}$ be a basis of $\mathcal{B}(\mathcal{H})$ in which the matrix representation of the Lindblad superoperator becomes a Jordan normal form as follows:
\begin{equation}\label{eq2-36}
    \mathscr{L} = 
  \mymat{
    J_{0} & & & \\
     & J_{1} & & \\
     & & \ddots & \\
     & & & J_{m}
  },\ 
  J_{k} = 
  \mymat{
    \lambda_{k} & 1 & & & \\
     & \lambda_{k} & 1 & & \\
     & & \ddots & \ddots & \\
     & & & \lambda_{k} & 1\\
     & & & & \lambda_{k}
  },
\end{equation}
where $\mathscr{L}$ is the matrix representation of the Lindblad superoperator $\mathcal{L}$, and $J_{k}\ (0 \leq k \leq m)$ is a Jordan cell of dimension $n_{k}$. The real part of any eigenvalue of the Lindblad superoperator (i.e., $\Re(\lambda_{k})$) is non-positive. Let $\lambda_{0}$ be zero, as there always exists a zero eigenvalue, which corresponds to the steady state. Then, due to the assumptions, $n_{0} = 1$, $\hat{Q}_{0}^{(1)}/\Tr[\hat{Q}_{0}^{(1)}] =  \hat{\rho}_{\rm SS}$ \footnote{$\Tr[\hat{Q}_{0}^{(1)}]$ is always non-zero.}, and $\Re(\lambda_{k}) < 0$ for $k\geq 1$. The initial state $\hat{\rho}(0)$ can be expanded in terms of the basis as follows:
\begin{align}\label{eq2-37}
    \hat{\rho}(0) = \hat{\rho}_{\rm SS} + \sum_{k=1}^{m}\sum_{j=1}^{n_{k}}c_{k}^{(j)}\hat{P}_{k}^{(j)}\quad (c_{k}^{(j)}\in \C).
\end{align}
Then, the solution of the GKSL equation is written as
\begin{align}\label{eq2-38}
  \hat{\rho}(t) = \hat{\rho}_{\rm SS} + \sum_{k=1}^{m}\sum_{j=1}^{n_{k}} \sum_{i=j}^{n_{k}}\frac{1}{(i-j)!}c_{k}^{(i)}t^{i-j}e^{\lambda_{k}t}\hat{P}_{k}^{(j)}.
\end{align}
Since $\Re(\lambda_{k}) < 0$ for $k\geq 1$, the density matrix converges exponentially in time to the steady state. Hence follows Corollary \ref{cor:expdecay}. \qed

\section{Dynamical stability of a set of quantum i.i.d. states}\label{sec3}
\subsection{Equivalent conditions for a system to maintain the quantum i.i.d. form}\label{sec3A}
In this section, we consider a system which maintains the quantum i.i.d. form throughout the time evolution if its initial state is a quantum i.i.d. state. In other words, once the state becomes a quantum i.i.d. state, the time evolution is closed within a set of quantum i.i.d. states. For spin systems, the equivalent condition for a system to maintain the quantum i.i.d. form can be formulated as Theorem \ref{thm:equivalent_sec3}.

\begin{theorem}\label{thm:equivalent_sec3}
    \textit{Equivalent conditions for a system to maintain the quantum i.i.d. form throughout its time evolution.}\\
   The following two conditions are equivalent.
    \begin{enumerate}[(I)]
        \item If the initial state is a quantum i.i.d. state, then the state maintains the quantum i.i.d. form throughout the time evolution of the system.
        \item $\hat{H}_{ij} \in \mathcal{B}_{\rm com}$ and the single-site Lindblad superoperator $\mathcal{L}_{i}$ is uniform. 
    \end{enumerate}
    Here, we say the single-site Lindblad superoperator is uniform if $\mathcal{L}_{i}(\hat{\rho}_{\rm loc}) = \mathcal{L}_{j}(\hat{\rho}_{\rm loc})$ holds as an operator equality on $\mathcal{H}_{\rm loc}$ for every single-site density matrix. Note that $\hat{\rho}_{\rm loc}^{\otimes n}$ should be a physical density matrix. Therefore, while $\hat{\rho}_{\rm loc}$ can be any operator in $\mathcal{S}(\mathcal{H}_{\rm loc})$ for spin systems and massless boson systems, $\hat{\rho}_{\rm loc}$ should commute with the single-site number operator $\hat{n}$ for massive fermion and boson systems, which are subject to the number-superselection rule. The following identities hold on the span of a set of single-site density matrices:
    \begin{align}
        &{\rm span}\ \mathcal{S}(\mathcal{H}_{\rm loc}) = \mathcal{B}(\mathcal{H}_{\rm loc}),\label{eq3-1-1}\\
        &{\rm span}\{\hat{\rho}_{\rm loc}\in \mathcal{S}(\mathcal{H}_{\rm loc}) | [\hat{\rho}_{\rm loc}, \hat{n}] = 0\}= \mathcal{N},\label{eq3-1-2}\\
        &\mathcal{N} \coloneq \{\hat{X}\in \mathcal{B}_{\rm loc} | [\hat{X}, \hat{n}] = 0\}.\label{eq3-1-3}
    \end{align}
    Accordingly, the uniformity of the single-site Lindblad superoperator means that $\mathcal{L}_{i}$ and $\mathcal{L}_{j}$ are equal as operators on $\mathcal{B}(\mathcal{H}_{\rm loc})$ (or its subalgebra $\mathcal{N}$) for systems without (or with) the number-superselection rule.
    See Lemma \ref{lem:commutable} for the definition and details of $\mathcal{B}_{\rm com}$. For example, since $\mathcal{B}_{\rm com} = {\rm span}\{\hat{P}_{\sigma} | \sigma \in \mathfrak{S}_{n}\}$, $\hat{H}_{ij}$ should belong to ${\rm span}\{\hat{I}, \hat{P}_{ij}\}$.
\end{theorem}

\noindent\textit{Proof of Theorem \ref{thm:equivalent_sec3}}
\begin{enumerate}[(1)]
    \item (II) $\Rightarrow$ (I)\\
    Let us assume that $\hat{H}_{ij} \in \mathcal{B}_{\rm com}$ and denote the initial density matrix as $\hat{\rho}(0) = \hat{\rho}_{\rm loc}(0)^{\otimes n}$. Let us also assume that the single-site Lindblad superoperator $\mathcal{L}_{i}$ is uniform and denote it as $\ell$ ($\forall i\ \mathcal{L}_{i} = \ell$ as a single-site superoperator). Then, it can straightforwardly be verified that the following solution satisfies the GKSL equation.
    \begin{align}\label{eq3-2}
        \hat{\rho}(t) = \hat{\rho}_{\rm loc}(t)^{\otimes n}, \ \ \hat{\rho}_{\rm loc}(t) = e^{\ell t}\qty(\hat{\rho}_{\rm loc}(0)). 
    \end{align}
    One might wonder if the solution given in Eq.~\eqref{eq3-2} is a ``physical'' density matrix. For spin systems and massless boson systems, which are not subject to the number-superselection rule, it is obvious that $\mathcal{\rho}_{\rm loc}(t)^{\otimes n}$ is a density matrix, since $\dv{\hat{\rho}}{t} = \ell(\hat{\rho}(t))$ can be seen as the GKSL equation on a single site. Due to the complete positivity, trace preservation of $e^{\ell t}$, and the preservation of Hermiticity (i.e., $\qty(\ell(\hat{X}))^{\dagger} = \ell(\hat{X}^{\dagger})$), $\hat{\rho}_{\rm loc}(t) = e^{\ell t}\hat{\rho}_{\rm loc}(0)$ is also a positive semidefinite Hermitian operator with a unit trace.

    \indent On the other hand, for fermion systems and massive boson systems, which are subject to the number-superselection rule, it is unclear whether or not the solution given in Eq.~\eqref{eq3-2} also commutes with the total number operator $\hat{N} = \sum_{i}\hat{n}_{i}$. Since
    \begin{align}\label{eq3-3}
        [e^{\mathcal{L}t}(\hat{\rho}), \hat{N}] = 0
    \end{align}
    should hold for every density matrix $\hat{\rho}$ that commutes with $\hat{N}$, we obtain $[\mathcal{L}(\hat{\rho}), \hat{N}] = 0$ by taking the derivative of Eq.~\eqref{eq3-3} with respect to $t$ and setting $t=0$. Using Eq.~\eqref{eq3-1-2}, we can conclude that
    \begin{align}\label{eq3-4}
        [\hat{X}, \hat{N}] = 0\ \ \Rightarrow \ \ [\mathcal{L}(\hat{X}), \hat{N}] = 0.
    \end{align}
    If the condition (II) is satisfied, we obtain
    \begin{align}\label{eq3-5}
        [\hat{\rho}_{\rm loc}, \hat{n}] = 0\ \ \Rightarrow \ \ [\ell(\hat{\rho}_{\rm loc}), \hat{n}] = 0,
    \end{align}
    by substituting $\hat{X} = \hat{\rho}_{\rm loc}^{\otimes n}$ into Eq.~\eqref{eq3-4} and taking ${\rm Tr}_{\overline{i}}$, where $\hat{n}$ is the single-site number operator. Therefore, $\hat{\rho}_{\rm loc}(t) = e^{\ell t}(\hat{\rho}_{\rm loc}(0))$ commutes with $\hat{n}$ and $\hat{\rho}(t) = \hat{\rho}_{\rm loc}(t)^{\otimes n}$ commutes with $\hat{N}$, which means that Eq.~\eqref{eq3-2} gives a physical density matrix. 
    
    \item (I) $\Rightarrow$ (II)\\
    It follows from condition (I) that there should exist $\hat{\rho}_{\rm loc}(t)$ that satisfies the GKSL equation $\dv{\hat{\rho}_{\rm loc}(t)^{\otimes n}}{t} = \mathcal{L}(\hat{\rho}_{\rm loc}(t)^{\otimes n})$ for arbitrary $\hat{\rho}_{\rm loc}(0)$. By calculating ${\rm Tr}_{\overline{ij}}$ of the GKSL equation and setting $t = 0$, we obtain
     \begin{align}
      -{\rm i}[\hat{H}_{ij}&, \hat{\rho}_{\rm loc}(0)\otimes \hat{\rho}_{\rm loc}(0)]\notag \\
      &+ \hat{\sigma}_{i}\otimes \hat{\rho}_{\rm loc}(0) + \hat{\rho}_{\rm loc}(0)\otimes \hat{\sigma}_{j} = 0,\label{eq3-6-1}\\
      \hat{\sigma}_{i}\coloneq &-\left.\dv{\hat{\rho}_{\rm loc}(t)}{t}\right|_{t = 0} + \mathcal{L}_{i}(\hat{\rho}_{\rm loc}(0))\notag \\
      &- {\rm i}\sum_{k\neq i, j}{\rm Tr}_{k}[\hat{H}_{ik}, \hat{\rho}_{\rm loc}(0)\otimes \hat{\rho}_{\rm loc}(0)],\label{eq3-6-2}\\
      \hat{\sigma}_{j}\coloneq &-\left.\dv{\hat{\rho}_{\rm loc}(t)}{t}\right|_{t = 0} +\mathcal{L}_{j}(\hat{\rho}_{\rm loc}(0)) \notag \\
      &- {\rm i}\sum_{k\neq i, j}{\rm Tr}_{k}[\hat{H}_{jk}, \hat{\rho}_{\rm loc}(0)\otimes \hat{\rho}_{\rm loc}(0)].\label{eq3-6-3}
    \end{align}
    Here, both $\hat{\sigma}_{i}$ and $\hat{\sigma}_{j}$ are traceless matrices. Just as we have shown in the proof of Theorem \ref{thm:equivalent_SS}, $\hat{\sigma}_{i} = \hat{\sigma}_{j} = 0$ can be derived as long as $\hat{\rho}_{\rm loc}(0)$ is a regular matrix. Substituting it in Eq.~\eqref{eq3-6-1},  we have $[\hat{H}_{ij}, \hat{\rho}_{\rm loc}(0)\otimes \hat{\rho}_{\rm loc}(0)] = 0$ for arbitrary regular matrix $\hat{\rho}_{\rm loc}(0)$. Therefore, it follows from Lemma \ref{lem:commutable} that $\hat{H}_{ij}\in \mathcal{B}_{\rm com}$. Moreover, using $\hat{H}_{ij} \in \mathcal{B}_{\rm com}$ and taking ${\rm Tr}_{\overline{i}}$ of the GKSL equation, we obtain
    \begin{align}\label{eq3-7}
      \forall i\ \ \mathcal{L}_{i}(\hat{\rho}_{\rm loc}(t)) = \dv{\hat{\rho}_{\rm loc}(t)}{t}.
    \end{align}
    Since $\hat{\rho}_{\rm loc}(t)$ can be chosen arbitrarily from $\mathcal{S}(\mathcal{H}_{\rm loc})$, or $\{\hat{\rho}_{\rm loc}\in \mathcal{S}(\mathcal{H}_{\rm loc}) | [\hat{\rho}_{\rm loc}, \hat{n}] = 0\}$ for systems with the number-superselection rule, the single-site Lindblad operator $\mathcal{L}_{i}$ should be uniform.
      \qed
\end{enumerate}

The mathematical structure of this theorem reminds us of a decoherence-free subspace \cite{PhysRevLett.81.2594}. The decoherence-free subspace is a subspace of density matrices in which the dissipative term of the Lindbladian vanish, i.e., $\sum_{k}\mathcal{D}_{\hat{L}_{k}}(\hat{\rho}) = 0$. On the other hand, our theorem states that the set of quantum i.i.d. states constitutes a subspace where the contribution to the Lindbladian from $\hat{H}_{ij}$ vanishes iff $\hat{H}_{ij} \in \mathcal{B}_{\rm com}$.\\

We also note that if an open quantum many-body system whose Hamiltonian consists of at most 2-body terms satisfies the sufficient condition in Theorem \ref{thm:sufficient}, the equivalent condition in Theorem \ref{thm:equivalent_sec3} is satisfied. See Appendix \ref{sec_SM_C} for the proof.



\subsection{Analytical results for dynamical properties}\label{sec3B}
For an open quantum many-body system that satisfies the condition (II) in Theorem \ref{thm:equivalent_sec3}, some dynamical properties such as time-correlation functions and response functions can be calculated analytically. 
For a spin system or a massless boson system, where $\mathcal{B}_{\rm com} = {\rm span}\{\hat{P}_{\sigma \in \mathfrak{S}_{n}}\}$, $\hat{H}_{ij}\in \mathcal{B}_{\rm com}$ reduces to $\hat{H}_{ij} \in {\rm span}\{\hat{I}, \hat{P}_{ij}\}$. We prove that time-correlation function of a physical quantity, which can be expressed as a sum of uniform 1-local operators, can be calculated analytically in the following theorem.

\begin{theorem}\label{thm:timecorrelation}
    \textit{Expression of time-correlation functions}\\
    Let us consider an open quantum many-body system of a spin system or a massless boson system with ${\rm dim}\mathcal{H}_{\rm loc} = d$.
    Suppose the system Hamiltonian satisfies $\hat{H}_{ij} \in {\rm span}\{\hat{I}, \hat{P}_{ij}\}$ and the single-site Lindblad operator $\mathcal{L}_{i}$ is uniform. Let $\{ \hat{X}_{i}^{\alpha} \}_{\alpha \in \Lambda}$ be an orthonormal basis of $\mathfrak{su}(d)$. Since the single-site superoperator $\mathcal{L}_{i}$ is trace-preserving and preserves Hermiticity, $\mathfrak{su}(d)$ is an invariant subspace of $\mathcal{L}_{i}$. Let $\mathfrak{L}$ be a $(d^2 -1)\times (d^2-1)$ matrix that represents the action of $\mathcal{L}_{i}$ on $\mathfrak{su}(d)$, i.e.,
    $\mathfrak{L}_{\alpha\beta} = {\rm Tr}[\hat{X}_{i}^{\alpha}\mathcal{L}_{i}(\hat{X}_{i}^{\beta})]$. Note that the matrix representation $\mathfrak{L}$ does not depend on $i$ since $\mathcal{L}_{i}$ is uniform. Then, the time-correlation function of $\hat{X}^{\alpha} = \sum_{i}\hat{X}_{i}^{\alpha}$ and $\hat{X}^{\beta} = \sum_{i}\hat{X}_{i}^{\beta}$, which is defined as
  \begin{align}\label{eq3-8}
    C_{\alpha\beta}(t + \tau, t) &\coloneq {\rm Tr}\left[ \hat{X}^{\alpha}e^{\mathcal{L}\tau}\hat{X}^{\beta}\hat{\rho}(t) \right],
  \end{align}
  can be expressed as
  \begin{align}\label{eq3-9}
    C_{\alpha \gamma}(t + \tau, t) = \qty(e^{\mathfrak{L}\tau})_{\alpha\beta}C_{\beta\gamma}(t, t).
  \end{align}
  Here and henceforth, the superoperators are assumed to act on the right of themselves, unless explicitly indicated by parentheses. For example, r.h.s. of Eq.~\eqref{eq3-8} is ${\rm Tr}\left[ \hat{X}^{\alpha}e^{\mathcal{L}\tau}\qty(\hat{X}^{\beta}\hat{\rho}(t)) \right]$.
\end{theorem}
\noindent \textit{Proof of Theorem \ref{thm:timecorrelation}}\\
This Theorem is essentially an application of the quantum regression theorem \cite{PhysRev.129.2342, gardiner2000quantum, breuer2002theory}. First of all, the derivative of the time-correlation function is written as
  \begin{align}\label{eq3-10}
    \dv{\tau}C_{\alpha\gamma}(t + \tau, t) &= {\rm Tr}\left[\hat{X}^{\alpha}\mathcal{L}e^{\mathcal{L}\tau}\hat{X}^{\gamma} \hat{\rho}(t) \right]\notag \\
    &={\rm Tr}\left[ \mathcal{L}^{\dagger}(\hat{X}^{\alpha})e^{\mathcal{L}\tau}\hat{X}^{\gamma} \hat{\rho}(t) \right].
  \end{align}
  Since $\hat{X}^{\alpha}$ has a permutation symmetry and commutes with $\hat{P}_{ij}$, we have
  \begin{align}\label{eq3-11}
    \mathcal{L}^{\dagger}(\hat{X}^{\alpha}) &= {\rm i}\sum_{(i, j)}\left[ \hat{H}_{ij}, \hat{X}^{\alpha} \right] + \sum_{i}\mathcal{L}_{i}^{\dagger}\qty(\hat{X}^{\alpha}) \notag \\
    &= \sum_{i, j}\mathcal{L}_{i}^{\dagger}\qty(\hat{X}_{j}^{\alpha}) = \sum_{i} \mathcal{L}_{i}^{\dagger}\qty(\hat{X}_{i}^{\alpha}).
  \end{align}
  Since $\{ \hat{X}_{i}^{\alpha} \}_{\alpha \in \Lambda}$ is the orthonormal basis, the matrix representation of $\mathcal{L}_{i}^{\dagger}$ as a superoperator that acts on $\mathfrak{su}(d)$ is $\mathfrak{L}^{\dagger}$. Since $\{ \hat{X}_{i}^{\alpha} \}_{\alpha \in \Lambda}$ is a set of Hermitian operators, we obtain
  \begin{align}\label{eq3-12}
    \mathfrak{L}_{\alpha\beta}^{\ast} &= {\rm Tr}\left[\qty(\hat{X}_{i}^{\alpha}\mathcal{L}_{i}(\hat{X}_{i}^{\beta}))^{\dagger}\right] = {\rm Tr}\left[\mathcal{L}_{i}(\hat{X}_{i}^{\beta})\hat{X}_{i}^{\alpha}\right] \notag \\
    &= {\rm Tr}[\hat{X}_{i}^{\alpha}\mathcal{L}_{i}(\hat{X}_{i}^{\beta})] = \mathfrak{L}_{\alpha\beta},
  \end{align}
  which means that $\mathfrak{L}$ is a real matrix. Therefore, $\mathcal{L}^{\dagger}(\hat{X}^{\alpha})$ can be rewritten as
  \begin{align}\label{eq3-13}
    \mathcal{L}^{\dagger}(\hat{X}^{\alpha}) = \sum_{i} \mathfrak{L}_{\beta \alpha}^{\dagger}\hat{X}_{i}^{\beta} = \mathfrak{L}_{\alpha\beta}\hat{X}^{\beta}.
  \end{align}
  Substituting Eq.~\eqref{eq3-13} into Eq.~\eqref{eq3-10}, we obtain the following first-order differential equation of the time-correlation function:
  \begin{align}\label{eq3-14}
    \dv{\tau}C_{\alpha \gamma}(t + \tau, t) = \mathfrak{L}_{\alpha\beta}C_{\beta\gamma}(t + \tau, t).
  \end{align}
  Thus, we obtain
  \begin{align}\label{eq3-15}
    C_{\alpha\gamma} (t + \tau, t) = \qty(e^{\mathfrak{L}\tau})_{\alpha\beta}C_{\beta\gamma}(t, t).
  \end{align}
  \qed
\\

We can find the analytical expression of time-correlation functions for fermion and massive boson systems that satisfy condition (II) in Theorem \ref{thm:equivalent_sec3} as well. The only difference is that we need to consider the following Lie algebra instead of $\mathfrak{su}(d)$:
\begin{align}\label{eq3-16}
    \mathfrak{n} \coloneq \{\hat{X}\in \mathcal{B}(\mathcal{H}_{\rm loc})\ |\  {\rm Tr}[\hat{X}] = 0,\ \hat{X}^{\dagger} = \hat{X},\ [\hat{X}, \hat{n}] = 0\},
\end{align}
where $\hat{n}$ is the single-site number operator.
To be precise, as in the case of $\mathfrak{su}(d)$, it is necessary to multiply the elements of $\mathfrak{n}$ by the imaginary unit in order for the Lie bracket $[\hat{X}, \hat{Y}]$ to be closed within the algebra. As we have seen in Eq.~\eqref{eq3-4} in the proof of Theorem \ref{thm:equivalent_sec3}, $\mathcal{L}_{i}(\hat{X})$ commutes with $\hat{n}$ if $\hat{X}\in \mathcal{B}(\mathcal{H}_{\rm loc})$ commutes with $\hat{n}$. Therefore, $\mathfrak{n}$ is an invariant subspace of $\mathcal{L}_{i}$. Therefore, by taking an orthonormal basis of $\mathfrak{n}$ denoted as $\{\hat{X}_{i}^{\alpha}\}_{\alpha\in \Lambda}$, and representing the single-site Lindbladian $\mathcal{L}_{i}$ as a matrix $\mathfrak{L}$ in this basis, which gives a ${\rm dim}\ \mathfrak{n} \times {\rm dim}\ \mathfrak{n}$ matrix, we obtain the same expression of the time-correlation function as given in Eq.~\eqref{eq3-9}.
\\
 
According to the linear response theory for open quantum systems \cite{PhysRevA.93.032101, PhysRevA.95.022126, PhysRevResearch.1.033156}, the response functions can be related to time-correlation functions. We consider the dynamics by the perturbed Lindbladian $\mathcal{L}(t) = \mathcal{L}_{0} + \xi(t)\mathcal{L}^{\prime}$, where the perturbation of the Lindbladian originates from the perturbation of the Hamiltonian $-\xi(t)\hat{A}$, i.e. $\mathcal{L}^{\prime}(\bullet) = {\rm i}[\hat{A}, \bullet]$. Let $\hat{\rho}(0)$ be the initial density matrix of the system, which is not necessarily a steady state, and let $\hat{\rho}(t)$ be the density matrix at time $t$ evolved under the perturbed Lindbladian. Then, the difference of the expectation value of an observable $\hat{B}$ at time $t$ with and without the perturbation is expressed as 
\begin{align}\label{eq3-17}
  \delta B(t) &= {\rm Tr}[\hat{B}\hat{\rho}(t)] - {\rm Tr}[\hat{B}e^{\mathcal{L}_{0}t}\hat{\rho}(0)]\notag \\
  &=\int_{0}^{t}{\rm d}\tau\ \xi(\tau)\phi_{BA}(t, \tau) + O(\xi^2),\\
  \phi_{BA}(t, \tau) &= -2{\rm Im}\left[ C_{BA}(t, \tau)\right],
\end{align}
where $C_{BA}(t, \tau)$ is the time-correlation function of a system without perturbation, defined as follows:
\begin{align}\label{eq3-18}
    C_{BA}(t, \tau) &\coloneq {\rm Tr}\left[ \hat{A}e^{\mathcal{L}_{0}(t-\tau)}\hat{B}e^{\mathcal{L}_{0}\tau}\hat{\rho}(0) \right].
\end{align}
Thus, the response function can also be calculated analytically if $\hat{A}$ and $\hat{B}$ are expressed as sums of 1-local operators.

\section{Specific Examples}\label{sec4}

There are various models for open quantum many-body systems in which the steady state becomes a quantum i.i.d. state. In spin-1/2 systems, a dissipative isotropic Heisenberg model has a quantum i.i.d. steady state, if the magnetic field and the dissipation are spatially uniform (Example \ref{ex1}). Even if the Heisenberg interaction is anisotropic, the dissipative system may still possess a quantum i.i.d. steady state at a particular magnetic field, as illustrated in Example \ref{ex2}. A dissipative spinless fermion model (Example \ref{ex3}) and a dissipative hard-core boson model (Example \ref{ex6}) with quite general Hamiltonian, including the hopping term and the interaction term ($\hat{n}_{i}\hat{n}_{j}$), also turn out to have a quantum i.i.d. steady state. Furthermore, in spin-1/2 fermion systems, both a dissipative $t-J$ model (Example \ref{ex4}) and a dissipative Hubbard model (Example \ref{ex5}) have a quantum i.i.d. steady state, regardless of their parameters. A key observation is that the steady state is determined by the single-site Lindbladian $\mathcal{L}_{i}$, as evidenced by condition (iii) of Theorem \ref{thm:equivalent_SS} or condition (iii$^{\prime}$) of Lemma \ref{lem:equivalent_SS}. Therefore, the quantum i.i.d. steady state is independent of the parameters of $\hat{H}_{ij}$, as well as the geometry and dimensionality of the underlying lattice.

\subsection{Spin Systems}
\begin{example}\label{ex1} dissipative spin-1/2 isotropic Heisenberg model\\
\indent For a spin-1/2 system, the permutation operator can be written in terms of the spin operators as $\hat{P}_{ij} = 2\bm{\hat{S}}_{i}\cdot \bm{\hat{S}}_{j} + \frac{1}{2}\hat{I}$.
Therefore, the isotropic Heisenberg interaction $\bm{\hat{S}}_{i}\cdot \bm{\hat{S}}_{j}$ can be written as a linear combination of a permutation operator and the identity operator. Thus, a dissipative spin-1/2 isotropic Heisenberg model with a spatially uniform magnetic field and a spatially uniform local dissipation satisfies the sufficient condition in Theorem \ref{thm:sufficient}, and consequently has a quantum i.i.d. steady state.
Since it also satisfies condition (II) in Theorem \ref{thm:equivalent_sec3}, the state maintains the quantum i.i.d. form throughout the time evolution, as long as the initial state is a quantum i.i.d. state. Moreover, from Theorem \ref{thm:timecorrelation}, the analytical expressions of time-correlation functions and response functions can be obtained.\\

For example, let us consider the following $n$-site dissipative spin-1/2 isotropic Heisenberg model with the local dissipation of $\hat{S}_{i}^{-}$ and the magnetic field uniformly applied on every site.
\begin{align}
    \hat{H} &= \sum_{(i, j)}J_{ij}\hat{\bm{S}}_{i}\cdot \hat{\bm{S}}_{j} - B\sum_{i}\hat{S}_{i}^{x},\label{eq4-1-1} \\
    \mathcal{L}(\hat{\rho}(t)) &= -{\rm i}[\hat{H}, \hat{\rho}(t)] + \gamma \sum_{i}\mathcal{D}_{\hat{S}_{i}^{-}}(\hat{\rho}(t)).\label{eq4-1-2}
\end{align}
Then, the single-site Lindbladian $\mathcal{L}_{i}$ is given as follows:
\begin{align}\label{eq4-2}
    \mathcal{L}_{i}\qty(\bullet) = {\rm i}B[\hat{S}_{i}^{x}, \bullet] + \gamma\mathcal{D}_{\hat{S}_{i}^{-}}(\bullet).
\end{align}
By solving $\mathcal{L}_{i}(\hat{\rho}_{\rm loc}) = 0$, the quantum i.i.d. steady state can be obtained as follows:
\begin{align}\label{eq4-3}
    \hat{\rho}_{\rm SS} &= \qty(\frac{1}{2B^2 + \gamma^2}\mymat{B^2 & {\rm i}B\gamma \\
    -{\rm i}B\gamma & B^2 + \gamma^2})^{\otimes n}.
\end{align}
Thus, the total magnetization $\bm{M}_{\rm SS} = {\rm Tr}\left[\hat{\bm{S}}\hat{\rho}_{\rm SS}\right]$ can be calculated as follows:
\begin{align}\label{eq4-4}
    \bm{M}_{\rm SS} &= \frac{1}{2}n\mymat{s_{x} & s_{y} & s_{z}}= -\frac{1}{2}n \mymat{0 & \frac{2\gamma B}{2B^2 + \gamma^2}& \frac{\gamma^2}{2B^2 + \gamma^2}}.
\end{align}
Here, $s_{x}, s_{y}, s_{z}\in \R$ are the expansion coefficients of $\hat{\rho}_{\rm loc}$ in terms of Pauli matrices, i.e., $\hat{\rho}_{\rm loc} = \frac{1}{2}\qty(\hat{I} + s_{x}\hat{X}_{i} + s_{y}\hat{Y}_{i} + s_{z}\hat{Z}_{i})\ \ (s_{x}, s_{y}, s_{z} \in \R,\ 0 \leq s_{x}^2 + s_{y}^2 + s_{z}^2 \leq 1)$. The time-correlation function of the total magnetization in the quantum i.i.d. steady state, defined by $\tilde{C}_{\alpha \beta}(t) = {\rm Tr}\left[ \hat{S}^{\alpha}e^{\mathcal{L}t}\hat{S}^{\beta}\hat{\rho}_{\rm SS} \right] - {\rm Tr}[\hat{S}^{\alpha}\hat{\rho}_{\rm SS}]\cdot{\rm Tr}[\hat{S}^{\beta}\hat{\rho}_{\rm SS}]$, can be calculated as follows:
  \begin{align}
    &\tilde{C}_{\alpha\gamma}(t) =  \frac{n}{4}\qty(e^{\mathfrak{L}t})_{\alpha\beta}\qty(\delta_{\beta\gamma}-s_{\beta}s_{\gamma}+{\rm i}\epsilon_{\beta\gamma\delta}s_{\delta}),\label{eq4-5-1}\\
    &e^{\mathfrak{L}t} = P\ {\rm diag}\qty(e^{-\frac{1}{2}\gamma t}, e^{\lambda_{+}t}, e^{\lambda_{-}t})P^{-1},\label{eq4-5-2}\\
    &\lambda_{\pm} = -\frac{3}{4}\gamma\pm \sqrt{\frac{1}{16}\gamma^2-B^2},\label{eq4-5-3}\\
    &P = \mymat{1 & 0 & 0\\
    0 & B & -\frac{1}{4}\gamma + \sqrt{\frac{1}{16}\gamma^2-B^2}\\
    0 & -\frac{1}{4}\gamma + \sqrt{\frac{1}{16}\gamma^2-B^2} & B}\label{eq4-5-4}.
  \end{align}
Note that all the above results depend neither on the geometry and the spatial dimension of the lattice, nor on the coupling constants $J_{ij}$'s.\\    
\end{example}

\begin{example}\label{ex2} dissipative spin-1/2 model with YZ+ZY interactions\\
\indent Let us consider the following $n$-site dissipative spin-1/2 model with YZ+ZY interactions, and utilize Corollary \ref{cor:equivalent_SS_2dim} to find when the system has a quantum i.i.d. steady state.
\begin{align}
    &\hat{H} = \sum_{(i, j)}J_{ij}\qty(r\hat{S}_{i}^{y} + \hat{S}_{i}^{z})\qty(r\hat{S}_{j}^{y} + \hat{S}_{j}^{z})  - B\sum_{i}\hat{S}_{i}^{x}, \label{eq4-6-1}\\
    &\mathcal{L}(\hat{\rho}(t)) = -{\rm i}[\hat{H}, \hat{\rho}(t)] + \gamma \sum_{i}\mathcal{D}_{\hat{S}_{i}^{-}}(\hat{\rho}(t)).\label{eq4-6-2}
\end{align}
First, let us examine the possibility of having a quantum i.i.d. steady state $\hat{\rho}_{\rm loc}^{\otimes n}$, where $\hat{\rho}_{\rm loc}$ is a pure state (Case 1 of Corollary \ref{cor:equivalent_SS_2dim}). We find from condition (i) in Corollary \ref{cor:equivalent_SS_2dim} that the only possibility is $\hat{\rho}_{\rm loc} = \ketbra{\downarrow}{\downarrow}$. However, since $\ket{\downarrow}^{\otimes n}$ is not an eigenstate of $\hat{H}_{\rm eff}$, such a quantum i.i.d. steady state does not exist. Next, let us examine the possibility of having a quantum i.i.d. steady state $\hat{\rho}_{\rm loc}^{\otimes n}$, where $\hat{\rho}_{\rm loc}$ is a regular matrix (Case 2 of Corollary \ref{cor:equivalent_SS_2dim}). We use condition (iii) in Corollary \ref{cor:equivalent_SS_2dim} to solve $\mathcal{L}_{i}(\hat{\rho}_{\rm loc}) = 0$, obtaining
\begin{align}\label{eq4-7}
    \hat{\rho}_{\rm loc} &= \frac{1}{2B^2 + \gamma^2}\mymat{B^2 & {\rm i}\gamma B\\
    -{\rm i}\gamma B & B^2 + \gamma^2}\notag \\
    &= \frac{1}{2}\qty(\hat{I}-\frac{2\gamma B}{2B^2 + \gamma ^2}\hat{Y} -\frac{\gamma^2}{2B^2 + \gamma ^2}\hat{Z}).
  \end{align}
Since $\hat{H}_{ij}$ should satisfy Eq.~\eqref{eq2-27} because of condition (iv) in Corollary \ref{cor:equivalent_SS_2dim}, $B = r\gamma/2$ should be satisfied in order to have a quantum i.i.d. steady state. Therefore, the system has a quantum i.i.d. steady state only at the particular value of the magnetic field.
\end{example}


\subsection{Fermi Systems}
\begin{example}\label{ex3} dissipative spinless fermion model\\
\indent Let us consider the following $n$-site dissipative spinless fermion model with hopping and interaction terms.
\begin{align}
    \hat{H} &= -\sum_{(i, j)} t_{ij}\qty(\hat{c}_{i}^{\dagger}\hat{c}_{j} + \hat{c}_{j}^{\dagger}\hat{c}_{i}) + \sum_{(i, j)}V_{ij}\hat{n}_{i}\hat{n}_{j} +  
    \mu\sum_{i}\hat{n}_{i}, \label{eq4-8-1}\\
    \mathcal{L}(\hat{\rho}(t)) &= -{\rm i}[\hat{H}, \hat{\rho}(t)] + \gamma_{-}\sum_{i}\mathcal{D}_{\hat{c}_{i}}(\hat{\rho}(t)) + \gamma_{+}\sum_{i}\mathcal{D}_{\hat{c}_{i}^{\dagger}}(\hat{\rho}(t)),\label{eq4-8-2}
\end{align}
where $\hat{n}_{i} = \hat{c}_{i}^{\dagger}\hat{c}_{i}$ is the number operator. The permutation operator of a spinless fermion system is given as follows:
\begin{align}\label{eq4-9}
    \hat{P}_{ij} = 1 - \qty(\hat{c}_{i}^{\dagger}-\hat{c}_{j}^{\dagger})\qty(\hat{c}_{i} - \hat{c}_{j}).
\end{align}
Therefore, the Hamiltonian can be rewritten as 
\begin{align}\label{eq4-10}
    \hat{H} = -\sum_{(i, j)}t_{ij}\qty(\hat{P}_{ij} + \hat{n}_{i} + \hat{n}_{j}-1) + \sum_{(i, j)}V_{ij}\hat{n}_{i}\hat{n}_{j} + \mu\sum_{i}\hat{n}_{i},
\end{align}
which belongs to $\mathcal{B}_{\rm com}$. Since the Lindblad operators are uniform, this system has a quantum i.i.d. steady state as a consequence of Theorem \ref{thm:sufficient}. The quantum i.i.d. steady state can be obtained by solving
\begin{align}\label{eq4-11}
    \ell_{i}(\hat{\rho}_{\rm loc}) = \gamma_{-}\mathcal{D}_{\hat{c}_{i}}\qty(\hat{\rho}_{\rm loc}) + \gamma_{+}\mathcal{D}_{\hat{c}_{i}^{\dagger}}\qty(\hat{\rho}_{\rm loc}) = 0.
\end{align}
The solution to Eq.~\eqref{eq4-11} is given by
\begin{align}\label{eq4-12}
    \hat{\rho}_{\rm SS} = \qty(\frac{\gamma_{-}}{\gamma_{-} + \gamma_{+}}\ketbra{0}{0} + \frac{\gamma_{+}}{\gamma_{-} + \gamma_{+}}\hat{c}_{i}^{\dagger}\ketbra{0}{0}\hat{c}_{i})^{\otimes n},
\end{align}
where $\ket{0}$ denotes the vacuum state.\\
\end{example}

Examples \ref{ex4} and \ref{ex5} concern dissipative spin-1/2 fermion models.
\begin{example}\label{ex4} dissipative $t-J$ model\\
\indent Let us consider the strong coupling regime ($U \gg t$) of a Hubbard model whose Hamiltonian can be written as follows:
\begin{align}\label{eq4-13}
    \hat{H}_{\rm Hubbard} = -\sum_{(i, j),\alpha}t_{ij}\qty(\hat{c}_{i\alpha}^{\dagger}\hat{c}_{j\alpha} + \hat{c}_{j\alpha}^{\dagger}\hat{c}_{i\alpha}) + U\sum_{i}\hat{n}_{i\uparrow}\hat{n}_{i\downarrow}.
\end{align}
Due to the large on-site repulsion, the low-energy Hilbert space is restricted to states with no doubly occupied sites.
The effective Hamiltonian $\hat{H}_{\rm eff}$ is obtained by the second-order perturbation of the Hubbard Hamiltonian in $t/U$, which is written as
\begin{align}
    \hat{H}_{\rm eff} &= \hat{\Pi}_{0}\qty(\hat{H}_{\rm hop} + \hat{H}_{\rm HI} + \hat{H}_{\rm three})\hat{\Pi}_{0},\label{eq4-14-1}\\
    \hat{H}_{\rm hop} &= -\sum_{(i, j),\alpha}t_{ij}\qty(\hat{c}_{i\alpha}^{\dagger}\hat{c}_{j\alpha} + {\rm h.c.}),\label{eq4-14-2}\\
    \hat{H}_{\rm HI} &= \sum_{(i, j)}J_{ij}\qty(\hat{\bm{S}}_{i}\cdot \hat{\bm{S}}_{j}-\frac{1}{4}\hat{n}_{i}\hat{n}_{j}),\label{eq4-14-3}\\
    \hat{H}_{\rm three} &= \frac{1}{2U}\sum_{\substack{j, k, l, \alpha, \beta, \alpha^{\prime}, \beta^{\prime}\\j\neq k\neq l\neq j}}t_{jk}t_{kl}\qty(\hat{c}_{j\alpha}^{\dagger}\bm{\sigma}_{\alpha\beta}\hat{c}_{l\beta}\cdot \hat{c}_{k\alpha^\prime}^{\dagger}\bm{\sigma}_{\alpha^{\prime}\beta^{\prime}}\hat{c}_{k\beta^{\prime}}),\label{eq4-14-4}
\end{align}
where $\hat{\Pi}_{0} = \prod_{k}\qty(1-\hat{n}_{k\uparrow}\hat{n}_{k\downarrow})$ is a projection operator to the subspace with no doubly occupied sites, $\alpha, \beta, \alpha^{\prime}, \beta^{\prime}\in \{\uparrow, \downarrow\}$, $J_{ij} = t_{ij}^2/U$, and $\bm{\sigma} = (\sigma^{x}, \sigma^{y}, \sigma^{z})$ are Pauli matrices. Spin operator $\hat{\bm{S}}_{i}$ is defined from the fermionic creation/annihilation operator by $\hat{\bm{S}}_{i} = \frac{1}{2}\hat{c}_{i\alpha}^{\dagger}\bm{\sigma}_{\alpha\beta}\hat{c}_{i\beta}$. Here $\hat{H}_{\rm hop}$ denotes the hopping term, $\hat{H}_{\rm HI}$ describes the isotropic Heisenberg interaction, and $\hat{H}_{\rm three}$ denotes the three-site terms. The three-site terms are often ignored, and the resulting Hamiltonian is called the $t-J$ Hamiltonian, which is written as follows:
\begin{align}\label{eq4-15}
    \hat{H}_{t-J} &= -\sum_{(i, j)}\left(t_{ij}\sum_{\alpha}(1 - \hat{n}_{i\bar{\alpha}})(1 - \hat{n}_{j\bar{\alpha}})\qty(\hat{c}_{i\alpha}^{\dagger}\hat{c}_{j\alpha} + \hat{c}_{j\alpha}^{\dagger}\hat{c}_{i\alpha})\right.\notag \\
    &\quad + \left. J_{ij}\qty(\hat{\bm{S}}_{i}\cdot \hat{\bm{S}}_{j}-\frac{1}{4}\hat{n}_{i}\hat{n}_{j})\right)\hat{\Pi}_{0/ij},
\end{align}
where $\hat{\Pi}_{0/ij} = \prod_{k\neq i, j}(1-\hat{n}_{k\uparrow}\hat{n}_{k\downarrow})$, and $\bar{\alpha}$ denotes the spin antiparallel to $\alpha$ \cite{essler2005one}. Now, let us show that $\hat{H}_{t-J}$ belongs to $\mathcal{B}_{\rm com} = \{\hat{P}_{\sigma\in \mathfrak{S}_{n}}, \hat{n}_{i\in \Lambda}\}^{\prime\prime}$, which is a set of operators generated by the scalar multiplication, addition, and multiplication of $\hat{P}_{\sigma}$'s and $\hat{n}_{i}$'s. First, using $\hat{n}_{k} = \hat{n}_{k\uparrow} + \hat{n}_{k\downarrow}$ and $\hat{n}_{k\alpha}^{2} = \hat{n}_{k\alpha}\ \ (\alpha\in \{\uparrow, \downarrow\})$, the projection operator $\hat{\Pi}_{0/ij}$ can be rewritten as
\begin{align}\label{eq4-16}
    \hat{\Pi}_{0/ij} = \prod_{k\neq i, j}\qty(1-\frac{\hat{n}_{k}^2-\hat{n}_{k}}{2}).
\end{align}
By using the permutation operator $\hat{P}_{ij}$ which is written as
\begin{align}\label{eq4-17}
    \hat{P}_{ij} = \prod_{\alpha \in \{\uparrow, \downarrow\}}\qty(1 - \qty(\hat{c}_{i\alpha}^{\dagger}-\hat{c}_{j\alpha}^{\dagger})\qty(\hat{c}_{i\alpha} - \hat{c}_{j\alpha})),
\end{align}
the hopping term and the isotropic Heisenberg interaction term can be rewritten as
\begin{align}
    \hspace{-2mm}\sum_{\alpha}(1 - \hat{n}_{i\bar{\alpha}})(1 - \hat{n}_{j\bar{\alpha}})\qty(\hat{c}_{i\alpha}^{\dagger}\hat{c}_{j\alpha} + \hat{c}_{j\alpha}^{\dagger}\hat{c}_{i\alpha}) &= \hat{P}_{ij}\hat{\Pi}_{ij}^{(1)},\label{eq4-18-1}\\
    \hat{\bm{S}}_{i}\cdot\hat{\bm{S}}_{j} = \frac{1}{4}&\hat{P}_{ij}\hat{\Pi}_{ij}^{(1, 1)},\label{eq4-18-2}
\end{align}
where $\hat{\Pi}_{ij}^{(1)}$ is a projection operator onto the eigenspace where the eigenvalue of $\hat{n}_{i} + \hat{n}_{j}$ is $1$, and $\hat{\Pi}_{ij}^{(1, 1)}$ is a projection operator onto the eigenspace where the eigenvalues of $\hat{n}_{i}$ and $\hat{n}_{j}$ are both $1$. As long as the Hilbert space is finite-dimensional, a projection operator onto the eigenspace of $\hat{A}$ associated with the eigenvalue $\alpha_{m}$ can be written as a polynomial of $\hat{A}$ as follows:
\begin{align}\label{eq4-19}
    \hat{\Pi}^{(\hat{A} = \alpha_{m})} = \prod_{k\neq m}\frac{\hat{A}-\alpha_{k}}{\alpha_{m} - \alpha_{k}},
\end{align}
where $\{\alpha_{k}\}$ is the set of eigenvalues of $\hat{A}$. Therefore, the projection operators $\hat{\Pi}_{ij}^{(1)}$ and $\hat{\Pi}_{ij}^{(1, 1)}$ can be written as polynomials of $\hat{n}_{i}$ and $\hat{n}_{j}$. Thus, $\hat{H}_{t-J}$ can be written as a polynomial of $\hat{P}_{ij}, \hat{n}_{i}$, and $\hat{n}_{j}$, and hence $\hat{H}_{t-J} \in \mathcal{B}_{\rm com}$. Therefore, from Theorem \ref{thm:sufficient}, the dissipative $t-J$ model has a quantum i.i.d. steady state regardless of the parameters and the geometry of the lattice, if the Lindblad operators are $1$-local and uniform.\\

For example, let us consider the following $n$-site dissipative $t-J$ model:
\begin{align}
    \hat{H} &= \hat{H}_{t-J} -B_{x}\sum_{i}\hat{S}_{i}^{x} - B_{z}\sum_{i}\hat{S}_{i}^{z},\label{eq4-20-1}\\
    \mathcal{L}\qty(\hat{\rho}(t)) &= -{\rm i}[\hat{H}, \hat{\rho}(t)] + \gamma\sum_{i}\mathcal{D}_{\hat{S}_{i}^{-}}\qty(\hat{\rho}(t)),\label{eq4-20-2}
\end{align}
where $\hat{S}_{i}^{-} = \hat{c}_{i\downarrow}^{\dagger}\hat{c}_{i\uparrow}$. The quantum i.i.d. steady state $\hat{\rho}_{\rm SS} = \hat{\rho}_{\rm loc}^{\otimes n}$ can be obtained from the solution of the single-site equation $\ell_{i}(\hat{\rho}_{\rm loc}) = 0$, giving
\begin{align}\label{eq4-21}
    \hat{\rho}_{\rm loc} &= \frac{r}{\gamma^2 + 2B_{x}^2 + 4B_{z}^2}\left(B_{x}^2 \ketbra{\uparrow}{\uparrow} - B_{x}\qty(2B_{z}-{\rm i}\gamma)\ketbra{\uparrow}{\downarrow}\right.\notag\\
    &\quad \left.-B_{x}\qty(2B_{z} + {\rm i}\gamma)\ketbra{\downarrow}{\uparrow} + \qty(\gamma^2 + B_{x}^2 + 4B_{z}^2)\ketbra{\downarrow}{\downarrow}\right)\notag \\
    &\quad + (1-r)\ketbra{0}{0},
\end{align}
where $0 \leq r \leq 1$ can be arbitrarily chosen (the steady state is not unique), and $\ket{0}$ is a vacuum state. The total number of particles and the total magnetization of the quantum i.i.d. steady state $\hat{\rho}_{\rm SS}$ are given by
\begin{align}
    N_{\rm SS} &= {\rm Tr}\left[\hat{N}\hat{\rho}_{\rm SS}\right] = nr,\label{eq4-22-1}\\
    \bm{M}_{\rm SS} &= {\rm Tr}\left[\hat{\bm{S}}\hat{\rho}_{\rm SS}\right] \notag \\
    &= -\frac{nr}{2\qty(\gamma^2 + 2B_{x}^2 + 4B_{z}^2)}\mymat{4B_{x}B_{z} \\ 2B_{x}\gamma \\ \gamma^2 + 4B_{z}^2},\label{eq4-22-2}
\end{align}
where $\hat{N} = \sum_{i}\hat{n}_{i}$ and $\hat{\bm{S}} = \sum_{i}\hat{\bm{S}}_{i}$.\\
\end{example}

\begin{example}\label{ex5} dissipative Hubbard model\\
\indent The Hubbard Hamiltonian in Eq.~\eqref{eq4-13} does not belong to $\mathcal{B}_{\rm com}$. Although the on-site repulsive terms belong to $\mathcal{B}_{\rm com} = \{\hat{P}_{\sigma \in \mathfrak{S}_{n}}, \hat{n}_{i\in \Lambda}\}^{\prime\prime}$ since $\hat{n}_{i\uparrow}\hat{n}_{i\downarrow} = \qty(\hat{n}_{i}^2 - \hat{n}_{i})/2$, the hopping term does not commute with every single-site density matrix $\hat{\rho}_{\rm loc}$ that satisfies $[\hat{\rho}_{\rm loc}, \hat{n}] = 0$, where $\hat{n} $ is the number operator on $\mathcal{H}_{\rm loc}$. For simplicity, we introduce the following notation:
\begin{align}\label{eq4-23}
    \ket{\uparrow\downarrow} \coloneq \hat{c}_{\uparrow}^{\dagger}\hat{c}_{\downarrow}^{\dagger}\ket{0},\ \ 
    \ket{\uparrow} \coloneq \hat{c}_{\uparrow}^{\dagger}\ket{0},\ \ \ket{\downarrow}\coloneq \hat{c}_{\downarrow}\ket{0},
\end{align}
where $\ket{0}$ is the vacuum state. 
The single-site density matrix $\hat{\rho}_{\rm loc}$ that commutes with $\hat{n}$ can be parametrized as follows:
\begin{align}\label{eq4-24}
    \hat{\rho}_{\rm loc} &= r_{11}\ketbra{\uparrow\downarrow}{\uparrow\downarrow} + r_{22}\ketbra{\uparrow}{\uparrow} + r_{23}\ketbra{\uparrow}{\downarrow} + r_{23}^{\ast}\ketbra{\downarrow}{\uparrow}\notag \\
    &\quad + r_{33}\ketbra{\downarrow}{\downarrow} + r_{44}\ketbra{0}{0},
\end{align}
where $r_{11}, r_{22}, r_{33}$, and $r_{44}$ are real parameters that satisfy $r_{11} + r_{22} + r_{33} + r_{44} = 1$ and $r_{23}$ is a complex parameter. 
We also introduce the following notation to express the state of sites $i$ and $j$ $(i < j)$:
\begin{align}\label{eq4-25}
    \ket{\uparrow\downarrow, \uparrow\downarrow} &\coloneq \hat{c}_{i\uparrow}\hat{c}_{i\downarrow}\hat{c}_{j\uparrow}\hat{c}_{j\downarrow}\ket{0},\notag \\
    \ket{\uparrow\downarrow, \uparrow} &\coloneq \hat{c}_{i\uparrow}\hat{c}_{i\downarrow}\hat{c}_{j\uparrow}\ket{0},\quad \textrm{etc.}
\end{align}
We adopt the ordering $(\hat{c}_{i\uparrow}, \hat{c}_{i\downarrow}, \hat{c}_{j\uparrow}, \hat{c}_{j\downarrow})$ for the fermionic creation operators.
Then, the commutator of $\sum_{\alpha\in \{\uparrow, \downarrow\}}\qty(\hat{c}_{i\alpha}^{\dagger}\hat{c}_{j\alpha} + \hat{c}_{j\alpha}^{\dagger}\hat{c}_{i\alpha})$ and $\hat{\rho}_{\rm loc}\otimes \hat{\rho}_{\rm loc}$ is calculated as
\begin{align}\label{eq4-26}
    &\quad \left[ \sum_{\alpha\in \{\uparrow,\downarrow\}}\qty(\hat{c}_{i\alpha}^{\dagger}\hat{c}_{j\alpha} + \hat{c}_{j\alpha}^{\dagger}\hat{c}_{i\alpha}), \hat{\rho}_{\rm loc}\otimes \hat{\rho}_{\rm loc} \right]\notag \\
    &= (r_{11}r_{44}-r_{22}r_{33} + |r_{23}|^2)\cdot\notag \\
    &\quad\left(\ketbra{\uparrow, \downarrow}{\uparrow \downarrow, 0} + \ketbra{\uparrow, \downarrow}{0, \uparrow\downarrow} - \ketbra{\downarrow, \uparrow}{\uparrow\downarrow, 0} - \ketbra{\downarrow, \uparrow}{0, \uparrow\downarrow} \right.\notag\\
  &\quad\left. -\ketbra{\uparrow\downarrow, 0}{\uparrow, \downarrow} -\ketbra{0, \uparrow\downarrow}{\uparrow, \downarrow} + \ketbra{\uparrow\downarrow, 0}{\downarrow, \uparrow} + \ketbra{0, \uparrow\downarrow}{\downarrow, \uparrow}\right).
\end{align}
Therefore, $\hat{\rho}_{\rm loc}$ should satisfy 
\begin{align}\label{eq4-27}
    r_{11}r_{44}-r_{22}r_{33} + |r_{23}|^2 = 0,
\end{align}
in order for $\hat{\rho}_{\rm loc}\otimes \hat{\rho}_{\rm loc}$ to commute with $\sum_{\alpha\in \{\uparrow,\downarrow\}}\qty(\hat{c}_{i\alpha}^{\dagger}\hat{c}_{j\alpha} + \hat{c}_{j\alpha}^{\dagger}\hat{c}_{i\alpha})$. Thus, $\hat{H}_{\rm Hubbard}$ does not belong to $\mathcal{B}_{\rm com}$. However, $\hat{H}_{\rm Hubbard}\notin \mathcal{B}_{\rm com}$ does not mean that the dissipative Hubbard model does not have a quantum i.i.d. steady state. Recalling Lemma \ref{lem:equivalent_SS}, if there exists $\hat{\rho}_{\rm loc}$ that satisfies
 \begin{enumerate}[(i$^{\prime}$)]
    \setcounter{enumi}{2}
    \item $\forall i\in \Lambda\ \  \mathcal{L}_{i}(\hat{\rho}_{\rm loc}) = 0$.
    \item $\forall (i, j)\in \Lambda_{2}\ \ [\hat{H}_{ij}, \hat{\rho}_{\rm loc}\otimes \hat{\rho}_{\rm loc}] = 0$.
  \end{enumerate}
at the same time, the system has a quantum i.i.d. steady state. Since $\hat{H}_{ij} = -t_{ij}\qty(\sum_{\alpha\in \{\uparrow,\downarrow\}}\qty(\hat{c}_{i\alpha}^{\dagger}\hat{c}_{j\alpha} + \hat{c}_{j\alpha}^{\dagger}\hat{c}_{i\alpha})) + V_{ij}\qty(\hat{n}_{i}-1)\qty(\hat{n}_{j}-1)$ for the Hubbard Hamiltonian, condition (iv$^{\prime}$) reduces to Eq.~\eqref{eq4-27}. Therefore, if we adjust $\mathcal{L}_{i}$ so that the solution of $\mathcal{L}_{i}(\hat{\rho}_{\rm loc}) = 0$ satisfies Eq.~\eqref{eq4-27}, we can constitute a system with a quantum i.i.d. steady state.

Let us consider a dissipative Hubbard model described by following Hamiltonian and Lindbladian:
\begin{align}
    \hat{H} = \hat{H}&_{\rm Hubbard} + \sum_{(i, j)}V_{ij}\hat{n}_{i}\hat{n}_{j}- B_{x}\sum_{i}\hat{S}_{i}^{x} - B_{z}\sum_{i}\hat{S}_{i}^{z},\label{eq4-28-1}\\
    \mathcal{L}(\hat{\rho}(t)) &= -{\rm i}[\hat{H}, \hat{\rho}(t)] \notag \\
    &\ \  + \sum_{i, \alpha}\qty(\gamma_{+\alpha}\mathcal{D}_{\hat{c}_{i\alpha}^{\dagger}}(\hat{\rho}(t)) + \gamma_{-\alpha}\mathcal{D}_{\hat{c}_{i\alpha}}(\hat{\rho}(t))) .\label{eq4-28-2}
\end{align}
Then, the single-site Lindbladian $\mathcal{L}_{i}$ is given as follows:
\begin{align}\label{eq4-29}
    \mathcal{L}_{i}(\bullet) &= -{\rm i}U[(\hat{n}_{i}^2 - \hat{n}_{i})/2, \bullet] -{\rm i}\sum_{k\neq i}V_{ik}[\hat{n}_{i}, \bullet] \notag \\
    &\quad + {\rm i}B_{x}[\hat{S}_{i}^{x}, \bullet] + {\rm i}B_{z}[\hat{S}_{i}^{z}, \bullet]\notag \\
    &\quad + \sum_{\alpha}\qty(\gamma_{+\alpha}\mathcal{D}_{\hat{c}_{i\alpha}^{\dagger}}(\bullet) + \gamma_{-\alpha}\mathcal{D}_{\hat{c}_{i\alpha}}(\bullet)),
\end{align}
where $V_{ij} \coloneq V_{ji}$ for $i > j$. Recalling the number-superselection rule, $\hat{\rho}_{\rm loc}$ should commute with $\hat{n}_{i}$. Therefore, first two terms in Eq.~\eqref{eq4-29} can be ignored, and the solution to $\mathcal{L}_{i}(\hat{\rho}_{\rm loc}) = 0$ is given by
\begin{align}\label{eq4-30}
    \begin{aligned}
    r_{11} &= \frac{\gamma_{+\uparrow}\gamma_{+\downarrow}}{\gamma_{\uparrow}\gamma_{\downarrow}} -\frac{\gamma_{\uparrow}\gamma_{+\uparrow}-\gamma_{\downarrow}\gamma_{+\downarrow}}{\gamma \gamma_{\uparrow}\gamma_{\downarrow}}B_{x}{\rm Im}(r_{23}) ,\\
  r_{22} &= \frac{\gamma_{+\uparrow}\gamma_{-\downarrow}}{\gamma_{\uparrow}\gamma_{\downarrow}} + \frac{\gamma_{\uparrow}\qty(\gamma_{+\uparrow} + \gamma_{\downarrow}) + \gamma_{\downarrow}\gamma_{-\downarrow}}{\gamma\gamma_{\uparrow}\gamma_{\downarrow}}B_{x}{\rm Im}(r_{23}),\\
  r_{33} &= \frac{\gamma_{-\uparrow}\gamma_{+\downarrow}}{\gamma_{\uparrow}\gamma_{\downarrow}} - \frac{\gamma_{\downarrow}\qty(\gamma_{+\downarrow} + \gamma_{\uparrow}) + \gamma_{\uparrow}\gamma_{-\uparrow}}{\gamma\gamma_{\uparrow}\gamma_{\downarrow}}B_{x}{\rm Im}(r_{23}),\\
  r_{44} &= \frac{\gamma_{-\uparrow}\gamma_{-\downarrow}}{\gamma_{\uparrow}\gamma_{\downarrow}} - \frac{\gamma_{\downarrow}\gamma_{-\downarrow}-\gamma_{\uparrow}\gamma_{-\uparrow}}{\gamma\gamma_{\uparrow}\gamma_{\downarrow}}B_{x}{\rm Im}(r_{23}),\\
  r_{23} &= \frac{\qty(\gamma_{+\downarrow}\gamma_{-\uparrow}-\gamma_{+\uparrow}\gamma_{-\downarrow})B_{x}\qty(-2B_{z} + {\rm i}\gamma)}{\gamma^2\gamma_{\uparrow}\gamma_{\downarrow} + B_{x}^2\gamma^2 + 4B_{z}^2\gamma_{\uparrow}\gamma_{\downarrow}},
\end{aligned}
\end{align}
where $\hat{\rho}_{\rm loc}$ is parametrized by Eq.~\eqref{eq4-24}. The definitions of $\gamma_{\uparrow}, \gamma_{\downarrow}$, and $\gamma$ are given by
\begin{align}
    \gamma_{\uparrow} &\coloneq \gamma_{+\uparrow} + \gamma_{-\uparrow}, \label{eq4-31-1}\\
    \gamma_{\downarrow} &\coloneq \gamma_{+\downarrow} + \gamma_{-\downarrow},\label{eq4-31-2}\\
    \gamma &\coloneq \gamma_{\uparrow} + \gamma_{\downarrow.}\label{eq4-31-3}
\end{align}
Then, the parameters in Eq.~\eqref{eq4-30} satisfy Eq.~\eqref{eq4-27}, which is equivalent to condition (iv$^{\prime}$). Therefore, the $n$-site dissipative Hubbard model described by Eqs.~\eqref{eq4-28-1} and \eqref{eq4-28-2} always have a quantum i.i.d. steady state $\hat{\rho}_{\rm loc}^{\otimes n}$ where $\hat{\rho}_{\rm loc}$ is determined by Eq.~\eqref{eq4-30}, regardless of the parameters. The total number and the total magnetization of the quantum i.i.d. steady state are given by 
\begin{align}
    N_{\rm SS} &= {\rm Tr}\left[\hat{N}\hat{\rho}_{\rm SS}\right] = n\left( \frac{2\gamma_{+\uparrow}\gamma_{+\downarrow} + \gamma_{+\uparrow}\gamma_{-\downarrow} + \gamma_{+\downarrow}\gamma_{-\uparrow}}{\gamma_{\uparrow}\gamma_{\downarrow}}\right. \notag \\
    &\quad \left.- \frac{B_{x}^2\gamma\qty(\gamma_{\uparrow}-\gamma_{\downarrow})\qty(\gamma_{+\downarrow}\gamma_{-\uparrow} - \gamma_{+\uparrow}\gamma_{-\downarrow})}{\gamma_{\uparrow}\gamma_{\downarrow}\qty(\gamma^2\gamma_{\uparrow}\gamma_{\downarrow} + B_{x}^2\gamma^2 + 4B_{z}^2\gamma_{\uparrow}\gamma_{\downarrow})} \right),\label{eq4-32-1}\\
    \bm{M}_{\rm SS}&= {\rm Tr}\left[\hat{\bm{S}}\hat{\rho}_{\rm SS} \right]\notag \\
    &=\frac{n\qty(\gamma_{+\uparrow}\gamma_{-\downarrow}-\gamma_{+\downarrow}\gamma_{-\uparrow})}{2\qty(\gamma^2\gamma_{\uparrow}\gamma_{\downarrow} + B_{x}^2\gamma^2 + 4B_{z}^2\gamma_{\uparrow}\gamma_{\downarrow})}\mymat{4B_{x}B_{z} \\ 2B_{x}\gamma \\ \gamma^2 + 4B_{z}^2}, \label{eq4-32-2}
\end{align}
where $\hat{N} = \sum_{i}\hat{n}_{i}$ and $\hat{\bm{S}} = \sum_{i}\hat{\bm{S}}_{i}$.
\end{example}

\subsection{Bose Systems}
\begin{example}\label{ex6} dissipative hard-core boson model\\
\indent Let us consider a dissipative model of a hard-core boson system, where commutation/anticommutation relations are given as follows:
\begin{align}\label{eq4-33}
    &[\hat{b}_{i}, \hat{b}_{j}] = [\hat{b}_{i}, \hat{b}_{j}^{\dagger}] = 0\quad (i\neq j),\notag \\
    &\{\hat{b}_{i}, \hat{b}_{i}^{\dagger}\} = 1. 
\end{align}
It follows that $\{\hat{b}_{i}, \hat{b}_{i}\} = \{\hat{b}_{i}^{\dagger}, \hat{b}_{i}^{\dagger}\} = 0$. Now, consider an $n$-site model with the following Hamiltonian and Lindbladian:
\begin{align}
    \hat{H} &= -\sum_{(i, j)} t_{ij}\qty(\hat{b}_{i}^{\dagger}\hat{b}_{j} + \hat{b}_{i}\hat{b}_{j}^{\dagger}) + \sum_{(i, j)}V_{ij}\hat{n}_{i}\hat{n}_{j} +  
    \mu\sum_{i}\hat{n}_{i}, \label{eq4-34-1}\\
    \mathcal{L}(\hat{\rho}(t)) &= -{\rm i}[\hat{H}, \hat{\rho}(t)] + \gamma_{-}\sum_{i}\mathcal{D}_{\hat{b}_{i}}(\hat{\rho}(t)) + \gamma_{+}\sum_{i}\mathcal{D}_{\hat{b}_{i}^{\dagger}}(\hat{\rho}(t)),\label{eq4-34-2}
\end{align}
where $\hat{n}_{i} = \hat{b}_{i}\hat{b}_{j}$ is the number operator. The permutation operator for a hard-core boson system can be written as follows:
\begin{align}\label{eq4-35}
    \hat{P}_{ij} = 1 - \qty(\hat{b}_{i}^{\dagger}-\hat{b}_{j}^{\dagger})\qty(\hat{b}_{i}-\hat{b}_{j}) + 2\hat{b}_{i}^{\dagger}\hat{b}_{i}\hat{b}_{j}^{\dagger}\hat{b}_{j}.
\end{align}
Therefore, the Hamiltonian can be rewritten as follows:
\begin{align}\label{eq4-36}
    \hat{H} &= -\sum_{(i, j)}t_{ij}\qty(\hat{P}_{ij} + \hat{n}_{i} + \hat{n}_{j}-2\hat{n}_{i}\hat{n}_{j} - 1) \notag \\
    &\quad + \sum_{(i, j)}V_{ij}\hat{n}_{i}\hat{n}_{j} + \sum_{i}\mu\hat{n}_{i},
\end{align}
which belongs to $\mathcal{B}_{\rm com}$. Since the Lindblad operators are uniform, this system has a quantum i.i.d. steady state as a consequence of Theorem \ref{thm:sufficient}. Just as the dissipative spinless fermion model, the quantum i.i.d. steady state can be obtained as
\begin{align}\label{eq4-37}
    \hat{\rho}_{\rm SS} = \qty(\frac{\gamma_{-}}{\gamma_{-} + \gamma_{+}}\ketbra{0}{0} + \frac{\gamma_{+}}{\gamma_{-} + \gamma_{+}}\hat{b}_{i}^{\dagger}\ketbra{0}{0}\hat{b}_{i})^{\otimes n}.
\end{align}
\end{example}

\section{Discussion and Conclusion}\label{sec5}
In Theorem \ref{thm:equivalent_SS}, we have derived an equivalent condition for $\mathcal{L}(\hat{\rho}_{\rm loc}^{\otimes n}) = 0$, which means that the system has the quantum i.i.d. state $\hat{\rho}_{\rm loc}^{\otimes n}$ as a steady state. The equation $\mathcal{L}(\hat{\rho}_{\rm loc}^{\otimes n}) = 0$ has been reformulated as conditions on local terms such as $\hat{H}_{i}, \hat{H}_{ij}$, and $\hat{L}_{i}^{(\alpha)}$, thereby providing a practical criterion for determining whether or not a given open quantum many-body system has a quantum i.i.d. steady state. This theorem can also be applied to investigate what class of open quantum many-body system has a quantum i.i.d. steady state. Although Theorem \ref{thm:equivalent_SS} holds for an arbitrary dimension of local Hilbert space, we have restricted the dimension to $d = 2$, resulting in a simple statement in Corollary \ref{cor:equivalent_SS_2dim}. Then, $\mathcal{B}_{\rm com}$ has been defined as a set of operators that commute with every quantum i.i.d. state, and it is equivalent to ${\rm span}\{\hat{P}_{\sigma \in \mathfrak{S}_{n}}\}$ or $\{\hat{P}_{\sigma\in \mathfrak{S}_{n}}, \hat{n}_{i\in \Lambda}\}^{\prime\prime}$, as shown in Lemma \ref{lem:commutable}. If the Hamiltonian can be written as a sum of $H_{\rm com}\in \mathcal{B}_{\rm com}$ and spatially uniform 1-local Hermitian operators, and if the Lindblad operators are 1-local and spatially uniform, we can conclude that the system has a quantum i.i.d. steady state, regardless of the details of 1-local operators. This result gives a sufficient condition to have a quantum i.i.d. steady state in Theorem \ref{thm:sufficient}. These conditions for the existence of a quantum i.i.d. steady state can be regarded as a no-go theorem for spatial correlations and quantum entanglement in the steady state under the assumption that the steady state is unique. If we additionally assume the absence of purely imaginary eigenvalues for the Lindbladian, we can conclude that both spatial correlations and quantum entanglement decay exponentially in time.

We have also discussed a system whose density matrix maintains the quantum i.i.d. form if the quantum i.i.d. state is initially prepared. In Theorem \ref{thm:equivalent_sec3}, we have proved that a system possesses such a property if and only if $\hat{H}_{ij} \in \mathcal{B}_{\rm com}$ and the single-site Lindblad superoperator is spatially uniform. Furthermore, we have demonstrated in Theorem \ref{thm:timecorrelation} that the dynamical properties, including time-correlation functions and response functions, can be calculated analytically for such systems. This analytical tractability makes such systems an ideal platform for studying nonequilibrium statistical mechanics.

Examples of systems with quantum i.i.d. steady states have been presented across a wide range of systems including spin-1/2 systems, spinless fermion systems, spin-1/2 fermion systems, and hard-core boson systems. These examples show that systems with a quantum i.i.d. steady states are prevalent among open quantum many-body systems subject to local dissipation. Our study deepens the understanding of spatial correlations and quantum entanglement in steady states, providing insight into where to investigate dissipative quantum phase transitions.

\section*{Acknowledgment}
 We are grateful to Masaya Nakagawa and our lab members for fruitful discussions during the seminar. We appreciate Ryusuke Hamazaki for pointing out the relationship with permutation operators at the beginning of our research. We also appreciate Hosho Katsura and Zongping Gong for their insightful suggestions, which helped us to extend our theorems to a much general setting, and Marko Žnidarič for notifying us of the relationship with his earlier work. T.I. was supported by KAKENHI Grant No. JP25KJ0839 from the Japan Society for the Promotion of Science (JSPS) and the Forefront Physics and Mathematics Program to Drive Transformation (FoPM), a World-Leading Innovative Graduate Study (WINGS) Program, the University of Tokyo. M.U. was supported by KAKENHI Grant No. JP22H01152 from the JSPS. We gratefully acknowledge the support from the CREST program “Quantum Frontiers” (Grant No. JPMJCR23I1) by the Japan Science and Technology Agency.

\appendix

\section{Supplementary lemma to Corollary \ref{cor:equivalent_SS_2dim} for spin-1/2 systems}\label{sec_SM_A}
We prove the following statement for spin-1/2 systems.

\begin{lemma}\label{lem:Pauliexpression}
    A set of 2-body operators, whose partial trace on either site vanishes and belong to $\{\hat{P}_{ij}, \hat{\rho}_{\rm loc}\otimes \hat{\rho}_{\rm loc}\}^{\prime}$, is given by
    \begin{align}\label{eqA-1}
        &{\rm span}_{\R}\left\{\hat{X}_{i}\hat{X}_{j} + \hat{Y}_{i}\hat{Y}_{j} + \hat{Z}_{i}\hat{Z}_{j}, \right.\notag \\
        &\ \  \left.\qty(s_{x}\hat{X}_{i} + s_{y}\hat{Y}_{i} + s_{z}\hat{Z}_{i})\qty(s_{x}\hat{X}_{j} + s_{y}\hat{Y}_{j} + s_{z}\hat{Z}_{j})\right\},
    \end{align}
    where $\hat{P}_{ij}$ is a permutation operator, $\hat{\rho}_{\rm loc}\in \mathcal{S}(\mathcal{H}_{\rm loc})$ is a single-site density matrix, $\hat{X}, \hat{Y}$, and $\hat{Z}$ are Pauli matrices, and $i$ and $j$ are labels that indicate the sites on which the operators act. ${\rm span}_{\R}(S)$ denotes the set of all finite linear combinations over $\R$ of elements in $S$. The real numbers $s_{x}, s_{y}$, and $s_{z}$ are the parametrization of $\hat{\rho}_{\rm loc}$ in terms of Pauli matrices: $\hat{\rho}_{\rm loc} = \frac{1}{2}\qty(\hat{I} + s_{x}\hat{X} + s_{y}\hat{Y} + s_{z}\hat{Z})$.
\end{lemma}

\noindent\textit{Proof of Lemma \ref{lem:Pauliexpression}}\\
Let us express a set of 2-body operators that belong to $\{\hat{P}_{ij}, \hat{\rho}_{\rm loc}\otimes \hat{\rho}_{\rm loc}\}^{\prime}$ in terms of Pauli matrices. We first consider the following orthogonal basis of Hermitian operators on $\mathcal{H}_{\rm loc}\otimes \mathcal{H}_{\rm loc}$ that are symmetric with regard to sites $i$ and $j$:
\begin{widetext}
\begin{align}\label{eqA-2}
    \left\{ \frac{\hat{X}_{i} + \hat{X}_{j}}{\sqrt{2}}, \frac{\hat{Y}_{i} + \hat{Y}_{j}}{\sqrt{2}}, \frac{\hat{Z}_{i} + \hat{Z}_{j}}{\sqrt{2}}, \hat{X}_{i}\hat{X}_{j}, \hat{Y}_{i}\hat{Y}_{j}, \hat{Z}_{i}\hat{Z}_{j}, \frac{\hat{X}_{i}\hat{Y}_{j} + \hat{Y}_{i}\hat{X}_{j}}{\sqrt{2}}, \frac{\hat{Z}_{i}\hat{X}_{j} + \hat{X}_{i}\hat{Z}_{j}}{\sqrt{2}}, \frac{\hat{Y}_{i}\hat{Z}_{j} + \hat{Z}_{i}\hat{Y}_{j}}{\sqrt{2}} \right\}.
\end{align}
We parametrize a symmetric Hermitian operator $\hat{A}$ as

\begin{align}\label{eqA-3}
    \hat{A} &= \alpha_{x}\frac{\hat{X}_{i} + \hat{X}_{j}}{\sqrt{2}} + \alpha_{y}\frac{\hat{Y}_{i} + \hat{Y}_{j}}{\sqrt{2}} + \alpha_{z}\frac{\hat{Z}_{i} + \hat{Z}_{j}}{\sqrt{2}} + \alpha_{xx}\hat{X}_{i}\hat{X}_{j} + \alpha_{yy}\hat{Y}_{i}\hat{Y}_{j} + \alpha_{zz}\hat{Z}_{i}\hat{Z}_{j} \notag\\
    &\quad + \alpha_{xy}\frac{\hat{X}_{i}\hat{Y}_{j} + \hat{Y}_{i}\hat{X}_{j}}{\sqrt{2}} + \alpha_{zx}\frac{\hat{Z}_{i}\hat{X}_{j} + \hat{X}_{i}\hat{Z}_{j}}{\sqrt{2}} + \alpha_{yz}\frac{\hat{Y}_{i}\hat{Z}_{j} + \hat{Z}_{i}\hat{Y}_{j}}{\sqrt{2}},
\end{align}
where $\alpha$'s are real parameters, and define a real vector $\bm{\alpha} = \mymat{\alpha_{x} & \alpha_{y} & \alpha_{z} & \alpha_{xx} & \alpha_{yy} & \alpha_{zz} & \alpha_{xy} & \alpha_{zx} & \alpha_{yz}}^{\top}$.
Then, by expressing the equation $[\hat{A}, \hat{\rho}_{\rm loc}\otimes \hat{\rho}_{\rm loc}] = 0$ in terms of parameters $\bm{\alpha}$, $s_{x}, s_{y}$, and $s_{z}$, we obtain the matrix equation $M\bm{\alpha} = \bm{0}$, where $M$ is given by

\begin{align}\label{eqA-4}
    \frac{1}{2}\mymat{0 & s_{z} & -s_{y} & 0 & \sqrt{2}s_{y}s_{z} & -\sqrt{2}s_{y}s_{z} & s_{x}s_{z} & -s_{x}s_{y} & -s_{y}^2 + s_{z}^2\\
  -s_{z} & 0 & s_{x} & \sqrt{2}s_{x}s_{z} & 0 & \sqrt{2}s_{x}s_{z} & -s_{y}s_{z} & s_{x}^2-s_{z}^2 & s_{x}s_{y}\\
  s_{y} & -s_{x} & 0 & \sqrt{2}s_{x}s_{y} & -\sqrt{2}s_{x}s_{y} & 0 & -s_{x}^2 + s_{y}^2 & s_{y}s_{z} & -s_{x}s_{z}\\
  0 & \sqrt{2}s_{x}s_{z} & -\sqrt{2}s_{x}s_{y} & 0 & 0 & 0 & \sqrt{2}s_{z} & -\sqrt{2}s_{y} & 0\\
  -\sqrt{2}s_{y}s_{z} & 0 & \sqrt{2}s_{x}s_{y} & 0 & 0 & 0 & -\sqrt{2}s_{z} & 0 & \sqrt{2}s_{x}\\
  \sqrt{2}s_{y}s_{z} & -\sqrt{2}s_{x}s_{z} & 0 & 0 & 0 & 0 & 0 & \sqrt{2}s_{y} & -\sqrt{2}s_{x}\\
  -s_{x}s_{z} & s_{y}s_{z} & s_{x}^2-s_{y}^2 & -\sqrt{2}s_{z} & \sqrt{2}s_{z} & 0 & 0 & s_{x} & -s_{y}\\
  s_{x}s_{y} & -s_{x}^2 + s_{z}^2 & -s_{y}s_{z} & \sqrt{2}s_{y} & 0 & -\sqrt{2}s_{y} & -s_{x} & 0 & s_{z}\\
  s_{y}^2-s_{z}^2 & -s_{x}s_{y} & s_{x}s_{z} & 0 & -\sqrt{2}s_{x} & \sqrt{2}s_{x} & s_{y} & -s_{z} & 0}.
\end{align}
Therefore, we find
\begin{align}\label{eqA-5}
    &\quad\left\{ \hat{A} | \hat{A}^{\dagger} = \hat{A}, \hat{A} \in \{\hat{P}_{ij}, \hat{\rho}_{\rm loc}\otimes \hat{\rho}_{\rm loc}\}^{\prime}\right\}\notag\\
    &= {\rm span}_{\R}\left\{s_{x}(\hat{X}_{i} + \hat{X}_{j}) + s_{y}(\hat{Y}_{i} + \hat{Y}_{j}) + s_{z}(\hat{Z}_{i} + \hat{Z}_{j}), \hat{X}_{i}\hat{X}_{j} + \hat{Y}_{i}\hat{Y}_{j} + \hat{Z}_{i}\hat{Z}_{j}, (s_{x}\hat{X}_{i} + s_{y}\hat{Y}_{i} + s_{z}\hat{Z}_{i})(s_{x}\hat{X}_{j} + s_{y}\hat{Y}_{j} + s_{z}\hat{Z}_{j})\right\}.
\end{align}     
\end{widetext}
Restricting the above set of operators to operators whose partial trace on either site is zero, we obtain
\begin{align}\label{eqA-6}
    &{\rm span}_{\R}\left\{\hat{X}_{i}\hat{X}_{j} + \hat{Y}_{i}\hat{Y}_{j} + \hat{Z}_{i}\hat{Z}_{j}, \right.\notag \\
    &\quad \left.(s_{x}\hat{X}_{i} + s_{y}\hat{Y}_{i} + s_{z}\hat{Z}_{i})(s_{x}\hat{X}_{j} + s_{y}\hat{Y}_{j} + s_{z}\hat{Z}_{j})\right\}.
\end{align}
\qed

\section{Proof of Lemma \ref{lem:commutable}}\label{sec_SM_B}
We begin with the proof of Schur-Weyl duality. We then apply it to prove Lemma \ref{lem:commutable}.

\begin{theorem}\label{thm:Schur-Weyl}
    \textbf{Schur-Weyl duality for $GL(\mathcal{H}_{\rm loc})$}\\
    We consider a $d$-dimensional complex vector space $\mathcal{H}_{\rm loc}$ and its $n$-fold tensor product $\mathcal{H} = \mathcal{H}_{\rm loc}^{\otimes n}$. We denote the general linear group of $\mathcal{H}_{\rm loc}$ as $GL(\mathcal{H}_{\rm loc})$ and let $\hat{P}_{\sigma}$ for $\sigma\in \mathfrak{S}_{n}$ be a permutation operator on $\mathcal{H}$ defined by Eq.~\eqref{eq2-7}. Then, the following equalities on the subset of $\mathcal{B}(\mathcal{H})$ holds:
    \begin{align}
        {\rm span}\{\hat{G}^{\otimes n} | \hat{G}\in GL(\mathcal{H}_{\rm loc})\}^{\prime} = {\rm span}\{\hat{P}_{\sigma} | \sigma \in \mathfrak{S}_{n}\},\label{eqB-1-1}\\
        {\rm span}\{\hat{G}^{\otimes n} | \hat{G}\in GL(\mathcal{H}_{\rm loc})\} = {\rm span}\{\hat{P}_{\sigma} | \sigma \in \mathfrak{S}_{n}\}^{\prime}.\label{eqB-1-2}
    \end{align}
\end{theorem}
We prove Schur-Weyl duality in three steps \cite{fulton2013representation, watrous2018theory}. We first introduce the von Neumann bicommutant theorem, which states that $\mathcal{A} = \mathcal{A}^{\prime\prime}$ holds for von Neumann algebra $\mathcal{A}$. Since we are only concerned with finite-dimensional cases, we provide a proof of the von Neumann bicommutant theorem for the finite-dimensional case in Theorem \ref{thm:bicommutant}. Then, we prove ${\rm span}\{\hat{G}^{\otimes n} | \hat{G}\in GL(\mathcal{H}_{\rm loc})\}^{\prime} = {\rm span}\{\hat{P}_{\sigma} | \sigma \in \mathfrak{S}_{n}\}$ by showing ${\rm span}\{\hat{X}^{\otimes n} | \hat{X}\in \mathcal{B}(\mathcal{H}_{\rm loc})\}^{\prime} = {\rm span}\{\hat{P}_{\sigma} | \sigma \in \mathfrak{S}_{n}\}$ in Lemma \ref{lem:Schur-Weyl-1} and ${\rm span}\{\hat{X}^{\otimes n} | \hat{X}\in \mathcal{B}(\mathcal{H}_{\rm loc})\} = {\rm span}\{\hat{G}^{\otimes n} | \hat{G} \in GL(\mathcal{H}_{\rm loc})\}$ in Lemma \ref{lem:Schur-Weyl-2}.\\

\begin{theorem}\label{thm:bicommutant}
\textbf{Von Neumann bicommutant theorem for finite dimension}\\
    Let $\mathcal{H}$ be a finite-dimensional Hilbert space, and $\mathcal{A} \subset \mathcal{B}(\mathcal{H})$ be a subset of bounded linear operators that satisfies the following conditions:
    \begin{enumerate}[(i)]
        \item $\mathcal{A}$ is an algebra (i.e., it is closed under scalar multiplication over $\C$, addition and multiplication).
        \item $\mathcal{A}$ is closed under Hermitian conjugation.
        \item $\mathcal{A}$ includes $\hat{I}$, which is the identity operator on $\mathcal{H}$.
    \end{enumerate}
    Then, $\mathcal{A} = \mathcal{A}^{\prime\prime}$.
\end{theorem}
\noindent \textit{Proof of Theorem \ref{thm:bicommutant}}
\begin{enumerate}[(1)]
    \item First, let us prove that $\mathcal{A} \subset \mathcal{A}^{\prime\prime}$. From the definition of the commutant, $\hat{Y}\in \mathcal{A}^{\prime}$ commutes with an arbitrary element in $\mathcal{A}$. Thus, for $\hat{X} \in \mathcal{A}$, $\forall \hat{Y}\in \mathcal{A}^{\prime}\quad [\hat{X}, \hat{Y}] = 0$. Therefore, $\hat{X}\in \mathcal{A}^{\prime\prime}$, and thus $\mathcal{A} \subset \mathcal{A}^{\prime\prime}$.
    \item Next, let us prove that $\mathcal{A} \supset \mathcal{A}^{\prime\prime}$.\\
    For $\ket{\psi}\in \mathcal{H}$, we define $\mathcal{A}\ket{\psi}\subset \mathcal{H}$ as
    \begin{align}\label{eqB-2}
        \mathcal{A}\ket{\psi}&\coloneq \{\hat{X}\ket{\psi} | \hat{X}\in \mathcal{A}\}.
    \end{align}
    Clearly, $\mathcal{A}\ket{\psi}$ is a vector subspace of $\mathcal{H}$. Since $\mathcal{A}$ is closed under multiplication, we have
    \begin{align}\label{eqB-3}
        \forall\hat{X}\in \mathcal{A}\ \ \forall\ket{\varphi}\in \mathcal{A}\ket{\psi}\quad \hat{X}\ket{\varphi}\in \mathcal{A}\ket{\psi}.
    \end{align}
    Let us consider the orthogonal decomposition of Hilbert space $\mathcal{H} = \mathcal{A}\ket{\psi} \oplus \qty(\mathcal{A}\ket{\psi})^{\perp}$. For $\hat{X}\in \mathcal{A}$ and $\ket{\varphi} \in \mathcal{A}\ket{\psi}$, it follows from Eq.~\eqref{eqB-3} that $\hat{X}^{\dagger}\ket{\varphi} \in \mathcal{A}\ket{\psi}$, recalling that $\mathcal{A}$ is closed under conjugate transpose. Thus, we have $\mel{\varphi^{\prime}}{\hat{X}^{\dagger}}{\varphi} = 0$ for $\ket{\varphi^{\prime}}\in \qty(\mathcal{A}\ket{\psi})^{\perp}$, which means that $\hat{X}\ket{\varphi^{\prime}}$ is orthogonal to arbitrary $\ket{\varphi} \in \mathcal{A}\ket{\psi}$. Therefore, we have
    \begin{align}\label{eqB-4}
        \forall \hat{X}\in \mathcal{A}\ \ \forall \ket{\varphi^{\prime}} \in \qty(\mathcal{A}\ket{\psi})^{\perp}\quad \hat{X}\ket{\varphi^{\prime}} \in \qty(\mathcal{A}\ket{\psi})^{\perp}.
    \end{align}
    Let $\hat{\Pi}_{\psi}$ be a projection operator to $\mathcal{A}\ket{\psi}$. Then, for arbitrary $\hat{X}\in \mathcal{A}$, it follows from Eq.~\eqref{eqB-3} and \eqref{eqB-4} that
    \begin{align}
        &\forall\ket{\varphi}\in \mathcal{A}\ket{\psi}\quad   \hat{\Pi}_{\psi}\hat{X}\ket{\varphi} = \hat{X}\ket{\varphi} = \hat{X}\hat{\Pi}_{\psi}\ket{\varphi},\label{eqB-5-1}\\
        &\forall\ket{\varphi^{\prime}}\in \qty(\mathcal{A}\ket{\psi})^{\perp}\quad \hat{\Pi}_{\psi}\hat{X}\ket{\varphi^{\prime}} = 0 = \hat{X}\hat{\Pi}_{\psi}\ket{\varphi^{\prime}}. \label{eqB-5-2}
    \end{align}
    Therefore, $\hat{\Pi}_{\psi}\hat{X}\ket{\phi} = \hat{X}\hat{\Pi}_{\psi}\ket{\phi}$ for arbitrary $\ket{\phi}\in \mathcal{H}$, which implies that $[\hat{\Pi}_{\psi}, \hat{X}] = 0$. Since this holds for arbitrary $\hat{X}\in \mathcal{A}$, it follows that $\hat{\Pi}_{\psi}\in \mathcal{A}^{\prime}$. Consequently, for any $\hat{Z}\in \mathcal{A}^{\prime\prime}$, we have
    \begin{align}\label{eqB-6}
        \hat{Z}\ket{\psi} = \hat{Z}\hat{\Pi}_{\psi}\ket{\psi} = \hat{\Pi}_{\psi}\hat{Z}\ket{\psi}.
    \end{align}
    Thus, we obtain $\hat{Z}\ket{\psi}\in \mathcal{A}\ket{\psi}$, and therefore $\mathcal{A}\ket{\psi}\supset \mathcal{A}^{\prime\prime}\ket{\psi}$. On the other hand, from (1), it is obvious that $\mathcal{A}\ket{\psi}\subset \mathcal{A}^{\prime\prime}\ket{\psi}$. We therefore conclude that
    \begin{align}\label{eqB-7}
        \mathcal{A}\ket{\psi} = \mathcal{A}^{\prime\prime}\ket{\psi}.
    \end{align}
    For $\hat{X}\in \mathcal{B}(\mathcal{H})$, let us define an operator $\hat{X}^{(d)}$ on $\mathcal{H}\oplus \cdots \oplus \mathcal{H}$, which is $d$-fold direct sum of $\mathcal{H}$ as follows:
    \begin{align}\label{eqB-8}
        &\hat{X}^{(d)}\qty(\ket{\varphi_1}\oplus \ket{\varphi_2}\oplus \cdots \oplus \ket{\varphi_d}) \notag \\
        &= \hat{X}\ket{\varphi_1} \oplus \hat{X}\ket{\varphi_2} \oplus \cdots \oplus \hat{X}\ket{\varphi_d},
    \end{align}
    where $d= {\rm dim}\mathcal{H}$ and $\ket{\varphi_i} \in \mathcal{H}\ (1 \leq i \leq d)$. Let us also define sets of operators $\mathcal{A}^{(d)}, \mathcal{A}^{\prime \prime(d)}\in \mathcal{B}\qty(\mathcal{H}\oplus \cdots \oplus \mathcal{H})$ as follows:
    \begin{align}
        \mathcal{A}^{(d)} &= \{\hat{X}^{(d)} | \hat{X}\in \mathcal{A}\}, \label{eqB-9-1}\\
        \mathcal{A}^{\prime\prime(d)} &= \{ \hat{X}^{(d)} | \hat{X} \in \mathcal{A}^{\prime\prime}\}.\label{eqB-9-2}
    \end{align}
    Operators on $\mathcal{H} \oplus \cdots \oplus \mathcal{H}$ can be represented as $d\times d$ matrices, whose components are elements of $\mathcal{B}(\mathcal{H})$. For example, $\hat{X}^{(d)}$ can be represented as a diagonal matrix ${\rm diag}(\hat{X}, \dots, \hat{X})$. Matrices that commute with ${\rm diag}(\hat{X}, \dots, \hat{X})$ for any $\hat{X}\in \mathcal{A}$ are $d\times d$ matrices whose components belong to $\mathcal{A}^{\prime}$. Moreover, matrices that commute with all such matrices are ${\rm diag}(\hat{X}, \dots, \hat{X})$ for $\hat{X}\in \mathcal{A}^{\prime\prime}$. Thus, it follows that
    \begin{align}\label{eqB-10}
        \qty(\mathcal{A}^{(d)})^{\prime\prime} = \mathcal{A}^{\prime\prime(d)}.
    \end{align}
    Let $\{\ket{\psi_i}\}_{i=1}^{d}$ be a basis of $\mathcal{H}$. Using Eq.~\eqref{eqB-7}, we have
    \begin{align}\label{eqB-11}
        \mathcal{A}^{(d)}\qty(\ket{\psi_1}\oplus \cdots \oplus \ket{\psi_d}) = \mathcal{A}^{\prime\prime(d)}\qty(\ket{\psi_1}\oplus \cdots \oplus \ket{\psi_d}).
    \end{align}
    Thus, for $\hat{Z}\in \mathcal{A}^{\prime\prime}$, there exists $\hat{X}\in \mathcal{A}$ that satisfies
    \begin{align}\label{eqB-12}
        &\hat{X}\ket{\psi_1}\oplus \hat{X}\ket{\psi_2} \oplus \cdots \oplus \hat{X}\ket{\psi_d}\notag \\
        &= \hat{Z}\ket{\psi_1}\oplus \hat{Z}\ket{\psi_2} \oplus \cdots \oplus \hat{Z}\ket{\psi_d}.
    \end{align}
    Therefore, $\forall i\ \ \hat{X}\ket{\psi_i} = \hat{Z}\ket{\psi_i}$, which implies that $\hat{Z} = \hat{X}$. Thus, $\hat{Z}\in \mathcal{A}$, and we therefore conclude that $\mathcal{A} \supset \mathcal{A}^{\prime\prime}$.
\end{enumerate}
From (1) and (2), we conclude that $\mathcal{A} = \mathcal{A}^{\prime\prime}$.\qed

\begin{lemma}\label{lem:Schur-Weyl-1}$ $
\\
    The following equality on the subset of $\mathcal{B}(\mathcal{H})$ holds:
    \begin{align}\label{eqB-13}
        {\rm span}\{\hat{X}^{\otimes n} | \hat{X}\in \mathcal{B}(\mathcal{H}_{\rm loc})\}^{\prime} = {\rm span}\{\hat{P}_{\sigma} | \sigma \in \mathfrak{S}_{n}\}.
    \end{align}
\end{lemma}
\noindent \textit{Proof of Lemma \ref{lem:Schur-Weyl-1}}
\begin{enumerate}[(1)]
    \item Since $\left[\sum_{\sigma}c_{\sigma}\hat{P}_{\sigma}, \sum_{k}a_{k}\hat{X}_{k}^{\otimes n}  \right] = 0$ for arbitrary complex coefficients $c_{\sigma}$ and $a_{k}$, it is obvious that ${\rm span}\{\hat{P}_{\sigma} | \sigma \in \mathfrak{S}_{n}\} \subset {\rm span}\{\hat{X}^{\otimes n} | \hat{X}\in \mathcal{B}(\mathcal{H}_{\rm loc})\}^{\prime}$.
    \item Next, let us prove that ${\rm span}\{\hat{X}^{\otimes n} | \hat{X}\in \mathcal{B}(\mathcal{H}_{\rm loc})\}^{\prime} \subset {\rm span}\{\hat{P}_{\sigma} | \sigma \in \mathfrak{S}_{n}\}$. From $\mathcal{A}\subset \mathcal{B} \Rightarrow \mathcal{A}^{\prime}\supset \mathcal{B}^{\prime}$, $\qty({\rm span}\mathcal{A})^{\prime} = \mathcal{A}^{\prime}$ and the von Neumann bicommutant theorem (Theorem \ref{thm:bicommutant}), it suffices to show that ${\rm span}\{\hat{X}^{\otimes n} | \hat{X}\in \mathcal{B}(\mathcal{H}_{\rm loc})\} \supset \{\hat{P}_{\sigma} | \sigma \in \mathfrak{S}_{n}\}^{\prime}$. Now, we define a symmetrizing superopeator $\mathcal{P}_{\rm sym}: \mathcal{B}(\mathcal{H})\to \mathcal{B}(\mathcal{H})$, which makes an operator permutation symmetric as follows:
    \begin{align}\label{eqB-14}
        \mathcal{P}_{\rm sym}(\hat{X}) \coloneq \frac{1}{n!}\sum_{\sigma \in \mathfrak{S}_{n}}\hat{P}_{\sigma}\hat{X}\hat{P}_{\sigma}^{\dagger}.
    \end{align}
    It is straightforward to verify that $\forall \sigma\in \mathfrak{S}_{n}\ \forall \hat{X}\in \mathcal{B}(\mathcal{H})\ \ [\hat{P}_{\sigma}, \mathcal{P}_{\rm sym}(\hat{X})] = 0$ and $\hat{X} \in \{\hat{P}_{\sigma} | \sigma \in \mathfrak{S}_{n}\}^{\prime} \Rightarrow \mathcal{P}_{\rm sym}(\hat{X}) = \hat{X}$. Thus, we have
    \begin{align}\label{eqB-15}
        {\rm Im}(\mathcal{P}_{\rm sym}) = \{\hat{P}_{\sigma} | \sigma \in \mathfrak{S}_{n}\}^{\prime},
    \end{align}
    where the image of the superoperator $\mathcal{P}_{\rm sym}$ is defined as ${\rm Im}(\mathcal{P}_{\rm sym}) = \{\mathcal{P}_{\rm sym}(\hat{X}) | \hat{X}\in \mathcal{B}(\mathcal{H})\}$. Let $\{\hat{X}^{\alpha}\}_{\alpha = 1}^{d^2}$ be a basis of $\mathcal{B}(\mathcal{H}_{\rm loc})$, where $d = {\rm dim}\mathcal{H}_{\rm loc}$. When $\hat{X}$ can be expanded in terms of the basis as $\hat{X} = \sum_{\alpha}c_{\alpha}\hat{X}^{\alpha}$, $\hat{X}^{\otimes n}$ can be expressed as
    \begin{widetext}
        \begin{align}\label{eqB-16}
        \hat{X}^{\otimes n} &= \sum_{\alpha_{1}, \dots, \alpha_{n} = 1}^{d^2} c_{\alpha_1}\cdots c_{\alpha_n}\hat{X}^{\alpha_1}\otimes \cdots \otimes \hat{X}^{\alpha_n} \notag \\
        &= \sum_{m_{i};\sum_{i}m_{i} = n}\frac{n!}{m_{1}!\cdots m_{d^2}!}\qty(\prod_{i = 1}^{d^2}c_{i}^{m_i})\mathcal{P}_{\rm sym}\qty(\bigotimes_{i = 1}^{d^2}\qty(\hat{X}^{i})^{\otimes m_i}).
    \end{align}
    \end{widetext}
    Since any derivative of $\hat{X}^{\otimes n}$ belongs to ${\rm span}\{\hat{X}^{\otimes n} | \hat{X}\in \mathcal{B}(\mathcal{H}_{\rm loc})\}^{\prime}$, we have
    \begin{align}\label{eqB-17}
        \ \ \mathcal{P}_{\rm sym}\qty(\bigotimes_{i = 1}^{d^2}\qty(\hat{X}^{i})^{\otimes m_i}) &= \frac{1}{n!}\qty(\pdv{c_{1}})^{m_1}\hspace{-3mm}\cdots \qty(\pdv{c_{d^2}})^{m_{d^2}}\hat{X}^{\otimes n}\notag \\
        &\in {\rm span}\{\hat{X}^{\otimes n} | \hat{X}\in \mathcal{B}(\mathcal{H}_{\rm loc})\}^{\prime}.
    \end{align}
    Since $\mathcal{P}_{\rm sym}\qty(\bigotimes_{i = 1}^{d^2}\qty(\hat{X}^{i})^{\otimes m_i})$ generates ${\rm Im}(\mathcal{P}_{\rm sym})$ under scalar multiplication and addition, we obtain
    \begin{align}\label{eqB-18}
        {\rm Im}(\mathcal{P}_{\rm sym}) \subset {\rm span}\{\hat{X}^{\otimes n} | \hat{X}\in \mathcal{B}(\mathcal{H}_{\rm loc})\}^{\prime}.
    \end{align}
    Substituting Eq.~\eqref{eqB-15} into Eq.~\eqref{eqB-18}, we have $\{\hat{P}_{\sigma} | \sigma \in \mathfrak{S}_{n}\}^{\prime} \subset {\rm span}\{\hat{X}^{\otimes n} | \hat{X}\in \mathcal{B}(\mathcal{H}_{\rm loc})\}^{\prime}$.
\end{enumerate}
From (1) and (2), we conclude that ${\rm span}\{\hat{X}^{\otimes n} | \hat{X}\in \mathcal{B}(\mathcal{H}_{\rm loc})\}^{\prime} = {\rm span}\{\hat{P}_{\sigma} | \sigma \in \mathfrak{S}_{n}\}$. \qed

\begin{lemma}\label{lem:Schur-Weyl-2}$ $
    \\
    The following equality on the subset of $\mathcal{B}(\mathcal{H})$ holds:
    \begin{align}\label{eqB-19}
        {\rm span}\{\hat{X}^{\otimes n} | \hat{X}\in \mathcal{B}(\mathcal{H}_{\rm loc})\} = {\rm span}\{\hat{G}^{\otimes n} | \hat{G} \in GL(\mathcal{H}_{\rm loc})\}.
    \end{align}
\end{lemma}
\noindent \textit{Proof of Lemma \ref{lem:Schur-Weyl-2}}\\
It is obvious that ${\rm span}\{\hat{X}^{\otimes n} | \hat{X}\in \mathcal{B}(\mathcal{H}_{\rm loc})\} \supset {\rm span}\{\hat{G}^{\otimes n} | \hat{G} \in GL(\mathcal{H}_{\rm loc})\}$. Since $GL(\mathcal{H}_{\rm loc})$ is dense in $\mathcal{B}(\mathcal{H}_{\rm loc})$ and $\|\hat{X}_{1}-\hat{X}_{2}\| < \epsilon \Rightarrow \|\hat{X}_{1}^{\otimes n}-\hat{X}_{2}^{\otimes n}\| < \epsilon n$, ${\rm span}\{\hat{G}^{\otimes n} | \hat{G} \in GL(\mathcal{H}_{\rm loc})\}$ is also dense in ${\rm span}\{\hat{X}^{\otimes n} | \hat{X}\in \mathcal{B}(\mathcal{H}_{\rm loc})\}$. Thus, the closure of ${\rm span}\{\hat{G}^{\otimes n} | \hat{G} \in GL(\mathcal{H}_{\rm loc})\}$ equals ${\rm span}\{\hat{X}^{\otimes n} | \hat{X}\in \mathcal{B}(\mathcal{H}_{\rm loc})\}$. However, since ${\rm span}\{\hat{G}^{\otimes n} | \hat{G} \in GL(\mathcal{H}_{\rm loc})\}$ is a vector space and thus a closed set, ${\rm span}\{\hat{X}^{\otimes n} | \hat{X}\in \mathcal{B}(\mathcal{H}_{\rm loc})\} = {\rm span}\{\hat{G}^{\otimes n} | \hat{G} \in GL(\mathcal{H}_{\rm loc})\}$.\qed\\

Theorem \ref{thm:Schur-Weyl} follows immediately from Theorem \ref{thm:bicommutant}, Lemma \ref{lem:Schur-Weyl-1} and \ref{lem:Schur-Weyl-2}. Schur-Weyl duality can be extended using the following lemma.

\begin{lemma}\label{lem:Schur-Weyl-3}$ $
    \\
    The following subsets of $\mathcal{B}(\mathcal{H})$ are all equivalent:
    \begin{enumerate}
        \item ${\rm span}\{\hat{X}^{\otimes n} | \hat{X}\in \mathcal{B}(\mathcal{H}_{\rm loc})\}$;
        \item ${\rm span}\{\hat{U}^{\otimes n} | \hat{U} \in \mathcal{B}(\mathcal{H}_{\rm loc}), \hat{U}^{\dagger}\hat{U} = \hat{I}\}$;
        \item ${\rm span }\{\hat{H}^{\otimes n} | \hat{H}\in \mathcal{B}(\mathcal{H}_{\rm loc}), \hat{H}^{\dagger} = \hat{H}\}$;
        \item ${\rm span}\{\hat{\rho}^{\otimes n} | \hat{\rho}\in \mathcal{S}(\mathcal{H}_{\rm loc})\}$.
    \end{enumerate}
\end{lemma}
\noindent \textit{Proof of Lemma \ref{lem:Schur-Weyl-3}}
\begin{enumerate}[(1)]
    \item ${\rm span}\{\hat{X}^{\otimes n} | \hat{X}\in \mathcal{B}(\mathcal{H}_{\rm loc})\}$\\
    $ = {\rm span}\{\hat{U}^{\otimes n} | \hat{U} \in \mathcal{B}(\mathcal{H}_{\rm loc}), \hat{U}^{\dagger}\hat{U} = \hat{I}\}$.\\
    It is obvious that ${\rm span}\{\hat{X}^{\otimes n} | \hat{X}\in \mathcal{B}(\mathcal{H}_{\rm loc})\} \supset {\rm span}\{\hat{U}^{\otimes n} | \hat{U} \in \mathcal{B}(\mathcal{H}_{\rm loc}), \hat{U}^{\dagger}\hat{U} = \hat{I}\}$. Now, let us show that ${\rm span}\{\hat{X}^{\otimes n} | \hat{X}\in \mathcal{B}(\mathcal{H}_{\rm loc})\} \subset {\rm span}\{\hat{U}^{\otimes n} | \hat{U} \in \mathcal{B}(\mathcal{H}_{\rm loc}), \hat{U}^{\dagger}\hat{U} = \hat{I}\}$. Recalling Lemma \ref{lem:Schur-Weyl-2}, it suffices to show that ${\rm span}\{\hat{G}^{\otimes n} | \hat{G} \in GL(\mathcal{H}_{\rm loc})\} \subset {\rm span}\{\hat{U}^{\otimes n} | \hat{U} \in \mathcal{B}(\mathcal{H}_{\rm loc}), \hat{U}^{\dagger}\hat{U} = \hat{I}\}$. A regular matrix $\hat{G}\in GL(\mathcal{H}_{\rm loc})$ can be written as $\hat{G} = e^{\hat{X}}$, where $\hat{X}\in \mathcal{B}(\mathcal{H}_{\rm loc})$. Then,
        $\hat{G}^{\otimes n} = \exp\qty(\sum_{i = 1}^{n} \hat{X}_{i})$,
    where $\hat{X}_{i}$ is an operator that acts on site $i$ as $\hat{X}$, i.e., 
    \begin{align}\label{eqB-20}
        &\hat{X}_{i}\qty(\ket{\psi_{1}}\otimes \cdots \otimes \ket{\psi_{i}}\otimes \cdots \otimes \ket{\psi_{n}})\notag \\
        &= \ket{\psi_{1}}\otimes \cdots \otimes \qty(\hat{X}\ket{\psi_{i}})\otimes \cdots \otimes \ket{\psi_n}.
    \end{align}
    $\hat{X}$ can be rewritten as $\hat{X} = \hat{H}_{\rm R} + {\rm i}\hat{H}_{\rm I}$, where $\hat{H}_{\rm R}$ and $\hat{H}_{\rm I}$ are Hermitian matrices. Let us define a parametrized unitary matrix $\hat{U}(t)$ as $\hat{U}(t) = e^{{\rm i}\hat{H}_{\rm R}t}$. Since derivatives of $\hat{U}(t)^{\otimes n}$ belong to ${\rm span}\{\hat{U}^{\otimes n} | \hat{U} \in \mathcal{B}(\mathcal{H}_{\rm loc}), \hat{U}^{\dagger}\hat{U} = \hat{I}\}$, we have
    \begin{align}\label{eqB-21}
        \sum_{i = 1}^{n}\hat{H}_{{\rm R}, i} &= -{\rm i}\left.\dv{t}\hat{U}(t)^{\otimes n}\right|_{t = 0}\notag \\
        &\in {\rm span}\{\hat{U}^{\otimes n} | \hat{U} \in \mathcal{B}(\mathcal{H}_{\rm loc}), \hat{U}^{\dagger}\hat{U} = \hat{I}\}.
    \end{align}
    Similarly, $\sum_{i=1}^{n}\hat{H}_{{\rm I}, i}$ also belongs to ${\rm span}\{\hat{U}^{\otimes n} | \hat{U} \in \mathcal{B}(\mathcal{H}_{\rm loc}), \hat{U}^{\dagger}\hat{U} = \hat{I}\}$. Thus, $\sum_{i = 1}^{n}\hat{X}_{i} \in {\rm span}\{\hat{U}^{\otimes n} | \hat{U} \in \mathcal{B}(\mathcal{H}_{\rm loc}), \hat{U}^{\dagger}\hat{U} = \hat{I}\}$. Since ${\rm span}\{\hat{U}^{\otimes n} | \hat{U} \in \mathcal{B}(\mathcal{H}_{\rm loc}), \hat{U}^{\dagger}\hat{U} = \hat{I}\}$ is an algebra, we have
    \begin{align}\label{eqB-22}
        \hat{G}^{\otimes n} &= \exp\qty(\sum_{i = 1}^{n} \hat{X}_{i}) \notag \\
        &\in {\rm span}\{\hat{U}^{\otimes n} | \hat{U} \in \mathcal{B}(\mathcal{H}_{\rm loc}), \hat{U}^{\dagger}\hat{U} = \hat{I}\}.
    \end{align}
    Therefore, we obtain ${\rm span}\{\hat{G}^{\otimes n} | \hat{G} \in GL(\mathcal{H}_{\rm loc})\} \subset {\rm span}\{\hat{U}^{\otimes n} | \hat{U} \in \mathcal{B}(\mathcal{H}_{\rm loc}), \hat{U}^{\dagger}\hat{U} = \hat{I}\}$.
    \item ${\rm span}\{\hat{X}^{\otimes n} | \hat{X}\in \mathcal{B}(\mathcal{H}_{\rm loc})\}$\\$ = {\rm span }\{\hat{H}^{\otimes n} | \hat{H}\in \mathcal{B}(\mathcal{H}_{\rm loc}), \hat{H}^{\dagger} = \hat{H}\}$.\\
    It is obvious that ${\rm span}\{\hat{X}^{\otimes n} | \hat{X}\in \mathcal{B}(\mathcal{H}_{\rm loc})\} \supset {\rm span }\{\hat{H}^{\otimes n} | \hat{H}\in \mathcal{B}(\mathcal{H}_{\rm loc}), \hat{H}^{\dagger} = \hat{H}\}$. Now, let us show that ${\rm span}\{\hat{X}^{\otimes n} | \hat{X}\in \mathcal{B}(\mathcal{H}_{\rm loc})\} \subset {\rm span }\{\hat{H}^{\otimes n} | \hat{H}\in \mathcal{B}(\mathcal{H}_{\rm loc}), \hat{H}^{\dagger} = \hat{H}\}$. We decompose $\hat{X}\in \mathcal{B}(\mathcal{H}_{\rm loc})$ in terms of Hermitian matrices as $\hat{X} = \hat{H}_{\rm R} + {\rm i}\hat{H}_{\rm I}$, and consider a Hermitian matrix $\hat{H} = c_{\rm R}\hat{H}_{\rm R} + c_{\rm I}\hat{H}_{\rm I}\ (c_{\rm R}, c_{\rm I}\in \R)$. Then, $\hat{H}^{\otimes n}$ can be rewritten as
    \begin{align}\label{eqB-23}
        \hat{H}^{\otimes n} = \sum_{\substack{n_{\rm R}, n_{\rm I};\\n_{\rm R} + n_{\rm I} = n}} \frac{n!}{n_{\rm R}! n_{\rm I}!}c_{\rm R}^{n_{\rm R}}c_{\rm I}^{n_{\rm I}}\mathcal{P}_{\rm sym}\qty(\hat{H}_{\rm R}^{\otimes n_{\rm R}}\hat{H}_{\rm I}^{\otimes n_{\rm I}}).
    \end{align}
    Since any derivative of $\hat{H}^{\otimes n}$ belongs to ${\rm span }\{\hat{H}^{\otimes n} | \hat{H}\in \mathcal{B}(\mathcal{H}_{\rm loc}), \hat{H}^{\dagger} = \hat{H}\}$, we have
    \begin{align}\label{eqB-24}
        \mathcal{P}_{\rm sym}\qty(\hat{H}_{\rm R}^{\otimes n_{\rm R}}\hat{H}_{\rm I}^{\otimes n_{\rm I}}) &= \frac{1}{n!}\qty(\pdv{c_{\rm R}})^{n_{\rm R}}\qty(\pdv{c_{\rm I}})^{n_{\rm I}}\hat{H}^{\otimes n}\notag \\
        &\hspace{-18mm}\in {\rm span }\{\hat{H}^{\otimes n} | \hat{H}\in \mathcal{B}(\mathcal{H}_{\rm loc}), \hat{H}^{\dagger} = \hat{H}\}.
    \end{align}
    Rewriting $\hat{X}^{\otimes n}$ in terms of $\hat{H}_{\rm R}$ and $\hat{H}_{\rm I}$, we obtain
    \begin{align}\label{eqB-25}
        \hat{X}^{\otimes n} &= \sum_{\substack{n_{\rm R}, n_{\rm I};\\n_{\rm R} + n_{\rm I} = n}}\frac{n!}{n_{\rm R}! n_{\rm I}!}{\rm i}^{n_{\rm I}}\mathcal{P}_{\rm sym}\qty(\hat{H}_{\rm R}^{\otimes n_{\rm R}}\hat{H}_{\rm I}^{\otimes n_{\rm I}}) \notag \\
        &\in {\rm span }\{\hat{H}^{\otimes n} | \hat{H}\in \mathcal{B}(\mathcal{H}_{\rm loc}), \hat{H}^{\dagger} = \hat{H}\}.
    \end{align}
    Therefore, ${\rm span}\{\hat{X}^{\otimes n} | \hat{X}\in \mathcal{B}(\mathcal{H}_{\rm loc})\} \subset {\rm span }\{\hat{H}^{\otimes n} | \hat{H}\in \mathcal{B}(\mathcal{H}_{\rm loc}), \hat{H}^{\dagger} = \hat{H}\}$.
    \item ${\rm span}\{\hat{X}^{\otimes n} | \hat{X}\in \mathcal{B}(\mathcal{H}_{\rm loc})\} = {\rm span}\{\hat{\rho}^{\otimes n} | \hat{\rho}\in \mathcal{S}(\mathcal{H}_{\rm loc})\}$.\\
    It is obvious that ${\rm span}\{\hat{X}^{\otimes n} | \hat{X}\in \mathcal{B}(\mathcal{H}_{\rm loc})\} \supset {\rm span}\{\hat{\rho}^{\otimes n} | \hat{\rho}\in \mathcal{S}(\mathcal{H}_{\rm loc})\}$. Now let us show that ${\rm span}\{\hat{X}^{\otimes n} | \hat{X}\in \mathcal{B}(\mathcal{H}_{\rm loc})\} \subset {\rm span}\{\hat{\rho}^{\otimes n} | \hat{\rho}\in \mathcal{S}(\mathcal{H}_{\rm loc})\}$. From what we have shown so far, it suffices to show that ${\rm span }\{\hat{H}^{\otimes n} | \hat{H}\in \mathcal{B}(\mathcal{H}_{\rm loc}), \hat{H}^{\dagger} = \hat{H}\}\subset {\rm span}\{\hat{\rho}^{\otimes n} | \hat{\rho}\in \mathcal{S}(\mathcal{H}_{\rm loc})\}$. Let $\hat{H}$ be a Hermitian matrix, and $\{\ket{\lambda}\}$ be an orthonormal basis that diagonalizes the Hermitian matrix as $\hat{H} = \sum_{\lambda}h_{\lambda}\ketbra{\lambda}{\lambda}$. Then, the Hermitian matrix can be expressed as the sum of two density matrices as follows:
    \begin{align}
        \hat{H} &= \qty(\sum_{\lambda}{\rm max}(h_{\lambda}, 0))\hat{\rho}_{+} + \qty(\sum_{\lambda}{\rm min}(h_{\lambda}, 0))\hat{\rho}_{-}, \label{eqB-26-1}\\
        \hat{\rho}_{\rm +} &= \frac{\sum_{\lambda}{\rm max}(h_{\lambda}, 0)\ketbra{\lambda}{\lambda}}{\sum_{\lambda}{\rm max}(h_{\lambda}, 0)},\label{eqB-26-2}\\
        \hat{\rho}_{\rm -} &= \frac{\sum_{\lambda}{\rm min}(h_{\lambda}, 0)\ketbra{\lambda}{\lambda}}{\sum_{\lambda}{\rm min}(h_{\lambda}, 0)}.\label{eqB-26-3}
    \end{align}
    If $\sum_{\lambda}{\rm max}(h_{\lambda}, 0) = 0$ or $\sum_{\lambda}{\rm min}(h_{\lambda}, 0) = 0$, we define $\hat{\rho}_{+}$ or $\hat{\rho}_{-}$ to be zero. Consider a positive semidefinite matrix $\hat{\rho} = c_{+}\hat{\rho}_{+} + c_{-}\hat{\rho}_{-}\ (c_{+}, c_{-}\geq 0)$. Since any derivative of $\hat{\rho}^{\otimes n}$ belongs to ${\rm span}\{\hat{\rho}^{\otimes n} | \hat{\rho}\in \mathcal{S}(\mathcal{H}_{\rm loc})\}$, we have
    \begin{align}\label{eqB-27}
        \mathcal{P}_{\rm sym}\qty(\hat{\rho}_{+}^{\otimes n_{+}}\hat{\rho}_{-}^{\otimes n_{-}}) &= \frac{1}{n!}\qty(\pdv{c_{+}})^{n_{+}}\qty(\pdv{c_{-}})^{n_{-}}\hat{\rho}^{\otimes n}\notag \\
        &\in {\rm span}\{\hat{\rho}^{\otimes n} | \hat{\rho}\in \mathcal{S}(\mathcal{H}_{\rm loc})\}.
    \end{align}
    Therefore, ${\rm span }\{\hat{H}^{\otimes n} | \hat{H}\in \mathcal{B}(\mathcal{H}_{\rm loc}), \hat{H}^{\dagger} = \hat{H}\}\subset {\rm span}\{\hat{\rho}^{\otimes n} | \hat{\rho}\in \mathcal{S}(\mathcal{H}_{\rm loc})\}$. \qed
\end{enumerate}$ $\\

From Schur-Weyl duality (Theorem \ref{thm:Schur-Weyl}), Lemma \ref{lem:Schur-Weyl-2} and \ref{lem:Schur-Weyl-3}, and the fact that $\qty({\rm span}\mathcal{A})^{\prime} = \mathcal{A}^{\prime}$, we obtain
\begin{align}\label{eqB-28}
    \{\hat{\rho}^{\otimes n} | \hat{\rho}\in \mathcal{S}(\mathcal{H}_{\rm loc})\}^{\prime} = {\rm span}\{\hat{P}_{\sigma} | \sigma \in \mathfrak{S}_{n}\},
\end{align}
which is equal to $\mathcal{B}_{\rm com}$ for spin systems and massless boson systems. Thus, Eq.~\eqref{eq2-28} in Lemma \ref{lem:commutable} is proved. On the other hand, for fermion and massive boson systems, density matrices are restricted to matrices that commute with the total number operator $\hat{N}$ due to the number-superselection rule. Suppose that a quantum i.i.d. state $\hat{\rho}_{\rm loc}^{\otimes n}$ satisfies the number-superselection rule, i.e., $[\hat{\rho}_{\rm loc}^{\otimes n}, \hat{N}] = 0$, where $\hat{N} = \sum_{i}\hat{n}_{i}$. By taking ${\rm Tr}_{\overline{i}}$ of this equation, we obtain $[\hat{\rho}_{\rm loc}, \hat{n}] = 0$. Conversely, if $[\hat{\rho}_{\rm loc}, \hat{n}] = 0$ holds, $\hat{\rho}_{\rm loc}^{\otimes n}$ obeys the number-superselection rule. Therefore, Eq.~\eqref{eq2-29} in Lemma \ref{lem:commutable} is restated as follows:
\begin{align}\label{eqB-29}
    \{\hat{\rho}^{\otimes n} | \hat{\rho}\in \mathcal{S}(\mathcal{H}_{\rm loc}), [\hat{\rho}, \hat{n}] = 0\}^{\prime} = \{\hat{P}_{\sigma\in \mathfrak{S}_{n}}, \hat{n}_{i \in \Lambda}\}^{\prime \prime},
\end{align}
where $\hat{n}$ is the number operator on $\mathcal{H}_{\rm loc}$.\\

\noindent\textit{Proof of Eq.~\eqref{eqB-29}}\\
The proof goes in three steps. First, we prove that ${\rm span}\{\hat{X}^{\otimes n} | \hat{X} \in \mathcal{B}(\mathcal{H}_{\rm loc}), [\hat{X}, \hat{n}] = 0\} = \{\hat{P}_{\sigma \in \mathfrak{S}_{n}}, \hat{n}_{i \in \Lambda}\}^{\prime}$. Then, we show that ${\rm span}\{\hat{X}^{\otimes n} | \hat{X} \in \mathcal{B}(\mathcal{H}_{\rm loc}), [\hat{X}, \hat{n}] = 0\} = {\rm span}\{\hat{H}^{\otimes n} | \hat{H} \in \mathcal{B}(\mathcal{H}_{\rm loc}), \hat{H}^{\dagger} = \hat{H}, [\hat{H}, \hat{n}] = 0\}$. Finally, we show that ${\rm span}\{\hat{H}^{\otimes n} | \hat{H} \in \mathcal{B}(\mathcal{H}_{\rm loc}), \hat{H}^{\dagger} = \hat{H}, [\hat{H}, \hat{n}] = 0\} = {\rm span}\{\hat{\rho}^{\otimes n} | \hat{\rho} \in \mathcal{S}(\mathcal{H}_{\rm loc}), [\hat{\rho}, \hat{n}] = 0\}$.
\begin{enumerate}[(1)]
    \item ${\rm span}\{\hat{X}^{\otimes n} | \hat{X} \in \mathcal{B}(\mathcal{H}_{\rm loc}), [\hat{X}, \hat{n}] = 0\}$\\ $ = \{\hat{P}_{\sigma \in \mathfrak{S}_{n}}, \hat{n}_{i \in \Lambda}\}^{\prime}$\\
    It is obvious that ${\rm span}\{\hat{X}^{\otimes n} | \hat{X} \in \mathcal{B}(\mathcal{H}_{\rm loc}), [\hat{X}, \hat{n}] = 0\} \subset \{\hat{P}_{\sigma \in \mathfrak{S}_{n}}, \hat{n}_{i \in \Lambda}\}^{\prime}$. Now, let us show that ${\rm span}\{\hat{X}^{\otimes n} | \hat{X} \in \mathcal{B}(\mathcal{H}_{\rm loc}), [\hat{X}, \hat{n}] = 0\} \supset \{\hat{P}_{\sigma \in \mathfrak{S}_{n}}, \hat{n}_{i \in \Lambda}\}^{\prime}$. Let us define $\mathcal{N}\subset \mathcal{B}(\mathcal{H}_{\rm loc})$ as a set of operators that commute with the number operator $\hat{n}$, i.e.,
    \begin{align}\label{eqB-30}
        \mathcal{N} = \{\hat{X} | \hat{X}\in \mathcal{B}(\mathcal{H}_{\rm loc}), [\hat{X}, \hat{n}] = 0\}.
    \end{align}
    Since $\mathcal{N}$ is a vector space, we choose a basis of $\mathcal{B}\qty(\mathcal{H}_{\rm loc})$, $\{\hat{X}^{\alpha}\}_{\alpha = 1}^{d^2}$ so that $\{\hat{X}^{\alpha}\}_{\alpha = 1}^{r}$ becomes a basis of $\mathcal{N}$, where $r = {\rm dim}\mathcal{N}$. The operators on $\mathcal{H}$ that commute with $\hat{n}_{i}$ for all $i\in \Lambda$ are limited to $\mathcal{N}^{\otimes n}$. Using the symmetrizing superoperator $\mathcal{P}_{\rm sym}$ defined in Eq.~\eqref{eqB-14}, we have
    \begin{align}\label{eqB-31}
        \mathcal{P}_{\rm sym}(\mathcal{N}^{\otimes n})&\coloneq \{\mathcal{P}_{\rm sym}(\hat{X}) | \hat{X}\in \mathcal{N}^{\otimes n}\}\notag \\
        &= \{\hat{P}_{\sigma \in \mathfrak{S}_{n}}, \hat{n}_{i\in \Lambda}\}^{\prime}.
    \end{align}
    Let $\hat{X} \in \mathcal{B}(\mathcal{H}_{\rm loc})$ be an operator that commutes with $\hat{n}$ and can be expanded in terms of the basis as $\hat{X} = \sum_{\alpha = 1}^{r} c_{\alpha}\hat{X}^{\alpha}$. Since any derivative of $\hat{X}^{\otimes n}$ belongs to ${\rm span}\{\hat{X}^{\otimes n} | \hat{X} \in \mathcal{B}(\mathcal{H}_{\rm loc}), [\hat{X}, \hat{n}] = 0\}$, we have
    \begin{align}\label{eqB-32}
        \quad&\mathcal{P}_{\rm sym}\qty(\bigotimes_{i = 1}^{r}\qty(\hat{X}^{i})^{\otimes m_{i}}) = \frac{1}{n!}\qty(\pdv{c_{1}})^{m_{1}}\qty(\pdv{c_{r}})^{m_{r}}\hat{X}^{\otimes n}\notag \\
        &\quad \in {\rm span}\{\hat{X}^{\otimes n} | \hat{X} \in \mathcal{B}(\mathcal{H}_{\rm loc}), [\hat{X}, \hat{n}] = 0\},
    \end{align}
    where $\sum_{i}m_{i} = n$. Since $\mathcal{P}_{\rm sym}\qty(\bigotimes_{i = 1}^{r}\qty(\hat{X}^{i})^{\otimes m_{i}})$ generates $\mathcal{P}_{\rm sym}(\mathcal{N}^{\otimes n})$ under scalar multiplication and addition, 
    \begin{align}\label{eqB-33}
        \mathcal{P}_{\rm sym}(\mathcal{N}^{\otimes n}) \subset {\rm span}\{\hat{X}^{\otimes n} | \hat{X} \in \mathcal{B}(\mathcal{H}_{\rm loc}), [\hat{X}, \hat{n}] = 0\}.
    \end{align}
    Substituting Eq.~\eqref{eqB-31} into Eq.~\eqref{eqB-33}, we obtain $\{\hat{P}_{\sigma \in \mathfrak{S}_{n}}, \hat{n}_{i\in \Lambda}\}^{\prime} \subset {\rm span}\{\hat{X}^{\otimes n} | \hat{X} \in \mathcal{B}(\mathcal{H}_{\rm loc}), [\hat{X}, \hat{n}] = 0\}$.
    \item ${\rm span}\{\hat{X}^{\otimes n} | \hat{X} \in \mathcal{B}(\mathcal{H}_{\rm loc}), [\hat{X}, \hat{n}] = 0\}$\\$ = {\rm span}\{\hat{H}^{\otimes n} | \hat{H} \in \mathcal{B}(\mathcal{H}_{\rm loc}), \hat{H}^{\dagger} = \hat{H}, [\hat{H}, \hat{n}] = 0\}$.\\
    It is obvious that ${\rm span}\{\hat{X}^{\otimes n} | \hat{X} \in \mathcal{B}(\mathcal{H}_{\rm loc}), [\hat{X}, \hat{n}] = 0\} \supset {\rm span}\{\hat{H}^{\otimes n} | \hat{H} \in \mathcal{B}(\mathcal{H}_{\rm loc}), \hat{H}^{\dagger} = \hat{H}, [\hat{H}, \hat{n}] = 0\}$. Now, let us show that ${\rm span}\{\hat{X}^{\otimes n} | \hat{X} \in \mathcal{B}(\mathcal{H}_{\rm loc}), [\hat{X}, \hat{n}] = 0\} \subset {\rm span}\{\hat{H}^{\otimes n} | \hat{H} \in \mathcal{B}(\mathcal{H}_{\rm loc}), \hat{H}^{\dagger} = \hat{H}, [\hat{H}, \hat{n}] = 0\}$. We can decompose $\hat{X}\in \mathcal{B}(\mathcal{H}_{\rm loc})$ that commutes with the number operator $\hat{n}$ in terms of Hermitian operators as $\hat{X} = \hat{H}_{\rm R} + {\rm i}\hat{H}_{\rm I}$. Since $\hat{H}_{\rm R} = (\hat{X} + \hat{X}^{\dagger})/2$ and $\hat{H}_{\rm I} = (\hat{X} - \hat{X}^{\dagger})/2{\rm i}$, both $\hat{H}_{\rm R}$ and $\hat{H}_{\rm I}$ commute with $\hat{n}$. Then, by following the same procedure as the proof of Lemma \ref{lem:Schur-Weyl-3}, we obtain ${\rm span}\{\hat{X}^{\otimes n} | \hat{X} \in \mathcal{B}(\mathcal{H}_{\rm loc}), [\hat{X}, \hat{n}] = 0\} \subset {\rm span}\{\hat{H}^{\otimes n} | \hat{H} \in \mathcal{B}(\mathcal{H}_{\rm loc}), \hat{H}^{\dagger} = \hat{H}, [\hat{H}, \hat{n}] = 0\}$.
    \item ${\rm span}\{\hat{H}^{\otimes n} | \hat{H} \in \mathcal{B}(\mathcal{H}_{\rm loc}), \hat{H}^{\dagger} = \hat{H}, [\hat{H}, \hat{n}] = 0\}$\\$ = {\rm span}\{\hat{\rho}^{\otimes n} | \hat{\rho} \in \mathcal{S}(\mathcal{H}_{\rm loc}), [\hat{\rho}, \hat{n}] = 0\}$.\\
    It is obvious that ${\rm span}\{\hat{H}^{\otimes n} | \hat{H} \in \mathcal{B}(\mathcal{H}_{\rm loc}), \hat{H}^{\dagger} = \hat{H}, [\hat{H}, \hat{n}] = 0\} \supset {\rm span}\{\hat{\rho}^{\otimes n} | \hat{\rho} \in \mathcal{S}(\mathcal{H}_{\rm loc}), [\hat{\rho}, \hat{n}] = 0\}$. Now, let us show that ${\rm span}\{\hat{H}^{\otimes n} | \hat{H} \in \mathcal{B}(\mathcal{H}_{\rm loc}), \hat{H}^{\dagger} = \hat{H}, [\hat{H}, \hat{n}] = 0\} \subset {\rm span}\{\hat{\rho}^{\otimes n} | \hat{\rho} \in \mathcal{S}(\mathcal{H}_{\rm loc}), [\hat{\rho}, \hat{n}] = 0\}$. Since $\hat{H}$ and $\hat{n}$ commute each other, there exists an orthonormal basis $\{\ket{\lambda}\}$ that simultaneously diagonalizes $\hat{H}$ and $\hat{n}$ as $\hat{H} = \sum_{\lambda}h_{\lambda} \ketbra{\lambda}{\lambda}$ and $\hat{n} = \sum_{\lambda}n_{\lambda}\ketbra{\lambda}{\lambda}$. Then, the Hermitian matrix can be expressed as the sum of two density matrices as in Eqs.~\eqref{eqB-26-1} - \eqref{eqB-26-3}, where $\hat{\rho}_{+}$ and $\hat{\rho}_{-}$ both commute with $\hat{n}$. Following the same procedure as the proof of Lemma \ref{lem:Schur-Weyl-3}, we obtain ${\rm span}\{\hat{H}^{\otimes n} | \hat{H} \in \mathcal{B}(\mathcal{H}_{\rm loc}), \hat{H}^{\dagger} = \hat{H}, [\hat{H}, \hat{n}] = 0\} \subset {\rm span}\{\hat{\rho}^{\otimes n} | \hat{\rho} \in \mathcal{S}(\mathcal{H}_{\rm loc}), [\hat{\rho}, \hat{n}] = 0\}$.
\end{enumerate}
It follows from (1) to (3) that ${\rm span}\{\hat{\rho}^{\otimes n} | \hat{\rho} \in \mathcal{S}(\mathcal{H}_{\rm loc}), [\hat{\rho}, \hat{n}] = 0\} = \{\hat{P}_{\sigma \in \mathfrak{S}_{n}, \hat{n}_{i\in \Lambda}}\}^{\prime\prime}$. Taking the commutant of both side and using $\qty({\rm span}\mathcal{A})^{\prime} = \mathcal{A}^{\prime}$, we finally obtain $\{\hat{\rho}^{\otimes n} | \hat{\rho}\in \mathcal{S}(\mathcal{H}_{\rm loc}), [\hat{\rho}, \hat{n}] = 0\}^{\prime} = \{\hat{P}_{\sigma\in \mathfrak{S}_{n}}, \hat{n}_{i \in \Lambda}\}^{\prime \prime}$. \qed

\section{Relationship between the sufficient condition in Theorem \ref{thm:sufficient} and the equivalent condition in Theorem \ref{thm:equivalent_sec3}}\label{sec_SM_C}
We prove in this section that if an open quantum many-body system whose Hamiltonian consists of at most 2-body terms satisfies the sufficient condition in Theorem \ref{thm:sufficient}, the equivalent condition in Theorem \ref{thm:equivalent_sec3} is satisfied. We formulate this in the following lemma.

\begin{lemma}\label{lem:conditions}
    Suppose the Hamiltonian $\hat{H}$ consists of at most 2-body terms. 
    Consider an open quantum many-body system that satisfies the following conditions:
    \begin{enumerate}[(a)]
        \item The Hamiltonian can be written as
    \begin{align}\label{eqC-1}
        \hat{H} = \hat{H}_{\rm com} + \sum_{i}\hat{h}_{i},
    \end{align}
    where $\hat{H}_{\rm com}\in \mathcal{B}_{\rm com}$, which is defined in Lemma \ref{lem:commutable}, and $\hat{h}_{i}$'s are uniform 1-local terms.
    \item The 1-local Lindblad operators $\hat{L}_{i}^{(\alpha)}$'s are uniform.
    \end{enumerate}
    Then, the system also satisfies the following (c) and (d):
    \begin{enumerate}[(a)]
    \setcounter{enumi}{2}
        \item $\hat{H}_{ij} \in \mathcal{B}_{\rm com}$.
        \item The single-site Lindblad superoperator $\mathcal{L}_{i}$ is spatially uniform.
    \end{enumerate}
\end{lemma}

\noindent \textit{Proof of Lemma \ref{lem:conditions}}\\
Since $\hat{H}$ consists of at most 2-body terms, 
\begin{align}\label{eqC-2}
    \hat{H}_{\rm com}\in {\rm span}\{\hat{P}_{ij} | (i, j)\in \Lambda_{2}\}
\end{align}
for spin systems and massless boson systems, and
\begin{align}\label{eqC-3}
    \hat{H}_{\rm com} \in \qty(\{\hat{P}_{ij} | (i, j)\in \Lambda_{2}\}\cup\{ \hat{n}_{i\in \Lambda}\})^{\prime\prime}
\end{align}
for fermion and massive boson systems subject to the number-superselection rule. We now examine the terms that arise in the expression of $\hat{H}_{ij}$ defined by Eq.~\eqref{eq2-2}. Since we need to calculate the partial trace of $\hat{H}_{\rm com}$, let us first calculate the partial trace of the permutation operator.
For spin systems and Bose systems, where operators on different sites commute each other, the following equation holds for arbitrary $\ket{\psi}, \ket{\phi} \in \mathcal{H}_{\rm loc}$:
\begin{align}\label{eqC-4}
    \hat{P}_{ij}\qty(\ket{\psi}_{i}\otimes \ket{\phi}_{j}) = \ket{\phi}_{i}\otimes \ket{\psi}_{j},
\end{align}
where the subscripts $i, j$ in the kets are the site labels. We omit these site labels below. Let $\{\ket{\psi_{k}}\}_{1 \leq k \leq d}$ be an orthonormal basis of $\mathcal{H}_{\rm loc}$. Then, we have
\begin{align}\label{eqC-5}
    \mel{\psi_{m}}{{\rm Tr}_{i}[\hat{P}_{ij}]}{\psi_{n}} &= \sum_{k}\qty(\bra{\psi_{k}}\otimes \bra{\psi_{m}})\hat{P}_{ij}\qty(\ket{\psi_{k}}\otimes \ket{\psi_{n}})\notag \\
    &= \sum_{k}\delta_{kn}\delta_{mk} = \delta_{mn}.
\end{align}
Therefore, we obtain
\begin{align}\label{eqC-6}
    {\rm Tr}_{i}[\hat{P}_{ij}] = \hat{I},
\end{align}
for spin systems and Bose systems. On the other hand, Eq.~\eqref{eqC-4} does not hold for Fermi systems. For example, let us define the canonical ordering of the creation operators to be $(\hat{c}_{i\uparrow}^{\dagger}, \hat{c}_{i\downarrow}^{\dagger}, \hat{c}_{j\uparrow}^{\dagger}, \hat{c}_{j\downarrow}^{\dagger})$ for $i < j$. Then, recalling $\hat{P}_{ij}\hat{c}_{is}^{\dagger} = \hat{c}_{js}^{\dagger}\hat{P}_{ij}$, where $s$ represents the internal degrees of freedom such as spins, we have $\hat{P}_{ij}\qty(\hat{c}_{is}^{\dagger}
\hat{c}_{js}^{\dagger}\ket{0}) = -\hat{c}_{is}^{\dagger}\hat{c}_{js}^{\dagger}\ket{0}$. This negative sign causes the difference in the result of ${\rm Tr}_{i}[\hat{P}_{ij}]$. Let $\bm{n}_{i}$ be a vector whose component $n_{is} \in \{0, 1\}$ denotes the eigenvalue of $\hat{n}_{is}$, and let $n_{i}$ be the eigenvalue of $\hat{n}_{i} = \sum_{s}\hat{n}_{is}$, which is equal to the number of nonzero components in $\bm{n}_{i}$. Then, we have
\newpage
\begin{align}\label{eqC-7}
    &\quad \bra{0}\qty(\prod_{s = d}^{1}\qty(\hat{c}_{js})^{n_{js}^{\prime}}){\rm Tr}_{i}[\hat{P}_{ij}]\qty(\prod_{s = 1}^{d}\qty(\hat{c}_{js}^{\dagger})^{n_{js}})\ket{0} \notag \\
    &= \sum_{\bm{n}_{i}}\bra{0}\qty(\prod_{s = d}^{1}\qty(\hat{c}_{js})^{n_{js}^\prime})\qty(\prod_{s = d}^{1}\qty(\hat{c}_{is})^{n_{is}})\hat{P}_{ij}\notag \\
    &\quad \quad \quad \quad \quad \cdot \qty(\prod_{s = 1}^{d}\qty(\hat{c}_{is}^{\dagger})^{n_{is}})\qty(\prod_{s = 1}^{d}\qty(\hat{c}_{js}^{\dagger})^{n_{js}})\ket{0} \notag \\
    &= \sum_{\bm{n}_{i}}\bra{0}\qty(\prod_{s = d}^{1}\qty(\hat{c}_{js})^{n_{js}^{\prime}})\qty(\prod_{s = d}^{1}\qty(\hat{c}_{is})^{n_{is}})\notag \\
    &\quad \quad \quad \quad \quad \cdot\qty(\prod_{s = 1}^{d}\qty(\hat{c}_{js}^{\dagger})^{n_{is}})\qty(\prod_{s = 1}^{d}\qty(\hat{c}_{is}^{\dagger})^{n_{js}})\ket{0}\notag \\
    &=\delta_{\bm{n}_{j}\bm{n}_{j}^{\prime}}(-1)^{n_{j}^2} = \delta_{\bm{n}_{j}\bm{n}_{j}^{\prime}}(-1)^{n_{j}},
\end{align}
where $\delta_{\bm{n}_{j}\bm{n}_{j}^{\prime}} = 1$ if $\bm{n}_{j} = \bm{n}_{j}^{\prime}$ and zero otherwise. The sign of Eq.~\eqref{eqC-7} is not affected by the occupation of sites other than $j$ in the ket and bra states that sandwich ${\rm Tr}_{i}[\hat{P}_{ij}]$. Therefore, the partial trace of the permutation operator $\hat{P}_{ij}$ is equal to the parity operator, i.e.,
\begin{align}\label{eqC-8}
    {\rm Tr}_{i}[\hat{P}_{ij}] = (-1)^{\hat{n}_{j}},
\end{align}
for Fermi systems. Now, let us rewrite the Hamiltonian Eq.~\eqref{eqC-1} using $\hat{H}_{ij}$'s and $\hat{H}_{i}$'s, which are defined in Eqs.~\eqref{eq2-2} and \eqref{eq2-3}, respectively. For spin systems and massless boson systems, $\hat{H}_{\rm com}$ can generally be expressed as 
\begin{align}\label{eqC-9}
    \hat{H}_{\rm com} = \sum_{(i, j)}c_{ij}\hat{P}_{ij} + c_{I}\hat{I},
\end{align}
where $c_{ij}$'s and $c_{I}$ are real numbers. Therefore, we obtain
\begin{align}
    \hat{H}_{ij} &= c_{ij}\qty(\hat{P}_{ij} - \frac{1}{d}\hat{I}),\label{eqC-10-1}\\
    \hat{H}_{i} &= \hat{h}_{i} - \frac{{\rm Tr}_{i}[\hat{h}_{i}]}{d}.\label{eqC-10-2}
\end{align}
Thus, $\hat{H}_{ij}$ belongs to $\mathcal{B}_{\rm com}$ and the single-site Lindblad superoperator $\mathcal{L}_{i}$ is spatially uniform. For massive boson systems, using $\hat{P}_{ij}\hat{n}_{i} = \hat{n}_{j}\hat{P}_{ij}$, $\hat{H}_{\rm com}$ can generally be written as
\begin{align}\label{eqC-11}
    \hat{H}_{\rm com} = \sum_{\substack{(i, j)\\ 0 \leq x_{i}, x_{j} \leq n_{\rm max}\\
    z\in \{0, 1\}}}c_{ij}^{(x_i, x_j, z)}\hat{n}_{i}^{x_i}\hat{n}_{j}^{x_j}\hat{P}_{ij}^{z},
\end{align}
where $n_{\rm max}$ is the maximal eigenvalue of the single-site number operator $\hat{n}$ and $c_{ij}^{(x_i,x_j, z)}$'s are complex numbers. Therefore, we obtain
\begin{widetext}
    \begin{align}
    \hat{H}_{ij} &= \sum_{0 \leq x_{i}, x_{j} \leq n_{\rm max}} c_{ij}^{(x_i, x_j, 1)}\qty(\hat{n}_{i}^{x_i}\hat{n}_{j}^{x_j}\hat{P}_{ij} - \frac{\hat{n}_{i}^{x_i + x_j}}{d} - \frac{\hat{n}_{j}^{x_i + x_j}}{d} + \frac{{\rm tr}[\hat{n}^{x_i + x_j}]}{d^2})\notag \\
    &\quad + \sum_{1 \leq x_{i}, x_{j} \leq n_{\rm max}}c_{ij}^{(x_i, x_j, 0)}\qty(\hat{n}_{i}^{x_i}\hat{n}_{j}^{x_j} - \frac{{\rm tr}[\hat{n}^{x_j}]\hat{n}_{i}^{x_i}}{d} - \frac{{\rm tr}[\hat{n}^{x_i}]\hat{n}_{j}^{x_j}}{d} + \frac{{\rm tr}[\hat{n}^{x_i}]\cdot {\rm tr}[\hat{n}^{x_j}]}{d^2}),\label{eqC-12-1}\\
    \hat{h}_{i}^{\prime}  &=  \sum_{k\neq i}\left( -\frac{1}{d}\qty(\sum_{0 \leq x_{i}, x_{k} \leq n_{\rm max}}c_{ik}^{(x_i, x_k, 1)}\hat{n}_{i}^{x_i + x_k} + \sum_{1 \leq x_{i}, x_{k} \leq n_{\rm max}}\hspace{-2mm}c_{ik}^{(x_i, x_k, 0)}{\rm tr}[\hat{n}^{x_k}]\hat{n}_{i}^{x_i}) + \sum_{x_{i} = 1}^{n_{\rm max}}c_{ik}^{(x_i, 0, 0)}\hat{n}_{i}^{x_i}\right) +\hat{h}_{i},\label{eqC-12-2}\\
    \hat{H}_{i} &= \hat{h}_{i}^{\prime} - \frac{{\rm Tr}_{i}[\hat{h}_{i}^{\prime}]}{d}. \label{eqC-12-3}
\end{align}
\end{widetext}
Here, ${\rm tr}[\bullet]$ denotes the trace over the single-site Hilbert space $\mathcal{H}_{\rm loc}$, whereas ${\rm Tr}[\bullet]$ denotes the trace over the whole Hilbert space $\mathcal{H} = \mathcal{H}_{\rm loc}^{\otimes n}$. Thus, $\hat{H}_{ij}$ belongs to $\mathcal{B}_{\rm com}$. The single-site Lindblad superoperator $\mathcal{L}_{i}$ can be written as the sum of a spatially uniform single-site superoperator $\ell_{i}$ and the commutator with the polynomial of the number operators, where $\ell_{i}$ is defined as follows:
\begin{align}\label{eqC-13}
    \ell_{i}(\bullet) = -{\rm i}[\hat{h}_{i}, \bullet] + \sum_{\alpha \in \Gamma_{i}}\mathcal{D}_{\hat{L}_{i}^{\alpha}}(\bullet).
\end{align}
Recall that the uniformity of single-site Lindblad superoperator is defined by $\forall i, j\ \ \mathcal{L}_{i}(\hat{\rho}_{\rm loc}) = \mathcal{L}_{j}(\hat{\rho}_{\rm loc})$ for every single-site density matrix $\hat{\rho}_{\rm loc}$ that commutes with the single-site number operator $\hat{n}$, for systems subject to the number-superselection rule. If $[\hat{\rho}_{\rm loc}, \hat{n}] = 0$, $\mathcal{L}_{i}(\hat{\rho}_{\rm loc}) = \ell_{i}(\hat{\rho}_{\rm loc})$, since the difference between $\mathcal{L}_{i}$ and $\ell_{i}$ is the commutator with the polynomial of the number operator. Therefore, $\mathcal{L}_{i}$ is a spatially uniform single-site Lindblad operator. Finally, let us consider Fermi systems. The general description of $\hat{H}_{\rm com}$ is given in Eq.~\eqref{eqC-11}. Then, $\hat{H}_{ij}$ and $\hat{H}_{i}$ can be written as follows:
\begin{widetext}
    \begin{align}
        \hat{H}_{ij} &= \sum_{0\leq x_{i}, x_{j} \leq n_{\rm max}}c_{ij}^{(x_{i}, x_{j}, 1)}\qty(\hat{n}_{i}^{x_i}\hat{n}_{j}^{x_j}\hat{P}_{ij} - \frac{(-1)^{\hat{n}_i}\hat{n}_{i}^{x_{i} + x_{j}}}{d}- \frac{(-1)^{\hat{n}_j}\hat{n}_{j}^{x_{i} + x_{j}}}{d} + \frac{{\rm tr}[(-1)^{\hat{n}}\hat{n}^{x_{i} + x_{j}}]}{d^2})\notag \\
        &\quad + \sum_{1 \leq x_{i}, x_{j}\leq n_{\rm max}}c_{ij}^{(x_i, x_j, 0)}\qty(\hat{n}_{i}^{x_i}\hat{n}_{j}^{x_j} - \frac{{\rm tr}[\hat{n}^{x_j}]\hat{n}_{i}^{x_i}}{d} - \frac{{\rm tr}[\hat{n}^{x_i}]\hat{n}_{j}^{x_j}}{d} + \frac{{\rm tr}[\hat{n}^{x_i}]\cdot {\rm tr}[\hat{n}^{x_j}]}{d^2}),\label{eqC-14-1}\\
    \hat{h}_{i}^{\prime}  &=  \sum_{k\neq i}\qty( -\frac{1}{d}\qty(\sum_{0 \leq x_{i}, x_{k} \leq n_{\rm max}}\hspace{-5mm}c_{ik}^{(x_i, x_k, 1)}(-1)^{\hat{n}_{i}}\hat{n}_{i}^{x_i + x_k} + \hspace{-5mm} \sum_{1 \leq x_{i}, x_{k} \leq n_{\rm max}}\hspace{-5mm}c_{ik}^{(x_i, x_k, 0)}{\rm tr}[\hat{n}^{x_k}]\hat{n}_{i}^{x_i}) + \sum_{x_{i} = 1}^{n_{\rm max}}c_{ik}^{(x_i, 0, 0)}\hat{n}_{i}^{x_i})+\hat{h}_{i},\label{eqC-14-2}\\
    \hat{H}_{i} &= \hat{h}_{i}^{\prime} - \frac{{\rm Tr}_{i}[\hat{h}_{i}^{\prime}]}{d}. \label{eqC-14-3}
    \end{align}
\end{widetext}
Thus, $\hat{H}_{ij}$ belongs to $\mathcal{B}_{\rm com}$, and since the single-site Lindblad superoperator $\mathcal{L}_{i}$ can be written as the sum of spatially uniform single-site superoperator $\ell_{i}$ and the commutator of the polynomial of the number operators, $\mathcal{L}_{i}$ is spatially uniform. Therefore, the conditions (c) and (d) in Lemma \ref{lem:conditions} hold for any systems that satisfy conditions (a) and (b). \qed

\bibliography{paper}

\end{document}